\documentclass[a4paper,12pt]{report}
\usepackage[utf8]{inputenc}
\usepackage[brazil]{babel}
\usepackage{epsfig}
\usepackage{subfigure}
\usepackage{float}
\usepackage{latexsym}
\usepackage{graphicx}
\usepackage{amssymb}
\usepackage{amsmath}
\usepackage{epsfig}
\usepackage{amsfonts}
\begin{document}
\hyphenation{des-cri-tas fe-cha-das se-guin-tes a-ces-sar ne-ces-sa-ria-men-te ris-co Hawking Vishveshwara a-pre-sen-ta-da es-co-lher-mos Riemann Ricci des-co-nhe-ci-das re-pre-sen-ta-do ou-tros cau-sa-li-da-de mo-men-tum des-cre-ve cons-tan-tes po-si-ti-vo des-cre-vem des-cre-ven-do de-cla-ra-dos dia-go-nais co-nhe-ci-das ne-ga-ti-vo di-fe-ren-te-men-te a-pre-sen-ta-das lo-ca-li-za-das a-na-li-sar ins-ta-bi-li-da-de ins-ta-bi-li-da-des as-sin-to-ti-ca-men-te or-to-go-nal}
\thispagestyle{empty}
\begin{center}
{\large{Universidade de São Paulo\\
Instituto de Física}}

\vspace{2cm}
{\bf \Large{Análise de geometrias com curvas fechadas no tempo}}\\
\vspace{2cm}
{\bf Alan Bendasoli Pavan\\
Orientador: Prof. Dr. Elcio Abdalla}
\end{center}

\vspace{1cm}


\begin{flushright}
\begin{minipage}[t]{7.5cm}
\hrulefill\\
Tese de doutorado apresentada ao Instituto de Física da Universidade de São Paulo para a obtenção do título de Doutor em Ciências

\hrulefill
\end{minipage}

\end{flushright}

\vspace{1cm}
\noindent
{\bf Comissão Examinadora:}\\
Prof. Dr. Elcio Abdalla (IF - USP)\\
Prof. Dr. Alberto Vazquez Saa (IMECC - UNICAMP)\\
Prof. Dr. Jorge Ernesto Horvath (IAG - USP)\\
Prof. Dr. George Emanuel Avraam Matsas (IFT - UNESP)\\
Prof. Dr. Ivano Damião Soares (CBPF)
\vspace{1.0cm}
\begin{center}
São Paulo\\
2010
\end{center}
\newpage

\thispagestyle{empty}
\begin{figure}
\includegraphics[width=1.0\textwidth]{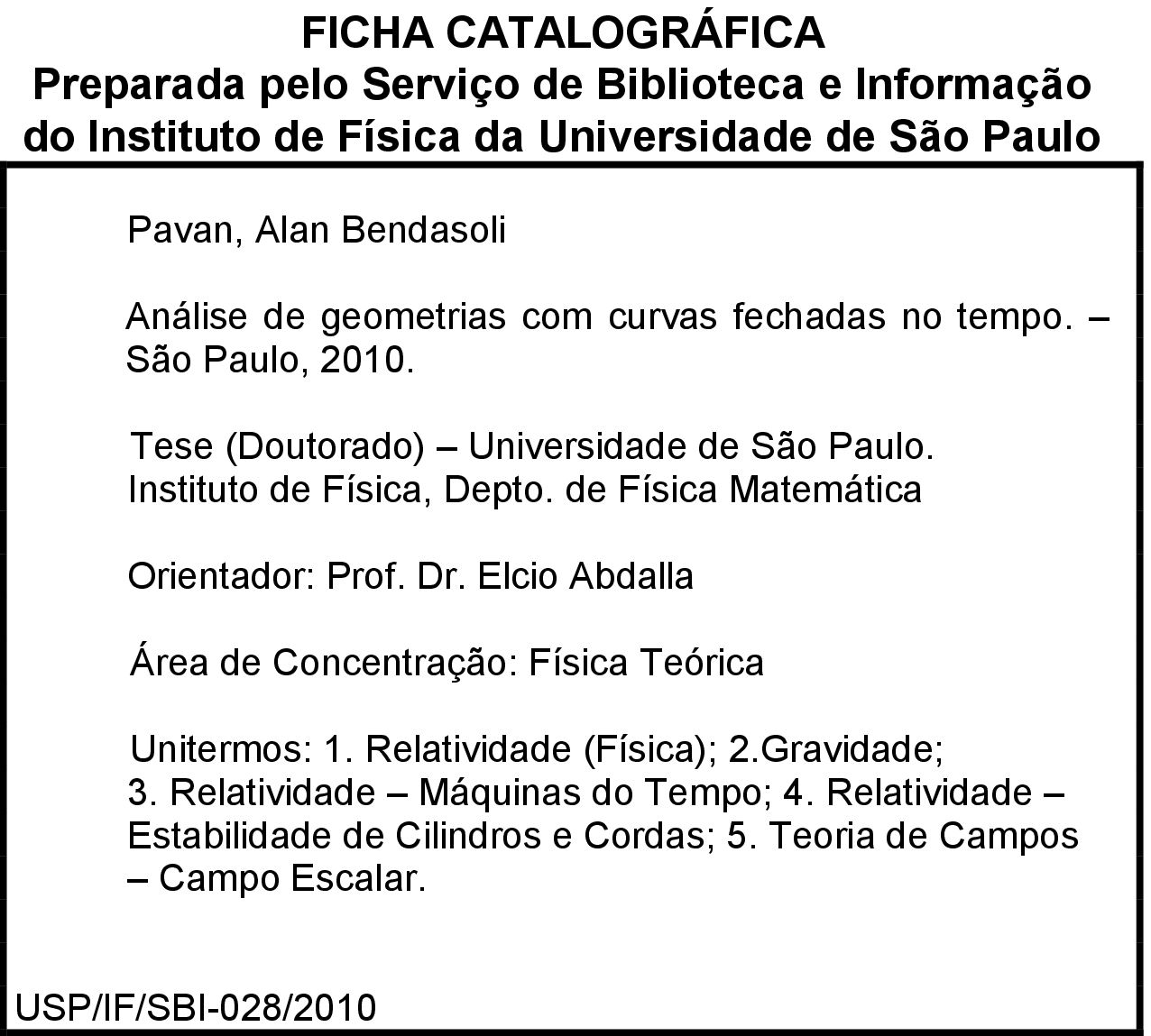}
\end{figure}

\chapter*{Resumo}

Neste trabalho dedicamo-nos à análise da presença de curvas fechadas tipo tempo em espaços-tempos cilindricamente simétricos e estacionários $\mathcal{C}_{t}$. A equação de movimento que descreve a evolução de um campo escalar massivo nos espaços-tempos $\mathcal{C}_{t}$ é apresentada.

Uma classe de espaços-tempos descrevendo cordas e cilindros cósmicos que admitem a presença de curvas fechadas tipo tempo é estudada em detalhes. A região não-causal desses espaços-tempos se revela acessível tanto a partículas massivas quanto a fótons. Curvas geodésicas e curvas fechadas tipo tempo são obtidas e investigadas. Essencialmente, dois tipos de órbitas descritas por partículas massivas e fótons foram observadas: órbitas confinadas e órbitas espalhadas. Os cones de luz evidenciaram claramente a intersecção de futuro e passado na região não-causal.

Soluções exatas das equações de movimento do campo escalar se propagando nos espaços-tempos das cordas e cilindros são apresentadas. No caso dos cilindros estático e em rotação os modos quasinormais de oscilação do campo escalar foram calculados. A presença de modos instáveis foi observada em alguns casos. Observamos, também, que tanto na corda estática quanto na corda em rotação o campo escalar não apresentam modos quasinormais de oscilação.

Concluimos com a proposta de uma conjectura relacionando curvas fechadas tipo tempo e instabilidades do espaço-tempo.


\chapter*{Abstract}

This work deals with the analysis of cylindrically symmetric and stationary space-times $\mathcal{C}_{t}$ with closed timelike curves. The equation of motion describing the evolution of a massive scalar field in a $\mathcal{C}_{t}$ space-time is obtained.

A class of space-times with closed timelike curves describing cosmic strings and cylinders is studied in detail. In such space-times, both massive particles as well as photons can reach the non-causal region. Geodesics and closed timelike curves are calculated and investigated.
We have observed that massive particles and photons describe, essentially, two kinds of trajectories: confined orbits and scattering states. The analysis of the light cones show us clearly  the intersection between future and past inside the non-causal region.

Exact solutions for the equation of motion of massive scalar field propagating in cosmic strings and cylinder space-times are presented. Quasinormal modes for the scalar field have been calculated in static and rotating cosmic cylinders. We found unstable modes in the rotating cases. Rotating as well as static cosmic strings, i.e., without regular interior solutions, do not display quasinormal modes for the scalar field.

We conclude presenting a conjecture relating closed timelike curves and space-time instability.

\chapter*{Dedicatória}

\vspace{5cm}
\vspace{8cm}
\hspace{1cm}\emph{À minha muito amada esposa Marcela, com quem começo a construir a minha família. }

\chapter*{Agradecimentos}

Ao amigo, orientador e colaborador Prof. Elcio Abdalla pela oportunidade de estudar a Natureza  e por auxiliar-me nessa aventura pelas fronteiras do conhecimento humano.
\\
\\
\noindent
Ao amigo e colaborador Prof. Carlos Molina pelas boas discussões sobre o nosso trabalho, sobre filosofia e sobre a vida.
\\
\\
\noindent
Aos grandes amigos da sala 319, Jeferson, Rodrigo, Carlos Eduardo e Danilo pela paciência, tolerância e pelo companheirismo. Esta sala foi palco de grandes discussões científicas e tribuna dos mais tolos debates. Obrigado meus amigos. Cuidem-se, afinal a regra zero ainda vale e \emph{o mundo dá voltas.}
\\
\\
\noindent
Aos meus pais, José Cláudio e Maria Auxiliadora, e à minha irmã Luiza pelo carinho e apoio desde do ínicio desta jornada.
\\
\\
\noindent
Às amigas, Amélia e Simone pela eficiência e pelo carinho e atenção que tiveram comigo durante esses 6 anos no departamento de Física-Matemática.
\\
\\
\noindent
À querida Dona Lurdes, nossa gaúcha, pelo café sempre fresquinho das manhãs e pelo sábio conselho: \emph{``Você deve fazer seu trabalho com amor".}
\\
\\
\newpage
\noindent
Aos amigos, Renann, Juliano, Pedro, Leandro e muitos outros, que participaram das nossas inúmeras discussões, regadas a muito café, sobre política, filosofia e sociedade. E também ao amigo João pela dica esperta de um programa que faz ótimos desenhos.
\\
\\
\noindent
À Fapesp pelo apoio financeiro desta e outras pesquisas científicas neste estado.
\\
\\
\noindent
À todos os espíritos que educam-se diariamente na Crosta terrestre, usando o espaço e o tempo como ferramentas para melhor compreender a obra de Deus.

\chapter*{Prólogo}

Resultado natural da trajetória evolutiva da humanidade é o desenvolvimento da ciência a fim de termos um certo controle ou melhor dizendo, uma melhor previsão dos eventos da Natureza.

Dominamos a terra, o mar e até o céu; aprendemos a controlar o espaço. Seguindo adiante nesse processo criamos a tecnologia, automatização, pois nos faltava tempo. Queríamos mais tempo.

Ponto marcante de nossa sociedade atual é a tentativa de controlar o tempo. A ciência corre na busca por meios de controlar nosso tempo, de entendê-lo e dominá-lo. Mas ele escapa, flui, evapora, enquanto tentamos compreendê-lo. Sempre um passo a frente, o tempo foge.

Recorramos então à Relatividade Geral. Ela diz que máquinas do tempo são, pelo menos, teoricamente possíveis. Bem, agora sim temos a possibilidade de algo concreto, mesmo que isso nos parece estranho afinal, ser concreto, absoluto isso o tempo não é.

Porém, para nossa infelicidade ou para nossa liberdade tais máquinas parecem ser instáveis. Relativas, inconstantes, assim como nós agora somos.

Vejamos que segredos podemos descobrir quando mergulhamos no oceano do tempo. Talvez não sejamos capazes de subir à tona novamente, mas e daí. Ser senhor do nosso tempo, não é necessariamente ser senhor do próprio tempo. Subir à tona é saber o que dizer a respeito do tempo. Eu não sei. Talvez não saiba nesta vida. Quem sabe na próxima então.
Não sei falar do tempo sem incluí-lo, necessariamente, na minha fala. Os verbos, fazem o tempo fluir para trás, ou para frente. Não tenho consciência do mundo sem usar o tempo.

Mesmo depois de tanta desinformação ou desconhecimento, quero tentar. Afinal, saber do tempo.... é não tomar conhecimento de limites.

\newpage
\tableofcontents
\newpage


\chapter{Introdução}
\label{int}

Uma das perguntas mais intrigantes que a Relatividade Geral trouxe à tona, com seu novo entendimento sobre a relação entre o espaço e o tempo, foi a respeito da possibilidade de viagens no tempo. A simples possibilidade de uma viagem ao futuro ou ao passado, contudo, nos obriga a lidar com certos obstáculos lógicos e filosóficos tão complicados de serem elucidados que geralmente acabamos por descartar soluções das equações de Einstein que tenham essa propriedade.

Se, por um lado, tal atitude ameniza nosso desconforto filosófico e previne nosso Universo de qualquer paradoxo causal, por outro, cria preconceitos que acabam prejudicando a investigação futura de fenômenos ainda não observados que poderiam ser relacionados a possíveis violações da causalidade. Não podemos, portanto, usar nossa ignorância em solucionar paradoxos causais como desculpa. Devemos sim, lembrarmo-nos de que situações paradoxais surgem, geralmente, quando ainda não somos capazes de formular adequadamente nossas perguntas a respeito do assunto em análise.

Em uma discussão apresentada em \cite{Hawking2}, Hawking argumenta que os postulados da Relatividade Geral nada dizem a respeito de possíveis violações globais da causalidade de modo que, a menos das nossas dificuldades lógicas, tais violações poderiam fazer parte do nosso Universo.

Partindo dessa hipótese, trataremos a existência de máquinas do tempo como uma realidade e procuraremos investigar seus possíveis efeitos em nosso mundo.
Deste modo, o foco principal desse trabalho é tentar ampliar nossa compreensão sobre máquinas do tempo questionando, de diferentes maneiras, algumas de suas propriedades.

As propriedades que mais nos interessam estudar são resumidas nos seguintes assuntos: garantias para o viajante do tempo realizar uma viagem segura e estabilidade da máquina do tempo.
Dentro desses dois assuntos procuraremos responder questões cruciais para uma viagem no tempo tais como:

\begin{enumerate}
\item \emph{Sob quais circunstâncias pode um candidato a viajante do tempo acessar a região da máquina do tempo onde a causalidade é violada?}

\item \emph{Existe alguma força de maré ou algo semelhante, que pode por em risco a vida de um viajante do tempo?}

\item \emph{O encontro de um viajante do tempo como seu avô, ainda jovem, implica necessariamente em uma contradição lógica futura?}

\item \emph{Máquinas do tempo se mantêm estáveis quando submetidas a algum tipo de perturbação?}
\end{enumerate}

Vejamos, então, que avanços já foram obtidos a respeito das propriedades das máquinas do tempo.

Embora a idéia de voltar no tempo dirigindo o \emph{De Lorean} do Dr. Brown\footnote{R. Zemeckis (diretor), {\it ``Back to the Future",} Universal Pictures, (1985).} nos pareça um tanto divertida e atraente, em Relatividade Geral, o estudo de máquinas do tempo pode ser matematicamente traduzido na análise de soluções das Equações de Einstein que possuem regiões onde curvas fechadas tipo-tempo são permitidas. Dentro dessas regiões, uma partícula massiva pode descrever uma trajétoria temporal circular voltando ao próprio passado.

A busca por soluções das equações de Einstein que admitem curvas fechadas tipo tempo não é recente. As primeiras soluções conhecidas descrevendo ``máquinas do tempo" foram encontradas por Van Stockum \cite{Van} em 1937 e Gödel \cite{Goedel} em 1949. Em termos do conteúdo material usado para gerá-las, ambas as soluções eram simples. Elas representavam um cilindro infinito de poeira em alta rotação, no primeiro caso e um universo com constante cosmológica em rotação, no segundo.

Aparentemente, a rotação dos objetos em Relatividade Geral é um fator importante na criação de uma máquina do tempo. A maioria das soluções que possuem curvas fechadas tipo-tempo são objetos em rotação e curiosamente, nem sempre essa rotação precisa ser muito alta. Hoje em dia já colecionamos uma boa quantidade de exemplos de espaços-tempos como essa propriedade. Alguns deles são, o buraco negro de Kerr \cite{Kerr}, o cilindro cósmico em rotação, obtido por Jensen \cite{Jensen}; a corda cósmica em rotação, obtida por Deser \cite{Deser}; o buraco negro cilíndrico em rotação, obtido por Lemos \cite{Lemos}; a corda em rotação cercada por um ``gás de cordas", obtida por Gron \cite{Gron}; uma classe geral de cordas em rotação, obtida por Özdemir \cite{Ozdemir} e o universo inomogêneo em rotação, obtido por Soares \cite{Damiao}. Duas outras soluções, importantes pelo seu conteúdo material, foram obtidas por Bonnor e colaboradores \cite{Bonnor,Bonnor2}. A primeira descreve uma régua finita, sem massa em rotação e a segunda descreve duas partículas em rotação presas por uma régua finita e sem massa.

Buscando um melhor entendimento sobre esses espaços-tempos apuramos que a maior parte dessas soluções já tiveram suas propriedades investigada por diversos pesquisadores.

As propriedades gerais da solução de Van Stockum foram bastante discutidas por Bonnor \cite{Bonnor3} e Tipler \cite{Tipler}. Enquanto Tipler estuda os três regimes de rotação do cilindro e prova a unicidade da solução, Bonnor calcula a massa e o momento angular do cilindro. Sobre o comportamento das geodésicas nessa solução, Steadman \cite{Steadman} mostrou a existência de geodésicas fechadas tipo tempo além de confirmar o resultado obtido por Opher e colaboradores \cite{Opher} sobre o confinamento de partículas massivas e fótons no exterior do cilindro. A estabilidade destas geodésicas fechadas tipo tempo foi analisada e confirmada por Letelier e Rosa \cite{Letelier}.

O universo de Gödel, por sua vez, teve algumas de suas propriedades discutidas por Hawking em \cite{Hawking2} e por Grave e colaboradores em \cite{Grave}. Hawking apresenta a solução e discute o comportamento dos seus cones de luz. As soluções exatas para as equações das geodésicas são obtidas e discutidas por Grave e também por Calvão, Soares e Tiomno \cite{Tiomno}. Em \cite{Letelier2}, Letelier analisa a estabilidade de algumas curvas fechadas tipo tempo presentes neste universo e mostra que elas são estáveis. A propagação de campos de matéria neste espaço-tempo também foi explorada e discutida por alguns pesquisadores. A evolução do campo eletromagnético e seu espectro de frequências foi analisada por Mashhoon \cite{Mashhoon} e Vishveshwara \cite{Vish}. Já, a propagação do campo escalar massivo e seu espectro de frequências foram discutidos por Pimentel \cite{Pimentel} e Hiscock \cite{Hiscock0}. O campo espinorial de Weyl foi estudado por Pimentel \cite{Pimentel} e a evolução das perturbações gravitacionais é discutida por Barrow \cite{Barrow}.

Como vimos, a corda cósmica em rotação foi estudada e discutida por Deser \cite{Deser}. A propagação de campos neste tipo de geometria também foi alvo da atenção de alguns pesquisadores há algum tempo. Resultados tanto clássicos quanto semi-clássicos foram obtidos e analisados. A evolução do campo escalar foi discutida tanto por Krori \cite{Krori} quanto por Aliev \cite{Aliev}. Perturbações eletromagnéticas e gravitacionais neste espaço-tempo foram obtidas por Aliev \cite{Aliev}. Neste trabalho ele também discute a presença de um efeito Aharonov-Bohm gravitacional neste espaço-tempo. Em \cite{Matsas}, Matsas obteve correções semi-clássicas na métrica da corda em rotação produzidas por flutuações quânticas de um campo escalar conformalmente acoplado.

Mais recentemente alguns buraco negros supersimétricos que admitem curvas fechadas tipo tempo foram analisados por Gibbons e colaboradores em \cite{Gibbons}. Nestes trabalhos eles discutiram o comportamento das geodésicas juntamente com a propagação de um campo escalar. Em dimensões mais altas Abdalla e Konoplya \cite{Abdalla} analisaram a evolução de um campo escalar em um buraco negro do tipo Schwarzschild-Gödel 5-dimensional em baixas rotações. Eles observaram que o campo escalar se comportava de maneira estável. Contudo, esse resultado não implicou na estabilidade de uma ``máquina do tempo" já que no regime de baixas rotações nenhuma curva fechada tipo tempo era permitida.

Outra solução bastante usada como máquina do tempo é o espaço-tempo de um ``buraco de minhoca". Em \cite{Thorne}, Thorne e colaboradores criaram uma máquina do tempo movimentando as gargantas do buraco de minhoca de modo a produzir curvas fechadas tipo tempo. O problema de Cauchy para um campo escalar se propagando em um buraco de minhoca também foi bastante discutido por Friedman \cite{Friedman}. Embora eles tenham obtido resultados interessantes não abordaremos tais espaços-tempos nesse trabalho.

Como se vê, muito tem sido estudado sobre máquinas do tempo. Desta forma, para oferecer alguma contribuição concreta nesse assunto apresentamos a proposta de estudar a classe de espaços-tempos das cordas e cilindros cósmicos. Esta classe foi escolhida devido, especialmente, à simplicidade das componentes das métricas além da facilidade em se obter soluções exatas para as equações de movimento de campos se propagando nessas geometrias. Investigaremos o espectro de frequências do campo escalar se propagando nessas geometrias buscando relacionar o comportamento deste espectro com a estabilidade dos espaços-tempos.



As idéias e resultados desse trabalho são organizados como se segue. Inicialmente apresentaremos definições e propriedades gerais de espaços-tempos cilindricamente simétricos e estacionários. No capítulo \ref{cap3escalar} mostraremos a equação mestra para a evolução de um campo escalar massivo em um espaço-tempo cilindricamente simétrico e estacionário. Em seguida, discutiremos a organização da estrutura causal de um espaço-tempo, a conjectura da proteção cronológica e os paradoxos causais em espaços-tempos que admitem curvas fechadas tipo tempo. No capítulo \ref{cap6spacetime} descreveremos as propriedades dos espaços-tempos analisados. Eles são classificados segundo suas estruturas internas em cordas cósmicas e cilindros cósmicos. No capítulo seguinte analisaremos quais são as condições necessárias para existência da uma região não-causal nos espaços-tempos estudados. Para entender o comportamento de partículas e luz viajando nesses espaços-tempos, estudaremos, no capítulo \ref{cap8geodesicas}, as características das geodésicas tipo tempo e tipo luz. Em seguida investigaremos alguns tipos de curvas fechadas tipo tempo que podem ocorrer nos espaços-tempos estudados. No capítulo \ref{cap10Evoescalar_et}, procuraremos soluções  exatas para a equação de movimento do campo escalar nos espaços-tempos estudados. Nos dois capítulos seguintes definiremos quais são as condições de contorno para o cálculo dos modos quasinormais de oscilação do campo escalar em espaços-tempos que admitem curvas fechadas tipo tempo e calculá-los-emos, quando possível, buscando possíveis valores que indiquem alguma instabilidade.

Concluímos esse trabalho apresentando uma síntese dos argumentos que consideramos dar suporte a nossa conjectura que relaciona curvas fechadas tipo tempo a instabilidades do espaço-tempo. Os principais resultados obtidos neste trabalho são apresentados no artigo \cite{Alan}.

\chapter{Propriedades de um espaço-tempo cilindricamente simétrico e estacionário}
\label{cap2geo}


Iniciaremos nossos estudos apresentando a definição de um espaço-tempo cilindricamente simétrico e estacionário. Em seguida, descreveremos algumas das suas principais propriedades e discutiremos como elas serão usadas para sustentar a argumentação na defesa de nossa tese.

Seja $\mathcal{M}$ um espaço-tempo arbitrário 4-dimensional que é solução das equações de Einstein
\begin{eqnarray}
R_{\mu\nu}-\frac{1}{2}g_{\mu\nu}R+g_{\mu\nu}\Lambda=8\pi T_{\mu\nu}\quad,
\end{eqnarray}
onde $T_{\mu\nu}$ é o tensor energia-momentum de matéria ou campos presentes no espaço-tempo e $\Lambda$ é a constante cosmológica. Se $\mathcal{M}$ permanece invariante sob rotações em relação a um eixo de simetria e reflexões por planos que contêm esse eixo ele é um espaço-tempo axialmente simétrico. Agora, se além da simetria axial, ele também for invariante por translações temporais então esse espaço-tempo é dito ser axialmente simétrico e estacionário. Por causa dessas simetrias, o elemento de linha que descreve este espaço-tempo, escrito no sistema de coordenadas $(t,\phi,x^{2},x^3)$, será
\begin{eqnarray}
\label{cap2geo1}
ds^2&=&-F dt^2+2M d\phi dt+L d\phi^2+H (dx^2)^2+S (dx^3)^2\quad,
\end{eqnarray}
onde as funções $F, M, L, H, S$ dependem apenas das coordenadas $(x^2,x^3)$. Um exemplo desse espaço-tempo é o buraco negro de Kerr escrito nas coordenadas de Boyer-Lindquist onde $x^2 =r$ e $x^3 =\theta$.

Além deste caso, a métrica (\ref{cap2geo1}) pode ainda ser usada para descrever inúmeros outros espaços-tempos se as coordenadas $(x^2, x^3)$ forem escolhidas convenientemente. Veremos, a seguir, que mesmo para espaços-tempos cilindricamente simétricos e estacionários essa descrição é possível.

Para justificar a afirmação acima precisamos, primeiramente, definir a noção de espaço-tempo cilindricamente simétrico. Usaremos a definição apresentada em \cite{Thorne_tese}.
Um espaço-tempo $\mathcal{C}$ é dito ser cilindricamente simétrico se ele for invariante sob rotações em relação a um eixo de simetria e sob translações ao longo desse eixo.

Se quisermos que o espaço-tempo $\mathcal{C}$ seja também estacionário ele deve ser invariante sob translações temporais. Portanto,
\begin{quote}
\emph{ um espaço-tempo $\mathcal{C}_{t}$ é dito ser cilindricamente simétrico e estacionário se for invariante sob rotações em relação a um eixo de simetria, sob translações ao longo desse eixo e ainda sob
trans-lações temporais}.
\end{quote}
Comparando o espaço-tempo $\mathcal{M}$ com $\mathcal{C}_{t}$ em relação à simetria axial e cilíndrica vemos que a imposição da condição de translação ao longo do eixo exige uma maior simetria do espaço-tempo $\mathcal{C}_{t}$ e impõe restrições à dependência das funções $F, M, L, H, S$ em relação às coordenadas $(x^2, x^3)$.

Para descrevermos os espaços-tempos $\mathcal{C}_{t}$ que serão aqui analisados usaremos o sistema de coordenadas cilíndricas dadas por $(t,\phi,r,z)$. Definiremos o eixo de simetria do espaço-tempo como sendo o eixo $z$. Neste sistema de coordenadas as funções que aparecem na métrica (\ref{cap2geo1}) devem depender apenas da coordenada radial $r$. Essa exigência é fruto da invariância de $\mathcal{C}_{t}$ em relação à translações ao longo do eixo de simetria.

Deste modo, reescrevendo a métrica (\ref{cap2geo1}) nesse sistema de coordenadas para espaços-tempos cilindricamente simétricos e estacionários  temos
\begin{eqnarray}
\label{cap2geo1a}
ds^2&=&-F\ dt^2 + 2M\ d\phi dt + L\ d\phi^2 + H\ dr^2 + S\ dz^2\quad,
\end{eqnarray}
onde $F, M, L, H, S$ são funções apenas da coordenada $r$. Como a métrica (\ref{cap2geo1a}) apresenta elementos fora da diagonal o cálculo de suas componentes contravariantes e o seu determinante é não trivial. Apresentamos abaixo cada uma das componentes e o determinante
\begin{eqnarray}
\label{cap2a2}
&&g^{tt}=-\frac{L}{FL+M^2}\quad, \quad g^{t\phi}=\frac{M}{FL+M^2}\quad, \\
\nonumber\\
&&g^{\phi\phi}=\frac{F}{FL+M^2}\quad,\quad g^{\phi t}=\frac{M}{FL+M^2}\quad, \\
\nonumber\\
&&\qquad \quad g^{rr}=\frac{1}{H}\quad, \qquad g^{zz}=\frac{1}{S}\quad,\\
\nonumber\\
&&g=det(g_{\mu\nu})=-(FL+M^2)HS=-\mathcal{X}HS\quad,
\end{eqnarray}
Estas expressões serão utilizadas muitas vezes ao longo do trabalho, especialmente nos capítulos sobre propagação de campos no espaço-tempo curvo.

Associados à cada uma dessas simetrias do espaço-tempo $\mathcal{C}_{t}$ teremos três vetores de Killing, $T=\partial_{t}$, $\Phi=\partial_{\phi}$ e $Z=\partial_{z}$. Cada um desses vetores será o gerador das simetrias do espaço-tempo. O vetor $T$ é o gerador das translações temporais, o vetor $\Phi$ é o gerador das rotações em torno do eixo de simetria e $Z$ é o gerador das translações ao longo do eixo $z$.

Além das suas relações com as simetrias, os vetores de Killing  guardam informações adicionais sobre espaço-tempo. Por exemplo, se analisarmos a norma quadrática do vetor $T$ quando a coordenada $r$ varia podemos descobrir se esse espaço-tempo em rotação produz ergoregiões como aquela que aparece no buraco negro de Kerr. As ergoregiões estão intimamente ligadas ao fenômeno de arraste do espaço-tempo. Dentro desta região o vetor $T$ passa a ser tipo espaço de modo que a trajetória de qualquer partícula deve obrigatoriamente seguir o sentido de rotação do espaço-tempo \cite{Carroll}.

Uma análise semelhante para o vetor $\Phi$ nos permite identificar possíveis regiões onde esse vetor possa tornar-se tipo tempo. Se existe no espaço-tempo uma região com essa característica então curvas fechadas tipo tempo são possíveis. Dentro dessa região a coordenada $\phi$ se comporta como uma coordenada temporal periódica permitindo esse tipo de curva. A essa região chamaremos região não-causal do espaço-tempo. Essa análise será amplamente discutida no capítulo \ref{cap7analisecausal}.

Outro aspecto importante dos vetores de Killing são sua relação com as geodésicas do espaço-tempo. Como se sabe, vetores de Killing e geodésicas estão intimamente ligados e ambos fornecem importantes informações sobre a estrutura causal do espaço-tempo. Uma curva $x^{\mu}(\tau)$ é dita ser um curva geodésica se ela satisfizer a equação
\begin{eqnarray}
\label{cap2k0}
\frac{d^2x^{\mu}}{d\tau^2}+\Gamma^{\mu\phantom{\beta}}_{\alpha\beta}\frac{dx^{\alpha}}{d\tau}\frac{dx^{\beta}}{d\tau}=0\quad.
\end{eqnarray}
Uma dessas relações nos diz que a projeção de um vetor de Killing se deslocando ao longo de uma geodésica é sempre constante, ou seja, se $K^{\mu}$ é um vetor de Killing e $x^{\mu}(\tau)$ é uma geodésica parametrizada pelo parâmetro afim $\tau$ vale a relação
\begin{eqnarray}
\label{cap2k}
g_{\mu\nu}K^{\mu}\dot{x}^{\nu}=C\quad,
\end{eqnarray}
onde $(\ \dot{}\ )$ representa a derivada ordinária em relação ao parâmetro afim. Portanto, usaremos a relação (\ref{cap2k}) para calcular quais são as constantes de movimento associadas aos vetores $T$, $\Phi$ e $Z$ para um espaço-tempo $\mathcal{C}_{t}$.  As componentes dos vetores de Killing $T$, $Z$ e $\Phi$ podem ser escritas como
\begin{eqnarray}
\label{cap2k1}
T^{\mu}=\delta^{\mu}_{t}\quad,\quad Z^{\mu}=\delta^{\mu}_{z}\quad \textrm{e}\quad \Phi^{\mu}=\delta^{\mu}_{\phi}\quad.
\end{eqnarray}
A substituição da métrica (\ref{cap2geo1a}) nas Eqs. (\ref{cap2k}, \ref{cap2k1}) resulta em
\begin{eqnarray}
\label{cap2k2}
E&=&-F\dot{t} + M\dot{\phi} \quad,\\
\nonumber\\
\label{cap2k3}
\ell&=&M\dot{t} + L\dot{\phi} \quad,\\
\nonumber\\
\label{cap2k4}
p_{z}&=&S\dot{z}\quad,
\end{eqnarray}
onde as constantes de movimento $E$, $\ell$ e $p_{z}$ são associadas aos vetores $T$, $\Phi$ e $Z$ respectivamente. Como as velocidades que aparecem nas Eqs.(\ref{cap2k2}, \ref{cap2k3}, \ref{cap2k4}) são relacionadas às curvas geodésicas podemos obter as primeiras integrais da Eq.(\ref{cap2k0}) tanto das geodésicas tipo tempo quanto das geodésicas tipo luz combinando essas equações com a métrica (\ref{cap2geo1a}). Para tal, derivaremos a métrica (\ref{cap2geo1a}) em relação ao parâmetro afim $\tau$ resultando
\begin{eqnarray}
\label{cap2k5}
-\epsilon=&-F\ \dot{t}^2 + 2M\ \dot{t}\dot{\phi} + L\ \dot{\phi}^2 + H\ \dot{r}^2 + S\ \dot{z}^2\quad,
\end{eqnarray}
onde $\epsilon=0,1$ indica a equação para geodésicas tipo luz e tipo tempo respectivamente. Rearranjando as Eqs.(\ref{cap2k2}, \ref{cap2k3}, \ref{cap2k4}) e substituindo-as na Eq. (\ref{cap2k5}) teremos
\begin{eqnarray}
\label{cap2geo2}
\dot{\phi}&=&\frac{\ell F-EM}{\mathcal{X}}\quad,\\
\nonumber\\
\label{cap2geo3}
\dot{t}&=&\frac{EL+\ell M}{\mathcal{X}}\quad,\\
\nonumber\\
\label{cap2geo4}
\dot{z}&=&\frac{p_{z}}{S}\quad,\\
\nonumber\\
\label{cap2geo5}
\dot{r}^2&=&-\frac{\epsilon}{H}-\frac{p_{z}^{2}}{HS}+\frac{2ME\ell-F\ell^2+LE^2}{\mathcal{X} H}\quad.
\end{eqnarray}
Estas são as primeiras integrais da Eq.(\ref{cap2k0}) neste espaço-tempo $\mathcal{C}_{t}$. Se escolhermos analisar a geodésica tipo tempo devemos lembrar que as constantes de movimento são todas normalizadas pela massa da partícula. As Eqs. (\ref{cap2geo2}, \ref{cap2geo3}, \ref{cap2geo4}, \ref{cap2geo5}) geralmente são combinadas afim de obtermos a dependência de uma coordenada em relação à outra. Por exemplo, se combinarmos as Eqs. (\ref{cap2geo2}, \ref{cap2geo5}) obteremos a dependencia da coordenada radial em função da coordenada angular. Isso nos permite descrever as órbitas de partículas no plano $(r,\phi)$.

Se quisermos conhecer o comportamento das acelerações sofridas pelas partículas ou fótons se propagando em $\mathcal{C}_{t}$ usaremos a Eq.(\ref{cap2k0}) juntamente com a métrica (\ref{cap2geo1a}) para obter
\begin{eqnarray}
\label{cap2geo6}
\ddot{r}&+&\frac{F'\dot{t}^2}{2H}-\frac{M'\dot{t}\dot{\phi}}{H}-\frac{L'\dot{\phi}^2}{2H}+\frac{H'\dot{r}^2}{2H}-\frac{S'\dot{z}^2}{2H}=0\quad,\\
\nonumber\\
\label{cap2geo7}
\ddot{z}&+&\frac{S'\dot{r}\dot{z}}{S}=0\quad,\\
\nonumber\\
\label{cap2geo8}
\ddot{t}&+&\frac{(LF'+MM')\dot{t}\dot{r}}{\chi}-\frac{(LM'-ML')\dot{\phi}\dot{r}}{\chi}=0\quad,\\
\nonumber\\
\label{cap2geo9}
\ddot{\phi}&+&\frac{(FM'-MF')\dot{t}\dot{r}}{\chi}+\frac{(MM'+FL')\dot{\phi}\dot{r}}{\chi}=0\quad.
\end{eqnarray}
Estas equações serão estudadas em grande detalhe nos próximos capítulos onde observaremos o comportamento de fótons e partículas massivas em espaços-tempos que admitem curvas fechadas tipo tempo.

Por fim, se quisermos compreender como se comporta a curvatura do espaço-tempo $\mathcal{C}_{t}$ precisamos conhecer as componentes dos tensores de Riemann e de Ricci.
Desta forma, as componentes não-nulas do tensor Riemann podem ser escrita de forma condensada como
\begin{eqnarray}
\label{Riemann1}
R_{ tr\phi r}&=&\frac{M''}{2}-\frac{M'H'}{4H}+\frac{1}{4\mathcal{X}}\left(M\bar{\mathcal{X}}-M'\mathcal{X}'\right)\quad,\\
\nonumber\\
R_{ r\phi r\phi}&=&\frac{L''}{2}-\frac{L'H'}{4H}+\frac{1}{4\mathcal{X}}\left(L\bar{\mathcal{X}}-L'\mathcal{X}'\right)\quad,\\
\nonumber\\
R_{ trt r}&=&-\frac{F''}{2}+\frac{F'H'}{4H}-\frac{1}{4\mathcal{X}}\left(F\bar{\mathcal{X}}-F'\mathcal{X}'\right)\quad,\\
\nonumber\\
R_{ tz\phi z}&=&\frac{M'S'}{4H}\quad, \qquad R_{ tzt z}=-\frac{F'S'}{4H}\quad, \qquad R_{ t\phi t\phi}=-\frac{\bar{\mathcal{X}}}{4H}\quad, \\
\nonumber\\
R_{ \phi z\phi z}&=&\frac{L'S'}{4H}\quad, \qquad R_{ r z r z}=\frac{S''}{2}-\frac{S'(SH)'}{4HS}\quad,
\end{eqnarray}
onde $'$ significa derivada com respeito à coordenada $r$ e a função $\bar{\mathcal{X}}$ é
\begin{eqnarray}
\bar{\mathcal{X}}=F'L'+(M')^2.
\end{eqnarray}
Estas componentes são usadas na análise de forças de maré próximas a região não-causal abordadas no Apêndice \ref{forcamare}. Para o tensor de Ricci as componentes não-nulas são
\begin{eqnarray}
\label{ricci1}
R_{ t t}&=&-\frac{1}{4\mathcal{X}H^2S}\left[\frac{}{}2FHS\bar{\mathcal{X}}-F'\left(HS\mathcal{X}\right)'+2H\mathcal{X}\left(F'S\right)'\right]\quad,\\
\nonumber\\
\label{ricci2}
R_{\phi \phi}&=&+\frac{1}{4\mathcal{X}H^2S}\left[\frac{}{}2LHS\bar{\mathcal{X}}-L'\left(HS\mathcal{X}\right)'+2H\mathcal{X}\left(L'S\right)'\right]\quad,\\
\nonumber\\
\label{ricci3}
R_{\phi t}&=&+\frac{1}{4\mathcal{X}H^2S}\left[\frac{}{}2MHS\bar{\mathcal{X}}-M'\left(HS\mathcal{X}\right)'+2H\mathcal{X}\left(M'S\right)'\right]\quad,\\
\nonumber\\
\label{ricci4}
R_{ z z}&=&+\frac{1}{4\mathcal{X}H^2S}\left[\frac{}{}2HS\left(S'\mathcal{X}\right)'-S'\left(HS\mathcal{X}\right)'\right]\quad, \\
\nonumber\\
\label{ricci5}
R_{ r r }&=&+\frac{1}{4\mathcal{X}^2HS^2}\left[\frac{}{}2HS\mathcal{X}\left(S''\mathcal{X}+S\mathcal{X}''\right)-2S^2H\mathcal{X}\bar{\mathcal{X}}-\right.\nonumber\\
\nonumber\\
&-&\left. S^2\mathcal{X}'\left(\mathcal{X}H\right)'-S'\mathcal{X}^2\left(SH\right)'\frac{}{}\right]\quad.
\end{eqnarray}

\chapter{Evolução de campos em um espaço-tempo cilindricamente simétrico e estacionário}
\label{cap3escalar}

A análise da evolução de campos de matéria se propagando em um espaço-tempo curvo tem se revelado uma importante fonte de informação sobre a própria dinâmica desse espaço-tempo. Representando uma ampla área de pesquisa, essa análise tem mostrado quão estreita é a relação entre estas propagações de campos e a estabilidade do próprio espaço-tempo \cite{Abdalla2,Ching,Cardoso0,Chandra}. Dentre os campos de matéria mais estudados destaca-se o campo escalar por ser matematicamente mais simples que os outros campos. Interessados nas informações que podem ser extraídas da propagação desse campo em um espaço-tempo cilindricamente simétrico e estacionário dedicaremos este capítulo ao cálculo da sua equação de movimento.

\section{Evolução do campo escalar }


O estudo da evolução de campos escalares em diversos espaços-tempos tem sido de grande utilidade para a compreensão de propriedades até então desconhecidas de buracos negros e estrelas, tais como superradiância \cite{Starobinski} e temperatura Hawking \cite{Hawking}. Embora estes campos não descrevam qualquer campo físico conhecido, sua evolução, geralmente mais simples e tratável, é capaz revelar as características essenciais sobre o comportamento geral da propagação de outros campos de spin mais alto. Todavia, esta idéia vem sendo reformulada. Com uma melhor compreensão da correspondência AdS/CFT, o estudo de campos escalares em espaços-tempos curvos adquiriu novo impulso e importância devido ao fato de que campos escalares se propagando em um espaço-tempo AdS podem ser interpretados como operadores $\mathcal{O}$ de uma teoria quântica de campos dual ao espaço-tempo AdS \cite{ADS}.




Quando se estudam espaços-tempos estacionários a equação de evolução do campo escalar se torna um pouco mais complicada de se tratar, pois ela nem sempre é separável e a presença do termo cruzado $\ g_{\phi t}\ d\phi dt\ $ altera o modo de calcular as componentes contravariantes da métrica.
Um bom exemplo ilustrando estas dificuldades é o estudo de campos escalares se propagando no exterior do buraco negro de Kerr. Neste caso as equações parecem ser não separáveis devido a uma dependência, não-trivial das coordenadas $r,\theta$, da métrica. Esta dificuldade foi contornada por Brill \cite{Brill} utilizando uma nova constante de movimento encontrada por Carter \cite{Carter} que separa as equações apesar da dependência complicada das funções que a descrevem.
No caso da métrica (\ref{cap2geo1a}) que possui alto grau de simetria, a equação da evolução do campo escalar é sempre separável. Exploraremos estas características para obter uma equação de evolução tão geral quanto possível.


A equação que descreve a evolução de um campo escalar massivo $\Psi$ em um espaço-tempo arbitrário é a equação de Klein-Gordon,
\begin{eqnarray}
\label{cap3a4}
\Box \ \Psi(t,\phi,x^2,x^3) = \beta^2 \ \Psi(t,\phi,x^2,x^3)\quad,
\end{eqnarray}
onde $\beta$ é a massa do campo $\Psi$ e
\begin{equation}
\Box = \frac{1}{\sqrt{-g}}\partial_{\mu}(\sqrt{g}g^{\mu\nu}\partial_{\nu})\quad.
\end{equation}
Usando a métrica (\ref{cap2geo1}) e expandindo a Eq.(\ref{cap3a4}) termo a termo teremos
\begin{eqnarray}
\label{cap3a5}
\frac{1}{\sqrt{-g}}\left\{ \frac{}{}\partial_t \left(\sqrt{-g}g^{tt}\partial_t\Psi+\sqrt{-g}g^{t\phi}\partial_\phi \Psi\right)+\right.\nonumber\\
\nonumber\\
+\ \partial_\phi \left(\sqrt{-g}g^{\phi t}\partial_t\Psi+\sqrt{-g}g^{\phi\phi}\partial_\phi \Psi\right)+\nonumber\\
\nonumber\\
\left. + \ \partial_2\left(\sqrt{-g}g^{22}\partial_2\Psi\right)+\partial_3\left(\sqrt{-g}g^{33}\partial_3\Psi\right)\frac{}{}\right\}=\beta^2\ \Psi\quad.
\end{eqnarray}
Se separarmos a Eq.(\ref{cap3a5}) em três partes e nos lembrarmos que a métrica depende apenas de $(x^2,x^3)$, os termos relacionados as coordenadas $(\phi , t )$ serão
\begin{eqnarray}
\label{cap3a6}
g^{tt}\partial^2_t\Psi+g^{\phi\phi}\partial^2_\phi\Psi+2g^{t\phi}\partial_t\partial_\phi\Psi=\mathcal{T}\Psi\quad,
\end{eqnarray}
e os termos relacionados as coordenadas $(x^2,x^3)$ serão
\begin{eqnarray}
\label{cap3a7}
\mathcal{K}\Psi=\frac{1}{\sqrt{-g}}\ \partial_2\left(\sqrt{-g}\right)g^{22}\partial_2\Psi +\partial_2\left(g^{22}\right)\partial_2\Psi+g^{22}\partial^2_2\Psi+\nonumber\\
\nonumber\\
+\frac{1}{\sqrt{-g}}\ \partial_3 \left(\sqrt{-g}\right)g^{33}\partial_3\Psi +\partial_3\left(g^{33}\right)\partial_3\Psi+g^{33}\partial^2_3\Psi\quad.
\end{eqnarray}
Desta forma, a Eq.(\ref{cap3a4}) pode ser reescrita na forma compacta
\begin{eqnarray}
\label{cap3a8}
\left(\mathcal{T}+\mathcal{K}\right)\Psi=\beta^2\ \Psi\quad.
\end{eqnarray}
Redefinindo as componentes contravariantes da métrica como sendo
\begin{eqnarray}
\label{cap3a9}
g^{tt}=\frac{-LHS}{-g}\quad,\quad g^{\phi t}=\frac{MHS}{-g}\quad, \quad g^{\phi\phi}=\frac{FHS}{-g}\quad,
\end{eqnarray}
e fazendo uso da relação
\begin{eqnarray}
\label{cap3a9a}
\frac{\partial_A(\sqrt{-g})}{\sqrt{-g}}=\frac{1}{2}\left(\frac{1}{-g}\right)(-g)_{,A}\quad,
\end{eqnarray}
onde $A=2,3 $ poderemos expandir a Eq.(\ref{cap3a5}) e iniciar a separação de variáveis. Assim procedendo, teremos
\begin{eqnarray}
\label{cap3a10}
\frac{-LHS}{-g}\partial^2_t\Psi+\frac{FHS}{-g}\partial^2_\phi\Psi+\frac{2MHS}{-g}\partial_t\partial_\phi\Psi+
\frac{1}{2}\left(\frac{-g_{,2}}{-g}\right)\frac{1}{H}\partial_{2}\Psi\nonumber\\
\nonumber\\
+\partial_{2}\left(\frac{\partial_{2}\Psi}{H}\right)+
\frac{1}{2}\left(\frac{-g_{,3}}{-g}\right)\frac{1}{S}\partial_{3}\Psi+\partial_{3}
\left(\frac{\partial_{3}\Psi}{S}\right)=\beta^2\ \Psi\quad.
\end{eqnarray}
Simplificando a Eq.(\ref{cap3a10}) teremos
\begin{eqnarray}
\label{cap3a11}
-L\partial^2_t\Psi+F\partial^2_\phi\Psi+2M\partial_t\partial_\phi\Psi+\frac{1}{2}\left(\mathcal{X} HS\right)_{,2}\frac{1}{H^2S}\partial_{2}\Psi\nonumber\\
\nonumber\\
+\mathcal{X} \partial_{2}\left(\frac{\partial_{2}\Psi}{H}\right)+\frac{1}{2}\left(\mathcal{X} HS\right)_{,3}\frac{1}{S^2H}\partial_{3}\Psi+\mathcal{X}\partial_{3}
\left(\frac{\partial_{3}\Psi}{S}\right)=\mathcal{X} \beta^2\ \Psi\quad,
\end{eqnarray}
onde usamos a função $\mathcal{X}=FL+M^2$, permitindo agora uma tentativa de separação de variáveis.

Usando como Ansatz a relação
\begin{eqnarray}
\label{cap3a12}
\Psi(t,\phi,x^2,x^3)=T(t)\Phi(\phi)R(x^2,x^3)\quad,
\end{eqnarray}
para a descrição do campo $\Psi$, a Eq.(\ref{cap3a11}) se torna
\begin{eqnarray}
\label{cap3a13}
-\frac{L}{T}\partial^2_tT+\frac{F}{\Phi}\partial^2_\phi\Phi+\frac{2M}{T\Phi}\partial_t(T)\partial_\phi(\Phi)+\frac{1}{2}\left(\mathcal{X} HS\right)_{,2}\frac{1}{R H^2S}\partial_{2}R\nonumber\\
\nonumber\\
+\frac{\mathcal{X}}{R} \partial_{2}\left(\frac{\partial_{2}R}{H}\right)+\frac{1}{2}\left(\mathcal{X} HS\right)_{,3}\frac{1}{RS^2H}\partial_{3}R+\frac{\mathcal{X}}{R}
\partial_{3}\left(\frac{\partial_{3}R}{S}\right)=\mathcal{X}\beta^2\quad.
\end{eqnarray}
Como nosso problema tem simetria axial, queremos garantir a periodicidade da solução identificando $\Phi(0)=\Phi(2\pi)$. Portanto, usaremos como Ansatz uma decomposição em modos de Fourier dada por
\begin{eqnarray}
\label{cap3a14}
\Phi(\phi)=\sum^{\infty}_{m=-\infty}\ c_{m}\  e^{i m\phi}\quad,
\end{eqnarray}
onde $m$ é um número inteiro devido à condição de periodicidade.


A dependência temporal do campo pode ser tratada de duas maneiras distintas. A primeira é fazer uma decomposição em modos de Fourier da função $T(t)$. Este tipo de decomposição é particularmente interessante pois permite-nos calcular os modos quasinormais de oscilação do campo escalar por meio, tanto do método WKB \cite{WKB} quanto do modo analítico. A segunda maneira é reescrever a função $R(x^2,x^3)$ de modo a incorporar a dependência temporal $\tilde{R}(t,x^2,x^3)$ para permitir uma integração númerica da equação de movimento. A vantagem desta decomposição é evidenciar detalhes do comportamento assintótico e decaimentos das oscilações do campo escalar \cite{Molina}.
Usando a primeira opção temos
\begin{eqnarray}
\label{cap3a15}
T(t)=\int^{\infty}_{-\infty}\ c(\omega) e^{-i\omega t}\ d\omega\quad,
\end{eqnarray}
com a equação do campo sendo dada por
\begin{eqnarray}
\label{cap3a16}
\frac{1}{2}\left(-g\right)_{,2}\frac{1}{ H^2S}\partial_{2}R&+&\mathcal{X} \partial_{2}\left(\frac{\partial_{2}R}{H}\right)+\frac{1}{2}\left(-g\right)_{,3}\frac{1}{S^2H}
\partial_{3}R+\nonumber\\
\nonumber\\
+\mathcal{X}\partial_{3}\left(\frac{\partial_{3}R}{S}\right)&=&R\left[\mathcal{X}\beta^2+Fm^2-2Mm\omega-L\omega^2\right]\quad,
\end{eqnarray}
e o campo $\Psi$ definido como
\begin{eqnarray}
\label{cap3a17}
\Psi(t,\phi,x^2,x^3)=\int^{\infty}_{-\infty}\sum_{m=-\infty}^{\infty}\  R_{m\omega}(x^2,x^3)e^{i(m\phi-\omega t)}\ d\omega\quad.
\end{eqnarray}

Usando a segunda opção a equação de campo se torna
\begin{eqnarray}
\label{cap3a18}
-L\partial^2_t\tilde{R}&+&2imM\partial_t(\tilde{R})+\frac{1}{2}\left(-g\right)_{,2}\frac{1}{ H^2S}\partial_{2}\tilde{R}+\frac{1}{2}\left(-g\right)_{,3}\frac{1}{S^2H}\partial_{3}\tilde{R}+\nonumber\\
\nonumber\\
&+&\mathcal{X}\partial_{2}\left(\frac{\partial_{2}\tilde{R}}{H}\right)+
\mathcal{X}\partial_{3}\left(\frac{\partial_{3}\tilde{R}}{S}\right)=\tilde{R}\left[\mathcal{X}\beta^2+Fm^2\right]\quad,
\end{eqnarray}
e o campo $\Psi$ será então
\begin{eqnarray}
\label{cap3a19}
\Psi(t,\phi,x^2,x^3)=\sum^{\infty}_{m=-\infty}\ \tilde{R}_{m}(t,x^2,x^3)e^{i(m\phi)}\quad.
\end{eqnarray}
Como as Eqs.(\ref{cap3a16}, \ref{cap3a18}) foram obtidas sem especificarmos a métrica e sua depêndencia em relação às coordenadas $(x^2, x^3)$ concluímos que elas são válidas para qualquer espaço-tempo axialmente simétrico e estacionário cuja métrica possa ser escrita na forma da Eq.(\ref{cap2geo1}).

Os espaços-tempos que estudaremos neste trabalho apresentam simetrias adicionais que nos auxiliam a simplificar um pouco mais essas equações de campo. Estes espaços-tempos são todos cilindricamente simétricos e suas métricas podem ser escritas no formato da Eq.(\ref{cap2geo1a}). As componentes das métricas dependem apenas da coordenada $r$. Portanto, a dependência do campo em relação à coordenada $z$ fica mais simples tornando a equação de movimento do campo sempre separável. Essa dependência pode ser expressa em termos da expansão de Fourier
\begin{eqnarray}
R(r,z)=R(r)\int^{\infty}_{-\infty}\ c(k)\ e^{i k z}\ dk\quad.
\end{eqnarray}
Deste modo, em todos os casos aqui estudados, o campo escalar será definido como
\begin{eqnarray}
\label{cap3a20}
\Psi(t,\phi,r,z)=\int^{\infty}_{-\infty}\int^{\infty}_{-\infty}\sum^{\infty}_{m=-\infty}\ \tilde{R}_{mk\omega}(r)e^{i(m\phi+ k z-\omega t)}d\omega dk\quad,
\end{eqnarray}
e a equação de campo será
\begin{eqnarray}
\label{cap3a21}
\frac{1}{2}\left(-g\right)_{,r}\frac{1}{ H^2S}\frac{dR}{dr}+\mathcal{X} \frac{d}{dr}\left(\frac{1}{H}\frac{dR}{dr}\right)=V_{km\omega\beta}^{esc}(r)\ R(r)\quad,
\end{eqnarray}
com o potencial $V^{esc}(r)$ dado pela expressão
\begin{eqnarray}
\label{cap3a22}
V_{km\omega\beta}^{esc}(r)=\left[\mathcal{X}\left(\beta^2+\frac{k^2}{S}\right)+Fm^2-2Mm\omega-L\omega^2\right]\quad.
\end{eqnarray}
Nos próximos capítulos, estas equações serão utilizadas para estudarmos a evolução do campo escalar em espaços-tempos com simetria cilíndrica e rotação. Esse tipo de configuração é de interesse pois parece possuir ca-racterísticas que favorecem o aparecimento de curvas fechadas tipo-tempo, além de ser simples o suficiente a ponto de obtermos resultados exatos para a evolução dos campos.

\chapter{Causalidade e Proteção Cronológica}
\label{cap5causal}

A segurança que adquirimos, com o tempo absoluto da gravitação Newtoniana, de que o nosso universo imutável percorre um linha definida, não nos protege mais. Com as revoluções trazidas pela Relatividade Geral, novas questões surgiram e o incerto passou a ser mais frequente. Nossa noção da relação entre espaço e tempo foi modificada e com ela também nosso entendimento sobre causalidade. De que maneira um viajante pode retornar no tempo e mudar seu próprio passado é uma pergunta possível apenas se admitirmos a posibilidade de que passado e futuro podem, de algum modo, se influenciarem mutuamente. Mas então, como é que a noção de passado, presente e futuro se organiza na Relatividade Geral?

Nossa noção a respeito do tempo e de que maneira sua dinâmica se dá ainda não é tão clara. Localmente, até conseguimos definir uma flecha do tempo através de um sistema termodinâmico. Mas relacionar essa flecha do tempo com aquela definida pela evolução do universo nos parece uma tarefa mais difícil. Esta dificuldade expõe também problemas maiores pois nossa noção de causalidade está intimamente ligada à dinâmica de tal flecha temporal. Portanto, o primeiro passo é saber de que maneira podemos atribuir uma orientação para essa flecha do tempo.

Isso pode ser feito se formos capazes de mapear, de maneira contínua, o espaço-tempo, dividindo-o em duas regiões, de modo que possamos distinguir em que direção apontam futuro e passado de um vetor nele contido.
Se esse mapeamento for possível diremos que o espaço-tempo é temporalmente orientado. Neste caso um vetor do tipo não-espacial pode ser classificado como futuro-direcionado ou passado-direcionado de maneira não ambígua.
Se não for possível, não poderemos definir de maneira inequívoca uma direção para a flecha do tempo e nem classificar a orientação de vetores do tipo não-espaciais. Esta é a situação quando o espaço-tempo admite curvas fechadas tipo tempo. A flecha do tempo, neste caso, é dinâmica, mudando de sentido e direção dependendo da região do espaço-tempo.

\section{Espaço-tempo temporalmente orientado}

Vejamos então, como a estrutura causal de um espaço-tempo temporalmente orientado se comporta. Usaremos, as algumas definições apresentadas por Hawking no capítulo 6 de \cite{Hawking2} para entendermos essa organização causal do espaço-tempo.

Sejam $\mathcal{S}$ e $\mathcal{U}$ dois conjuntos de pontos pertencentes ao espaço-tempo $\mathcal{M}$. Definiremos o futuro cronológico $I^{+}(\mathcal{S},\mathcal{U})$ de $\mathcal{S}$ relativo a $\mathcal{U}$ como o conjunto de todos os pontos que partem de $\mathcal{S}$ e podem alcançar $\mathcal{U}$ por meio de uma curva tipo tempo futuro-direcionada em $\mathcal{U}$.

De modo semelhante, definiremos o futuro causal de $\mathcal{S}$ relativo a $\mathcal{U}$ representado por $J^{+}(\mathcal{S},\mathcal{U})$ como a união de $\mathcal{S}\cap \mathcal{U}$ com o conjunto de todos os pontos em $\mathcal{U}$ que partem de $\mathcal{S}$ e podem alcançar $\mathcal{U}$ por uma curva não-espacial futuro-direcionada, ou seja, essa curva pode ser tipo tempo ou tipo luz.

As definições de passado cronológico $I^{-}(\mathcal{S},\mathcal{U})$ e passado causal $J^{-}(\mathcal{S},\mathcal{U})$ são diretas.

Deste modo, a estrutura causal de um espaço-tempo temporalmente orientado pode ser organizada usando os conceitos de passado e futuro causal do conjunto de pontos que compreende o espaço-tempo. Tomemos como exemplo o espaço-tempo de Minkowski representado por $\mathcal{M}_{I}$.
\begin{figure}[!ht]
\begin{center}
\epsfig{file =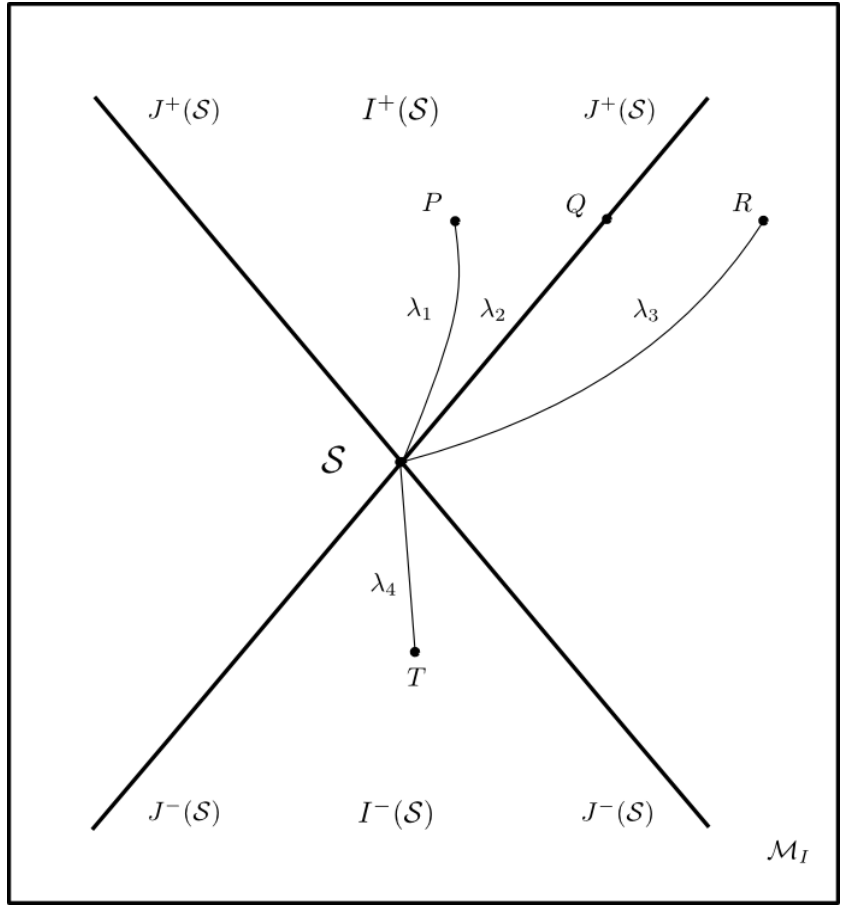, width=0.8\linewidth, clip=}
\end{center}
\caption{{\small Estrutura causal de $\mathcal{M}_{I}$}. As curvas $(\lambda_{1},\lambda_{2},\lambda_{3},\lambda_{4})$ são curvas tipo tempo, tipo luz, tipo espaço e tipo tempo respectivamente.}
\label{figcausalM}
\end{figure}

O conjunto de pontos $\mathcal{S}$ pertencente a $\mathcal{M}_{I}$ possui um futuro e um passado causal dados por $J^{+}(\mathcal{S},\mathcal{M}_{I})$ e $J^{-}(\mathcal{S},\mathcal{M}_{I})$ respectivamente. Na Fig. (\ref{figcausalM}) observamos que os eventos $P$ e $Q$ localizados no futuro causal de $\mathcal{S}$ podem ser causalmente afetados por eventos em $\mathcal{S}$, ou seja, existirá uma ou mais curvas tipo tempo ou tipo luz que ligam $P$ e $Q$ a $\mathcal{S}$. De modo análogo o evento $T$ localizado no passado causal de $\mathcal{S}$ afetará causalmente os eventos em $\mathcal{S}$. O evento $R$, por outro lado, não pode ser afetado por eventos em $\mathcal{S}$ a menos que liguemos os eventos por meio de uma curva não-causal do tipo espaço.

Embora a estrutura causal de espaços-tempos mais gerais nem sempre sejam tão simples quanto no caso do espaço-tempo de Minkowski elas podem ser analizadas de maneira semelhante a realizada.

Uma análise mais completa, além de outras definições, seriam necessárias para discutirmos espaços-tempos mais gerais onde as relações causais são mais complexas, como por exemplo, quando um ponto do espaço-tempo é removido. Porém, as definições aqui apresentadas já são suficientes aos nossos propósitos de estudar espaços-tempos cilindricamente simétricos e estacionários que admitem curvas fechadas.

\section{Proteção cronológica}

Até o presente momento temos falado apenas de espaços-tempos cuja orientação temporal é possível. Contudo, nosso interesse é justamente naqueles espaços-tempos que admitem curvas fechadas tipo tempo e cuja orientação temporal não é bem definida.

Em \cite{Hawking2}, Hawking argumenta que o postulado da Relatividade Geral sobre causalidade impõe apenas que ela seja preservada localmente nada dizendo a respeito da sua característica global. Portanto, a teoria permite que a causalidade seja violada ao menos globalmente. É preciso então, haver outros argumentos ou fatos para que possamos descartar um espaço-tempo que admita curvas fechadas tipo tempo como sendo fisicamente inviável.

Dentre os muitos argumentos utilizados para justificar esta exclusão o mais plausível é aquele que se apóia no fato de que uma violação da causalidade permitirá a existência de paradoxos lógicos. O mais conhecido deles é o paradoxo do avô. Esse paradoxo possui muitos outros nomes e muitas variações mas a idéia central é apresentada como se segue.

\begin{quote}

 \emph{Um jovem estudante de Física constrói uma máquina do tempo. Voltando ao passado esse jovem encontra seu avô e este encontro acaba modificando as condições necessárias para seu nascimento no futuro. Então, como é que esse jovem vai nascer no futuro se ele modificou as condições necessárias para seu próprio nascimento?}

\end{quote}

Essa possibilidade acaba criando uma contradição lógica já que não teremos mais uma causa (o encontro do avô e avó) para o efeito (o jovem). Por outro lado, essa contradição pode ser evitada se abrirmos mão do livre arbítrio como princípio básico da nossa filosofia da ciência. Sob essa condição o jovem até poderia voltar no tempo mas não poderia alterá-lo, ou poderia alterá-lo de uma maneira consistente de modo que sua existência no futuro não fosse afetada. Novikov e colaboradores analisam e discutem estas possibilidades em \cite{Novikov}. Eles chegam inclusive a enunciar um princípio de auto-consistência que permite a volta no tempo mas proíbe mudanças no passado de modo que a história desse universo se repita infinitamente. Assim, em um universo onde o princípio de auto-consistência é válido, os indivíduos perdem a liberdade de realizar experimentos livremente, repetindo as mesmas ações toda vez que retornam no tempo.

Discutiremos, novamente o princípio de auto-consistência no capítulo \ref{MQN} quando apresentarmos as condições de contorno para o estudo dos modos quasinormais de um campo escalar em um espaço-tempo cilindricamente simétrico e estacionário que admite curvas fechadas tipo tempo.

Baseado na contradição lógica que o paradoxo do avô apresenta, Hawking prefere preservar o livre arbítrio e descartar espaços-tempos que admitem curvas fechadas tipo tempo propondo a conjectura da proteção cronológica. Ela apresenta o seguinte enunciado \cite{Novikov,Hawkingcpc}

\begin{quote}

{\bf Conjectura da proteção cronológica:}\\

\emph{As leis da Física não permitem o aparecimento de curvas fechadas tipo tempo.}

\end{quote}

Embora esta conjectura não tenha sido provada, existem argumentos tanto clássicos quanto quânticos que a reforçam. Um exemplo de resultado a favor da conjectura, obtido por Hawking \cite{Hawkingcpc},  mostra que é impossível criar uma máquina do tempo em uma região finita do espaço-tempo sem violar a condição de energia nula média. Outro resultado usando campos escalares quantizados favorável à conjectura foi obtido por Frolov \cite{Frolov}. Ele mostra que o tensor energia-momentum renormalizado do vácuo diverge perto de um horizonte cronológico futuro\footnote{ Horizonte cronológico futuro é a superfície limite que marca o início da região acronal e o fim da região cronal do espaço-tempo.}. Isso significa que a retroação originada pelo campo escalar quântico previne a formação de máquinas do tempo ou as destroem criando uma singularidade.

É com o apoio desses e de outros resultados que a conjectura da proteção cronológica vai se fortalecendo. Portanto, nossa contribuição possível neste assunto, consiste em apresentar elementos que confirmem ou refutem a conjectura.


\section{Exemplos de violação da causalidade em Minkowski}

Na seção anterior apresentamos algumas dificuldades que são enfrentadas quando espaços-tempos que admitem curvas fechadas tipo tempo são aceitos como soluções fisicamente válidas. Para ilustrar algumas dessas discussões anteriormente apresentadas usaremos o espaço-tempo de Minkowski $\mathcal{M}_{1+1}$ (1+1) dimensões, onde identificaremos a coordenada temporal $t$  tal que a condição $(t_{f}=t_{i})$ é satisfeita. A coordenada $x$ é definida no intervalo $x\geq0$.

\begin{figure}[!ht]
\begin{center}
\epsfig{file =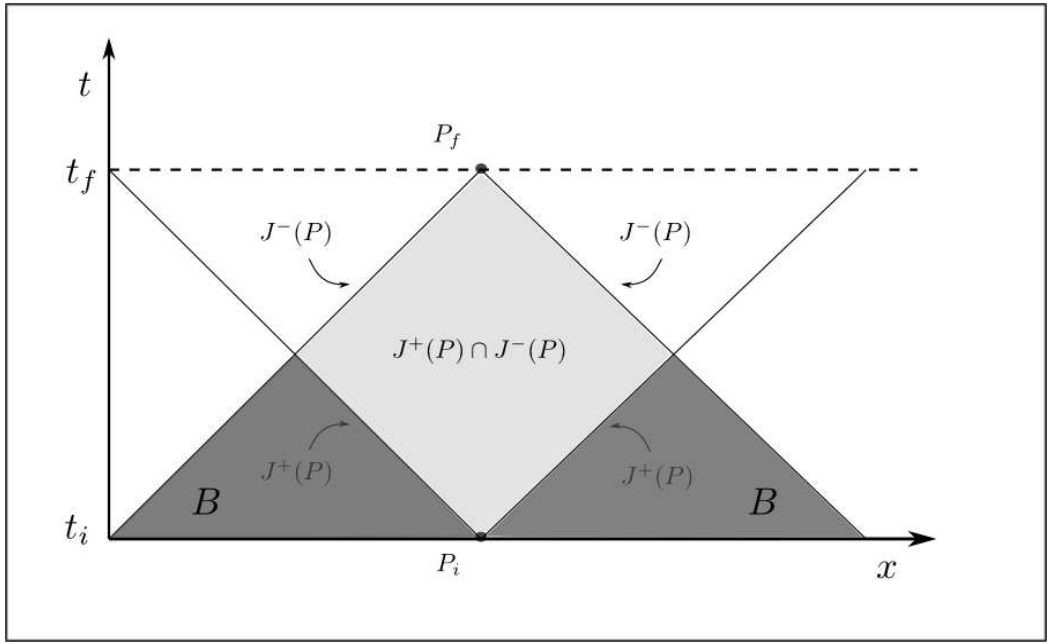, width=1.0\linewidth, clip=}
\end{center}
\caption{{\small Estrutura causal de um evento $P$ em $\mathcal{M}_{1+1}$. A linhas diagonais representam geodésicas nulas.}}
\label{figcausalM3}
\end{figure}

De modo geral, um espaço-tempo que admite curvas fechadas tipo tempo pode ser divido em regiões cronais onde não existem tais curvas e em regiões acronais que permitem tais curvas. Na região acronal, um evento pode ser influenciado tanto pelo seu passado quanto pelo seu futuro. Vejamos na Fig.(\ref{figcausalM3}) a estrutura causal de um evento $P$.

Como vemos, o evento $P$ acontece exatamente quando $t=t_{f}=t_{i}$ o que implica que $P_{i}=P_{f}$. O futuro causal do evento $P$, $J^{+}(P)$, emerge de $P_{i}$, enquanto seu passado causal $J^{-}(P)$ emerge de $P_{f}$. Por causa dessa identificação temporal, um evento localizado na região onde $J^{+}(P)\cap J^{-}(P)$ pertence tanto ao passado quanto ao futuro do evento $P$. Pensando a respeito do paradoxo do avô nesse espaço-tempo podemos observar algo interessante. Vamos dizer que o nascimento do jovem é representado pelo evento $P=P_{i}$. Então o futuro causal do jovem será $J^{+}(P)$ e seu passado causal $J^{-}(P)$. Obviamente, o avô do jovem se encontra no seu passado causal em $J^{-}(P)$ mas não necessariamente no seu futuro causal. Se o avô estiver na região $B$ que não se intercepta com o futuro causal do jovem, pode acontecer deles não se encontrarem. Isso permitiria que o jovem voltasse ao passado mas não mudasse as condições necessárias para o seu nascimento.

\begin{figure}[!ht]
\begin{center}
\epsfig{file =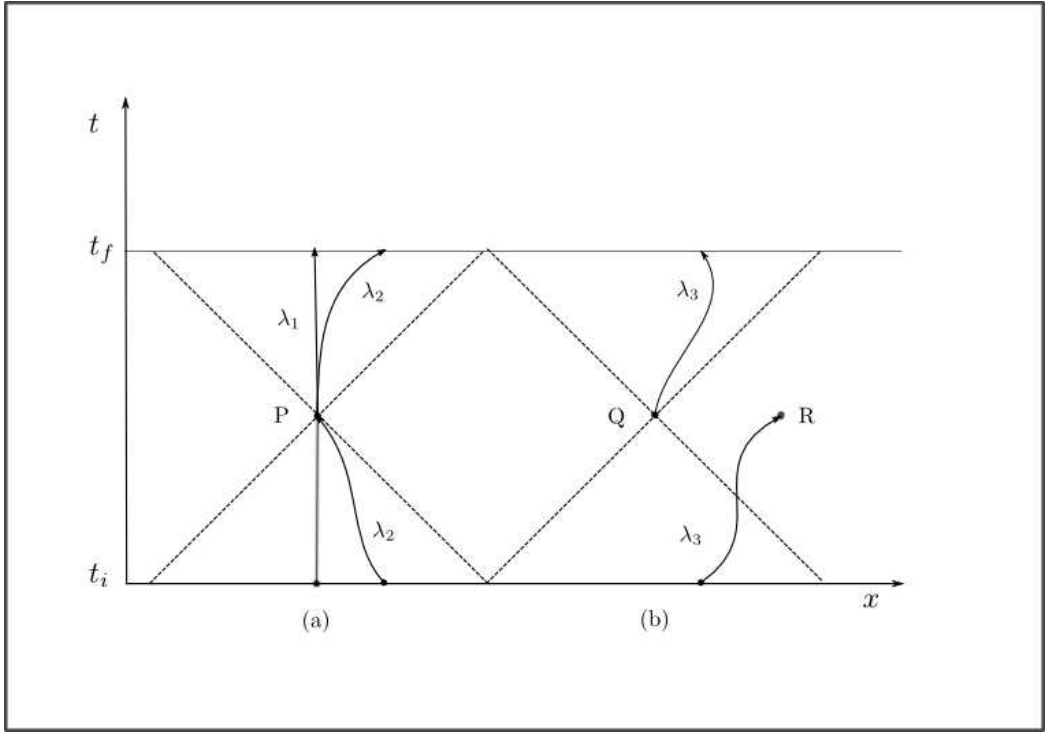, width=1.0\linewidth, clip=}
\end{center}
\caption{{\small Curvas fechadas tipo tempo em $\mathcal{M}_{1+1}$. A linhas diagonais representam geodésicas nulas.}}
\label{figcausalM2}
\end{figure}

Vejamos agora como se comportam curvas fechadas tipo tempo em um espaço-tempo onde o princípio de auto consistência é válido. A Fig.(\ref{figcausalM2}) representa as situações onde o princípio é aplicado $(a)$ e onde ele não é $(b)$.

A curva $\lambda_{1}$ que cruza o ponto $P$ representa a trajetória de uma partícula parada. Neste caso, onde vale o princípio de auto-consistência, a partícula volta ao passado e se encontra de maneira consistente pois sua posição final é igual a inicial. Isso implica que a partícula repete uma mesma história um número infinito de vezes. A curva $\lambda_{2}$ representa uma partícula que realiza um movimento oscilatório. Ela também satisfaz o princípio de auto-consistência retornando à sua posição inicial. Já a curva $\lambda_{3}$ se comporta de maneira distinta das anteriores. Ela viola o principio de auto-consistencia uma vez que sua posição final $R$ é distinta da posição inicial $Q$. Neste caso, a partícula pode modificar sua história passada.

Como vemos, o princípio de auto-consistência elimina o paradoxo do avô mas limita o movimento de partículas.

\chapter{Descrição dos espaços-tempos estudados}
\label{cap6spacetime}

Como vimos, as primeiras soluções das equações de Einstein onde curvas fechadas tipo tempo são permitidas foram obtidas por Van Stockum \cite{Van} e Goedel \cite{Goedel}. Estas duas soluções e algumas outras que citamos anteriormente são apresentadas no Apêndice \ref{solucao_CTC}.
Se levarmos em conta apenas o conteúdo material necessário à formação de tais espaços-tempos, parece-nos que os exemplos citados acima confirmam a possibilidade da existência e construção de uma máquina do tempo. Contudo ainda nos resta responder algumas questões importantes sobre esse assunto. Exploraremos duas dessas questões neste trabalho:
\begin{enumerate}

\item Estes espaços-tempos são estáveis quando submetidos à uma perturba-ção em sua geometria?

\item Esta região que permite curvas fechadas tipo-tempo é acessível, ou seja, partículas materiais podem adentrar ou escapar desta região?

\end{enumerate}

Na tentativa de responder estas duas questões escolhemos um conjunto de espaços-tempos com curvas fechadas tipo tempo que nos parece significativo o bastante.
\newpage
São eles,
\begin{itemize}

\item o Cilindro cósmico estático,

\item a Corda cósmica estática,

\item a Corda cósmica em rotação e

\item o Cilindro cósmico em rotação com
 \subitem solução interior  tipo ``vaso de flor" e
 \subitem solução interior  tipo ``caneta esferográfica".

\end{itemize}

\begin{figure}[!ht]
\begin{center}
\epsfig{file =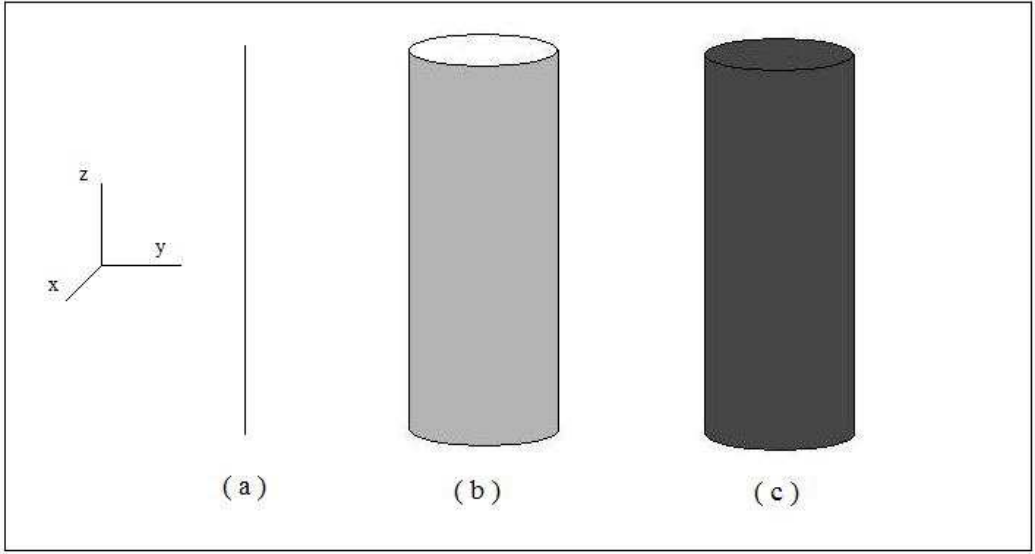, width=1.0 \linewidth, clip=}
\end{center}
\caption{{\small Representação dos espaços-tempos: (a) Corda cósmica, (b) cilindro cósmico: casca cilíndrica, (c) cilindro cósmico: cilindro de matéria. A coordenada $t$ foi suprimida nesta projeção.}}
\label{figmateria}
\end{figure}

Como pode-se observar, todos eles possuem simetria cilíndrica e são majoritariamente estacionários. Para fixarmos a nomenclatura utilizada ao longo deste trabalho chamaremos de cilindro cósmico todo espaço-tempo que possui uma região interna limitada, preenchida com algum conteúdo material, cuja fronteira repousa em $r_{s}$ e se estende infinitamente ao longo da direção $z$. Chamaremos de corda cósmica o espaço-tempo que preserva todas as características do cilindro cósmico descritas anteriormente exceto pelos fatos de que neste caso, todo conteúdo material se concentra na origem e a fronteira da corda localiza-se em $r_{s}=0$. A Fig.(\ref{figmateria}) mostra a distribuição de matéria dos espaços-tempos que estudaremos.

A corda cósmica estática e a corda em rotação são representadas pela Fig.(\ref{figmateria}-a), o cilindro em rotação com interior do tipo ``vaso de flor" pela Fig.(\ref{figmateria}-b) e o cilindro estático e o cilindro em rotação com interior do tipo ``caneta esferográfica", pela Fig.(\ref{figmateria}-c). A nomenclatura adotada para os tipos de interiores dos cilindros em rotação fica clara quando observamos a projeção da superfície $(t=constante, z=constante)$ dessas soluções na Fig.(\ref{figccr1}).
O cilindro estático e a corda cósmica estática foram as primeiras soluções a serem estudadas pela simplicidade e pela possibilidade de obtermos soluções exatas das equações de propagação de campos e das geodésicas. Apesar da simplicidade da métrica muito se pode aprender com elas. O cilindro e a corda cósmica em rotação são os primeiros espaços-tempos estudados onde aparecem curvas fechadas tipo tempo. Apesar da rotação, eles possuem métricas bem simples, fornecendo uma grande quantidade de informação sobre como se comportam campos e partículas em espaços-tempos com curvas fechadas tipo tempo. A seguir descreveremos as principais propriedades desses espaços-tempos que serão analisados neste trabalho.

\section{Cilindro cósmico estático}
\label{cordaestatica}

A solução para um cilindro cósmico estático foi obtida primeiramente por Gott III \cite{Gott} e logo em seguida por Hiscock \cite{Hiscock}.

A descrição deste espaço-tempo se resume a duas soluções das equações de Einstein que se conectam suavemente na fronteira do cilindro. A solução interior descreve a curvatura exercida pelo conteúdo de matéria do cilindro e a solução exterior descreve como o cilindro curva o espaço vazio ao seu redor.
O conteúdo de matéria será disposto de forma cilíndrica ao longo da coordenada $z$ com uma pressão longitudinal dada por $p_{z}=-\rho$. Portanto, o tensor energia-momentum que descreve essa distribuição de matéria será
\begin{eqnarray}
\label{c0}
T^{z}_{z}=T^{t}_{t}=-\frac{1}{8\pi \bar{r}_{0}^2}\quad.
\end{eqnarray}
Isso implica que a densidade de energia do cilindro será $\rho=\frac{1}{8\pi \bar{r}_{0}^2}$. A métrica que descreve a região interior será
\begin{eqnarray}
\label{c1}
ds^2=-dt^2+dr^2+dz^2+\bar{r}_{0}^2\sin(r/\bar{r}_{0})^{2}d\phi^2\quad,
\end{eqnarray}
onde os intervalos das coordenadas são $-\infty <t< \infty$, $0<r<\bar{r}_{0}\theta_{M}$, $0\leq \phi \leq 2\pi$, e $-\infty<z<\infty$. As constantes $\bar{r}_{0}$ e $\theta_{M}$ serão relacionadas aos parâmetros da solução exterior.
A métrica que descreve a região exterior será
\begin{eqnarray}
\label{c2}
ds^2=-dt^2+dr^2+dz^2+(1-4\mu)^2r^2d\phi^2\quad,
\end{eqnarray}
onde $\mu=\frac{1}{4}(1-\cos(\theta_{M}))$ é a densidade linear de massa do cilindro. Nesta região os intervalos das coordenadas são $-\infty <t< \infty$,  $0\leq \phi \leq 2\pi$, $r_{b}<r<\infty$ e $-\infty<z<\infty$. Veremos a seguir que o raio do cilindro $r_{b}$ no sistema de coordenadas exterior tem um valor diferente daquele mostrado no sistema de coordenadas interior, ou seja, ($\bar{r}_{0}\theta_{M}\ne r_{b})$ na fronteira do cilindro. Isso ocorre devido à imposição de que a circunferência descrita pela superficíe do cilindro deve ser igual nos dois sistemas de coordenadas. Os valores de $r_{b}$ dependem, portanto, desta condição e são classificados de acordo com o valor da densidade linear de massa.

Se $0\leq \mu <\frac{1}{4}$, e $\theta_{M}<\frac{\pi}{2}$ então a continuidade das componentes da métrica na fronteira imposta pelas condições de junção resulta no valor do raio do cilindro $r_{b}=\frac{\bar{r}_{0}\sin(\theta_{M})}{(1-4\mu)}$. Neste caso, a métrica interior se comporta como uma casca menor que um hemisfério. Ela é ligada à metrica exterior em $r=r_{b}$  que neste caso representa um cone de ângulo de déficit $\Delta\phi=8\pi\mu$. A condição imposta sobre $\theta_{M}$ acaba vinculando $\rho$ ao tamanho do cilindro. Escrevendo, de outro modo, temos
\begin{eqnarray}
\label{c3}
\theta_{M}=\frac{r_{s}}{\bar{r}_{0}}=r_{s}\sqrt{\rho 8 \pi}<\frac{\pi}{2}\quad,
\end{eqnarray}
onde $r_{s}$ é o raio do cilindro no sistema de coordenadas exterior. Portanto, não podemos ter um cilindro de raio $r_{s}$ com qualquer valor de densidade de energia pois a relação (\ref{c3}) deve ser preservada.

Gott ainda analisa os casos quando o parâmetro $\mu$ assume os valores $\frac{1}{4}\leq\mu\leq\frac{1}{2}$. Nestes casos, o comportamento da coordenada radial e da topologia da região exterior são alterados. Embora tenham características peculiares não os abordaremos aqui.


\section{Corda cósmica estática}

A solução da corda cósmica foi primeiramente obtida por Vilenkin \cite{Vilenkin} como uma solução aproximada das equações de Einstein no limite de campo fraco para uma distribuição singular de matéria representada pelo tensor energia-momentum
\begin{eqnarray}
\label{lc1}
T^{t}_{t}=T^{z}_{z}=-\mu \delta(x)\delta(y)\quad.
\end{eqnarray}
Nesta solução o conteúdo material da corda se concentra em $r=0$ e a densidade de energia $\mu$ é compensada por uma pressão $p$ na direção $z$ satisfazendo a equação de estado $p=-\mu$. A métrica que descreve a solução resultante, escrita em coordenadas cilíndricas, é
\begin{eqnarray}
\label{lc2}
ds^2=-dt^2+dr^2+dz^2+(1-8\mu)r^2d\phi^2\quad.
\end{eqnarray}
Contudo, esta métrica é válida apenas em primeira ordem de aproximação já que a grandes distâncias o limite de campo fraco é violado. A solução exata para a corda cósmica foi obtida por Hiscock \cite{Hiscock} como um caso limite do cilindro cósmico estático quando $r_{b}\rightarrow 0$ mantendo $\mu$ constante. Neste caso a métrica da corda será
\begin{eqnarray}
\label{lc3}
ds^2=-dt^2+dr^2+dz^2+(1-4\mu)^2r^2d\phi^2\quad.
\end{eqnarray}
Os intervalos das coordenadas serão $-\infty <t< \infty$, $0<r<\infty$, $0\leq \phi \leq 2\pi$, e $-\infty<z<\infty$. Vemos que, para pequenos valores de $\mu$, a métrica (\ref{lc3}) recupera o resultado apresentado por Vilenkin. Um aspecto importante que a métrica (\ref{lc3}) nos apresenta é que a corda não deforma o espaço-tempo ao seu redor, apenas muda sua topologia.

\begin{figure}[!ht]
\begin{center}
\epsfig{file =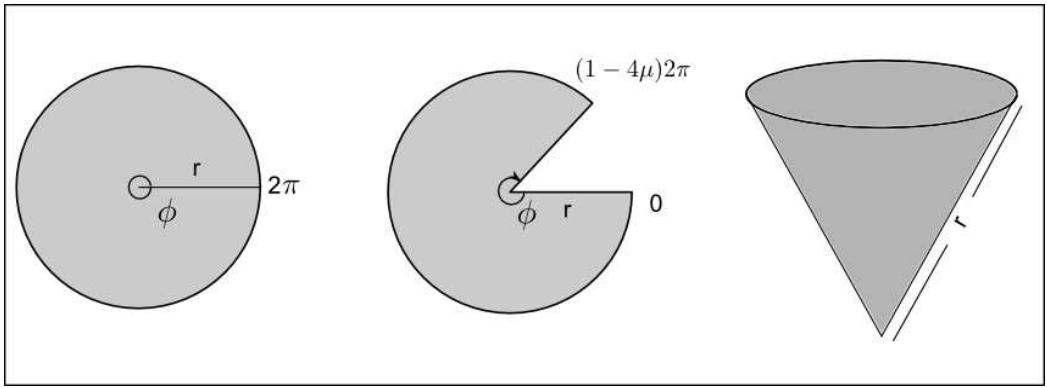, width=1.0 \linewidth, clip=}
\end{center}
\caption{{\small Disco unitário (esquerda). Disco recortado com identificação entre $0$ e $(1-4\mu)2\pi$ (centro). Cone formado pelo plano $(r,\phi)$ resultado da identificação (direita).} }
\label{figdisco}
\end{figure}

Observando-a vemos que ela se parece com a métrica de Minkowski em coordenadas cilíndricas, entretanto, por causa do fator $(1-4\mu)$ na componente $g_{\phi\phi}$, a identificação entre os pontos $0$ e $(1-4\mu)2\pi$ da coordenada angular $\phi$ acabam introduzindo um recorte no disco unitário, gerando um espaço-tempo em forma de cone. A representação desse recorte pode ser vista na Fig.(\ref{figdisco}).

A diferença entre $2\pi$ e o ângulo de identificação é dada pelo ângulo de deficit
\begin{eqnarray}
\label{lc4}
\Delta\phi=8\pi\mu\quad.
\end{eqnarray}
A presença desse ângulo de deficit foi usada por Gott \cite{Gott} no estudo de lentes gravitacionais. Esse mesmo comportamento para a coordenada $\phi$ estará presente em todos os outros espaços-tempos estudados aqui.

Usaremos a solução exata para a corda cósmica estática em nossos cálculos ao longo da tese.

\section{Corda cósmica em rotação}
\label{linharotacao}

A corda cósmica em rotação foi obtida por Deser e Jackiw \cite{Deser}. Ela se estende ao longo do eixo $z$. Sua massa por unidade de comprimento é $\mu$ e seu spin intrínseco por unidade de comprimento é $J$. A secção-transversal da linha é desprezivel, de modo que a densidade de massa é proporcional à função espacial $\delta$ bidimensional, enquanto a densidade do spin é mais singular, sendo dada por uma densidade de momentum proporcional à derivada de $\delta$.
Esta solução não tem métrica interna já que a densidade de massa só é definida em $r=0$. Portanto, a métrica da corda cósmica em rotação é
\begin{eqnarray}
\label{lr1}
ds^2=-dt^2+dr^2+dz^2-8Jd\phi dt+\left[(1-4\mu)^2r^2-16J^2\right]d\phi^2\quad,
\end{eqnarray}
onde fizemos $G=c=1$. Os intervalos das coordenadas são os mesmos apresentados para a corda cósmica estática.


Se observarmos a componente $g_{\phi\phi}$ veremos que para
\begin{eqnarray}
\label{lr2}
r<\frac{4|J|}{1-4\mu}\quad,
\end{eqnarray}
a métrica admite curvas fechadas tipo tempo.

Como na corda estática, essa métrica pode ser escrita localmente como Minkowski através da transformação de coordenadas
\begin{eqnarray}
\label{lr3}
\tilde{t}=t+4J\phi\quad, \qquad \tilde{\phi}=(1-4\mu)\phi\quad.
\end{eqnarray}

Todavia, isso também introduz um ângulo de déficit de $(1-4\mu)2\pi$ em $\tilde{\phi}$. Além disto, a coordenada $\tilde{t}$, ao invés de se comportar de modo suave e linear, dá saltos de $8\pi J$ sempre que a corda é circundada. O ângulo de déficit transforma a secção espacial do espaço-tempo em um cone que depende de $\mu$ enquanto a estrutura helicoidal do  tempo que depende de $J$ gera as curvas fechadas tipo tempo. Apesar da simplicidade da métrica essa solução é muito rica tendo muitas propriedades interessantes que serão mencionadas nos próximos capítulos.

\section{Cilindro cósmico em rotação}
\label{cordarotacao}

A solução do cilindro cósmico em rotação foi obtida por Jensen e Soleng \cite{Jensen} e difere da solução da corda em rotação pois ela apresenta soluções internas. Enquanto a região interior permite duas soluções distintas, a região exterior é descrita apenas por uma única solução.

A solução externa se assemelha bastante à solução externa da corda cósmi-ca, sendo dada pela métrica
\begin{eqnarray}
\label{cr1}
ds^2=-dt^2+dr^2+dz^2-8Jd\phi dt+\left[(1-4\mu)^2(r+r_{0})^2-16J^2\right]d\phi^2\quad,\ \ \
\end{eqnarray}
onde $\mu$ é a densidade linear de massa do cilindro e $J$ é o momento angular por unidade de comprimento. A constante $r_{0}$ determina a origem da coordenada radial exterior. Para este espaço-tempo, a condição de existência de curvas fechadas tipo tempo é
\begin{eqnarray}
\label{cr2}
r+r_{0}<\pm\frac{\sqrt{16J^2}}{(1-4\mu)}\quad.
\end{eqnarray}
Vemos, portanto, que esta condição será satisfeita dependendo do valor da constante $r_{0}$. Este aspecto será discutido no próximo capítulo.

\begin{figure}[!ht]
\begin{center}
\epsfig{file =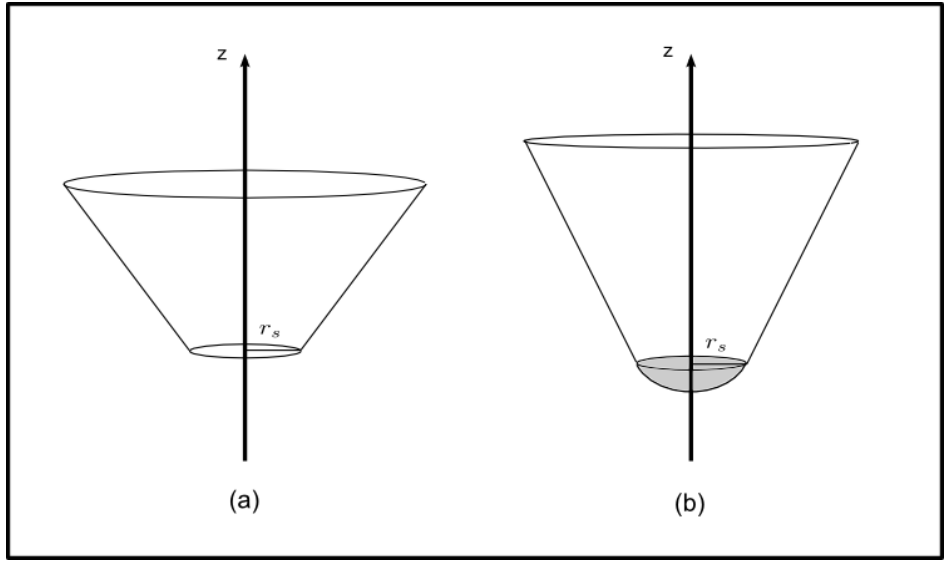, width=0.9 \linewidth, clip=}
\end{center}
\caption{{\small Projeção bidimensional das métricas interiores: (a) métrica tipo ``vaso de flor", (b) métrica tipo ``caneta esferográfica".}}
\label{figccr1}
\end{figure}

A primeira solução interior chamaremos, como no artigo original, de solução tipo ``vaso de flor". Este nome é devido ao formato da projeção da superfície $(t=constante, z=constante)$ da solução. Essa projeção é apresentada na Fig.(\ref{figccr1}-a). A solução ``vaso de flor" descreve uma casca infinitamente fina em rotação e sua métrica tem a forma
\begin{eqnarray}
\label{cr3}
ds^2=-\left[1+\frac{16J^{2}}{r^{4}_{s}}(r^{2}_{s}-r^{2})\right]dt^2-\frac{8Jr^{2}}{r^{2}_{s}}d\phi dt+r^{2}d\phi^2+dr^2+dz^2\quad,
\end{eqnarray}
onde $r_{s}$ descreve o raio onde as duas soluções são conectadas suavemente. Neste caso o raio $r_{s}$ se relaciona com $r_{0}$ da métrica exterior da forma
\begin{eqnarray}
\label{cr4}
r_{0}=\pm\sqrt{\frac{r^2_{s}+16J^2}{(1-4\mu)^2}}-r_{s}\quad.
\end{eqnarray}
Inspecionando a componente $g_{\phi\phi}$ desta solução interna observamos que curvas fechadas tipo tempo não são permitidas no interior da casca cilíndrica.

A segunda solução interna chamaremos de solução tipo ``caneta esfe-rográfica". Se observarmos a Fig.(\ref{figccr1}-b) fica evidente a razão de tal nome. A solução ``caneta esferográfica" descreve um cilindro em rotação preenchido com um conteúdo de material cuja densidade de energia é dada por
\begin{eqnarray}
\label{cr5}
8\pi \rho=\lambda+\Omega^2\quad,  \qquad \Omega=\tilde{\alpha}\lambda(r_{s}-r)\quad,
\end{eqnarray}
onde $\tilde{\alpha}\leq 1$ é uma constante relacionada ao momento angular por unidade de comprimento $J$ que podemos ajustar para termos ou não curvas fechadas tipo tempo nesta solução interior. Estas constantes apresentam relações da forma
\begin{eqnarray}
\label{cr6}
\mu=\frac{1-\cos(\sqrt{\lambda }r_{s})}{4}\quad,\nonumber\\
\nonumber\\
\label{cr6a}
J=\frac{\tilde{\alpha}}{2}\left(r_{s}-\frac{\sin(\sqrt{\lambda}r_{s})}{\sqrt{\lambda}}\right)\quad.
\end{eqnarray}
Portanto, o elemento de linha desta segunda solução é
\begin{eqnarray}
\label{cr7}
ds^2=-dt^2-2Md\phi dt+\left(\frac{\sin^2(\sqrt{\lambda}r)}{\lambda}-M^2\right)d\phi^2+dr^2+dz^2\quad,\\
\nonumber\\
M=2\tilde{\alpha}\left[(r-r_{s})\cos(\sqrt{\lambda}r)-\frac{\sin(\sqrt{\lambda}r)}{\sqrt{\lambda}}+r_{s}\right]\quad.\nonumber\ \ \ \ \
\end{eqnarray}
Neste caso, $r_{s}$ se relaciona com $r_{0}$ da métrica exterior como
\begin{eqnarray}
\label{cr8}
r_{0}=\left[\pm\frac{\sqrt{1-(1-4\mu)^2}}{(1-4\mu)\arccos{(1-4\mu)}}-1\right]r_{s}\quad.
\end{eqnarray}
Aparentemente, a raiz negativa da Eq.(\ref{cr8}) foi ignorada por Jensen, contudo ela será importante para nossa análise causal que vem a seguir.


\chapter{Análise da região não-causal}
\label{cap7analisecausal}

\begin{figure}[!ht]
\begin{center}
\epsfig{file =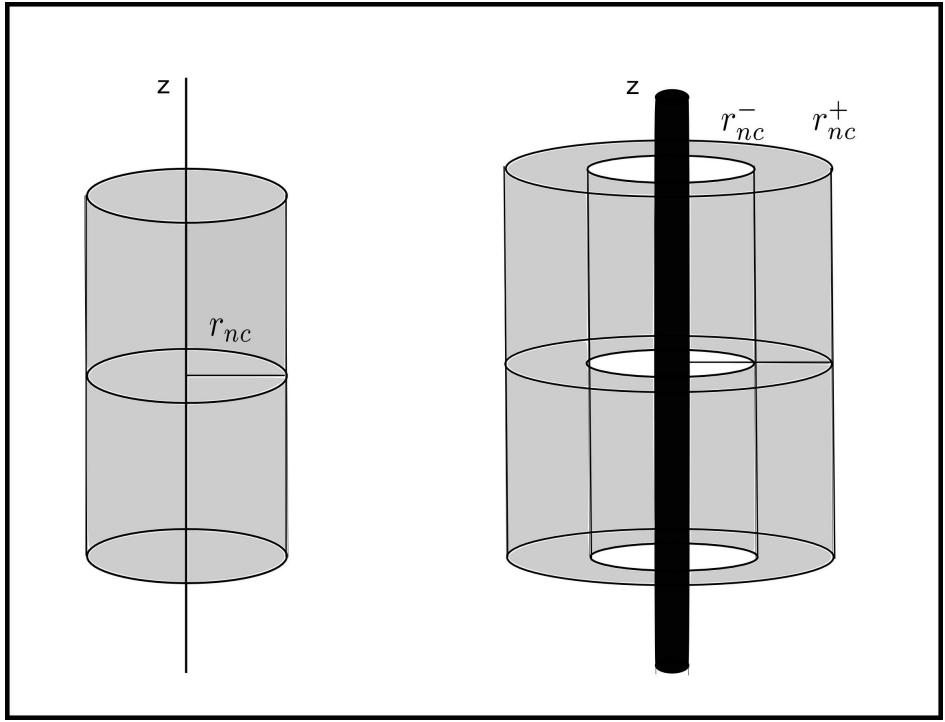, width=0.9 \linewidth, clip=}
\end{center}
\caption{{\small Formatos da região não-causal: Na corda $0<r\leq r_{nc}$ (esquerda) e nos cilindros $r_{nc}^{-}\leq r\leq r_{nc}^{+}$ (direita).}}
\label{fignaocausal}
\end{figure}

Até o presente momento, temos discutido somente propriedades gerais dos espaços-tempos descritos no capítulo anterior. Neste capítulo, apresentaremos o primeiro argumento que usaremos para defender nossa tese sobre a estabilidade de máquinas do tempo. Este primeiro argumento se relaciona com a existência de um parâmetro da solução capaz de controlar a presença de curvas fechadas tipo tempo e é resultado de uma análise mais detalhada da região do espaço-tempo que permite a existência de curvas fechadas tipo tempo. A essa região demos o nome de região não-causal. No caso dos cilindros em rotação essa região é limitada por duas superfícies tipo luz localizadas em $r_{nc}^{\pm}$. No caso da corda em rotação essa região compreende o intervalo $0<r\leq r_{nc}^{+}$.
Esses dois casos podem ser observados na Fig.(\ref{fignaocausal}).

A maior parte dos espaços-tempos aqui estudados podem ser divididos entre uma região causal, onde a noção de tempo é bem definida, e uma região não-causal onde essa noção é perdida ou alterada. Nos casos aqui estudados essas regiões são cascas cilíndricas infinitamente longas, como foi mostrado na Fig.(\ref{fignaocausal}). A existência e a extensão desta região dependerá do conjunto de parâmetros do espaço-tempo.

A presença de curvas fechadas tipo tempo em um espaço-tempo pode ser verificada através da análise da sua componente $g_{\phi\phi}$. Se existir algum lugar no espaço-tempo onde $g_{\phi\phi}<0$ para algum conjunto de parâmetros válidos, então esse espaço-tempo admite curvas fechadas tipo tempo.

Analisando a componente $g_{\phi\phi}$ tanto do cilindro cósmico estático quanto da corda estática vemos que ambas são positivas definidas. Desta forma, os casos estáticos, podem ser declarados causalmente bem comportados, ou seja, nenhuma curva fechadas tipo tempo é possível nesses espaços-tempos.

Os espaços-tempos que possuem rotação devem ser analisados caso a caso pois apresentam particularidades relacionadas à solução interior que os descreve.

\begin{figure}[!h]
\centering
\epsfig{file =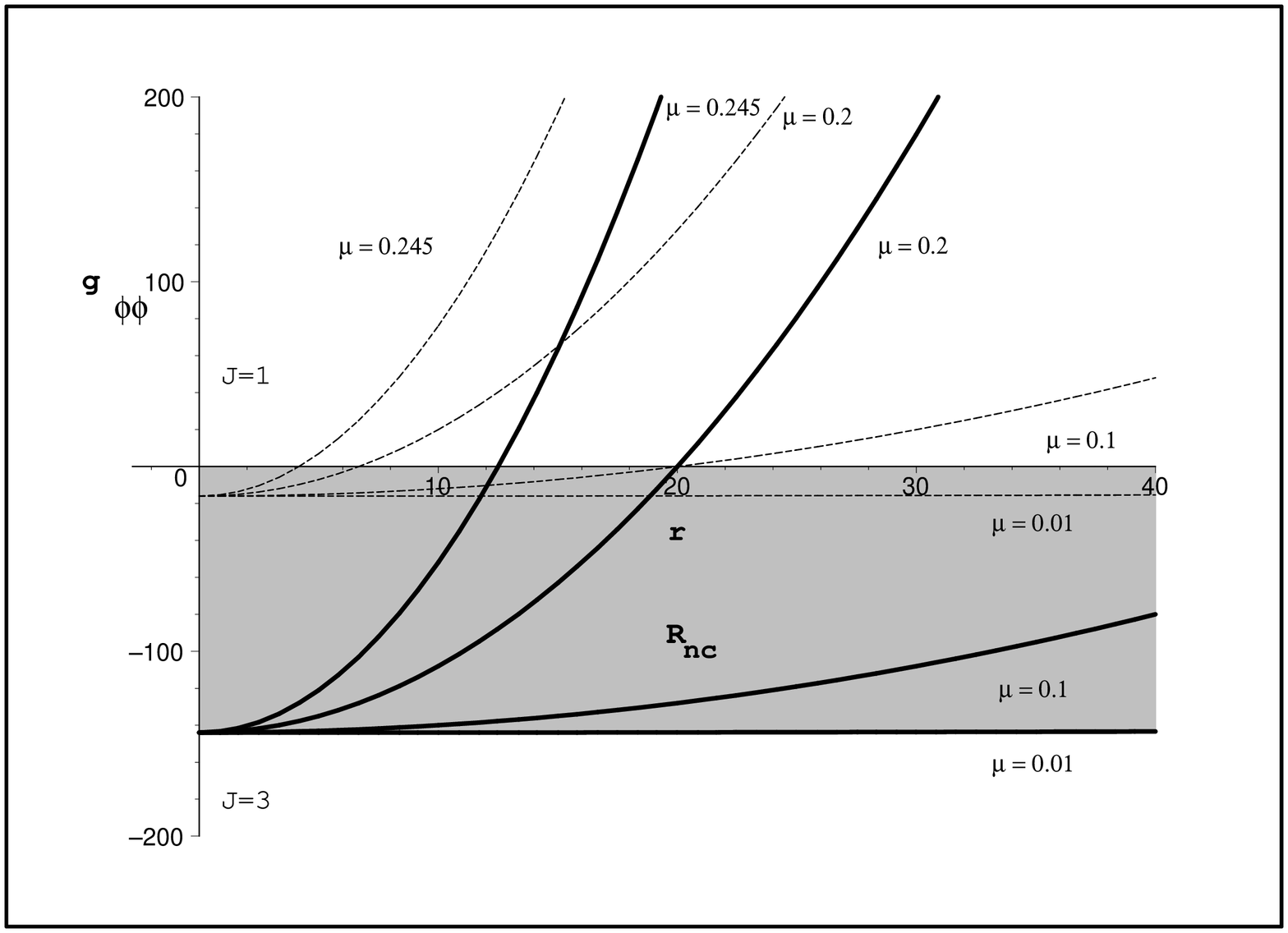, height=14cm, width=8.5cm, angle=-90} 
\epsfig{file =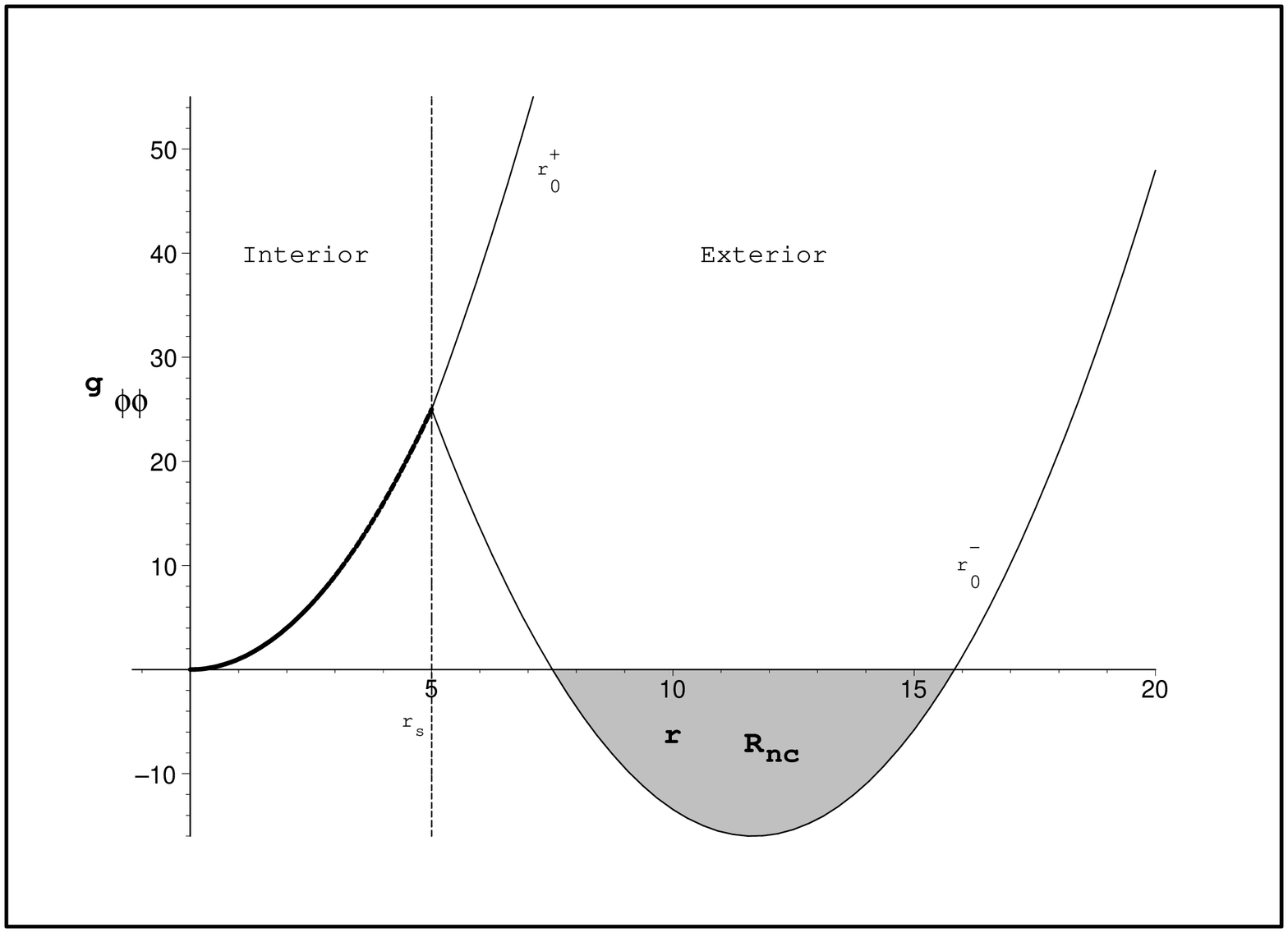, height=14cm, width=8.5cm, angle=-90}
\caption{ \small{Componente $g_{\phi\phi}$ da corda cósmica em rotação para diferentes valores de $J$ e $\mu$ (acima). Componente $g_{\phi\phi}$ da região interior e exterior do cilindro em rotação co, interior do tipo ``vaso de flor" para $r_{0}^{+}$ e $r_{0}^{-}$ satisfazendo (\ref{flcond}) com $r_{s}=5$ (abaixo).}}
\label{figomega1}
\end{figure}

Repetindo o processo de análise para a corda cósmica em rotação vemos que ela admite uma região não-causal se
\begin{eqnarray}
\label{causal1}
r_{nc}<\frac{\sqrt{16J^2}}{(1-4\mu)}\quad.
\end{eqnarray}
Portanto, a região não-causal compreende o intervalo $0<r\leq r_{nc}$. A superfície localizada em $r_{nc}$ é tipo-luz. A extensão dessa região é arbitrária, dependendo apenas das constantes $\mu$ e $J$.  Na Fig.(\ref{figomega1}) podemos observar como essas constantes mudam a extensão da região não-causal. A área em cinza entre a origem e cada uma das curvas em preto representa a região não-causal para cada par $(\mu,J)$.

Para esse espaço-tempo, nenhuma restrição é imposta aos valores de $J$. Já para densidade linear de massa impomos que $\mu<1/4$ para que o ângulo de déficit permaneça positivo.

Os dois outros casos dos cilindros cósmicos em rotação são mais complexos que a corda cósmica em rotação. Daqui para frente, manteremos a região interior de ambos os cilindros livre de curvas fechadas tipo tempo restringindo o espaço de parâmetros. Essa imposição é necessária pois queremos que a fonte material da máquina do tempo seja causalmente bem comportada. Portanto permitiremos a presença de curvas fechadas tipo tempo apenas na região exterior dos cilindros.

\begin{figure}[!h]
\centering
\epsfig{file =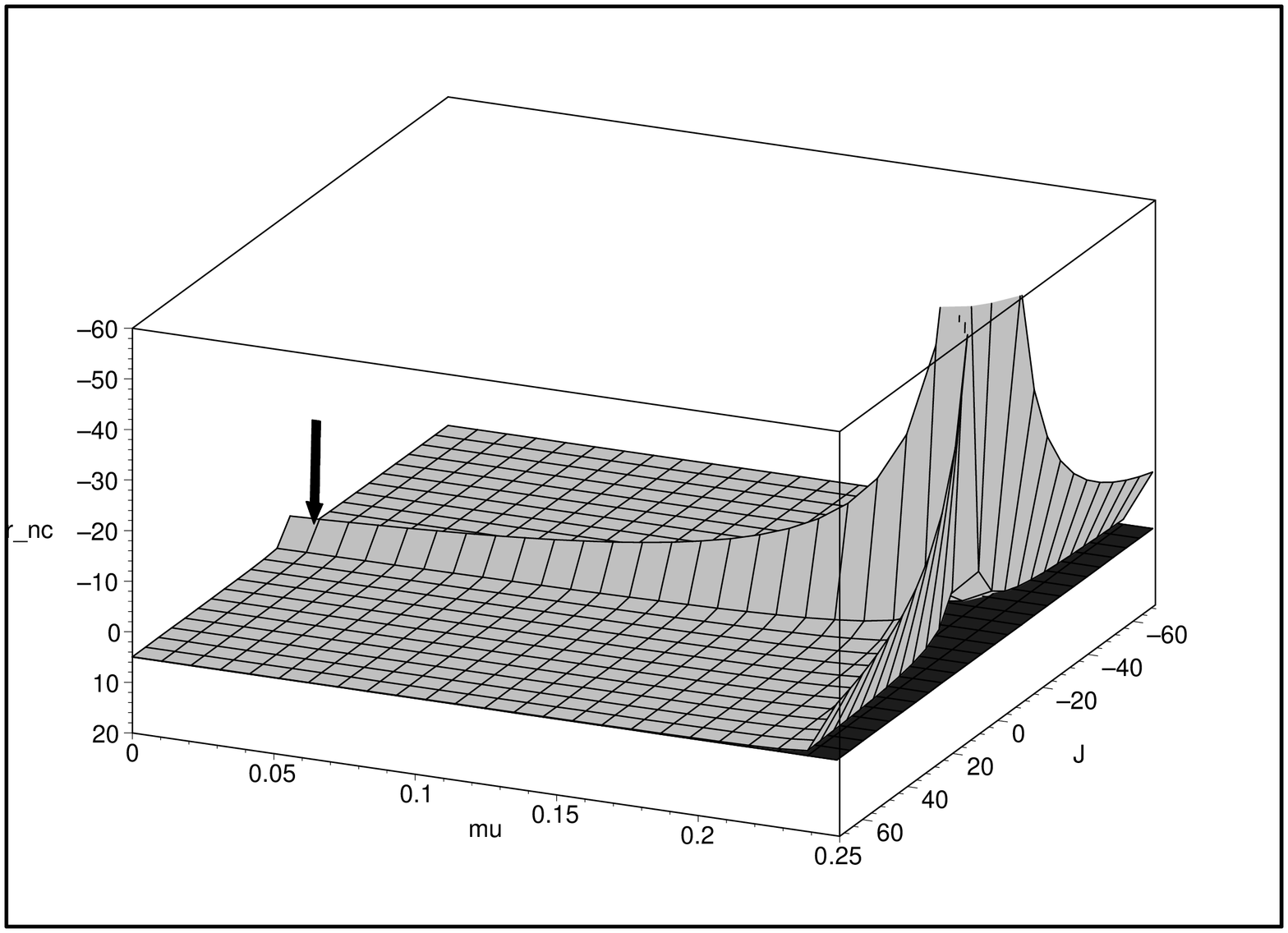,height=7cm,width=7cm, angle=-90}
\epsfig{file =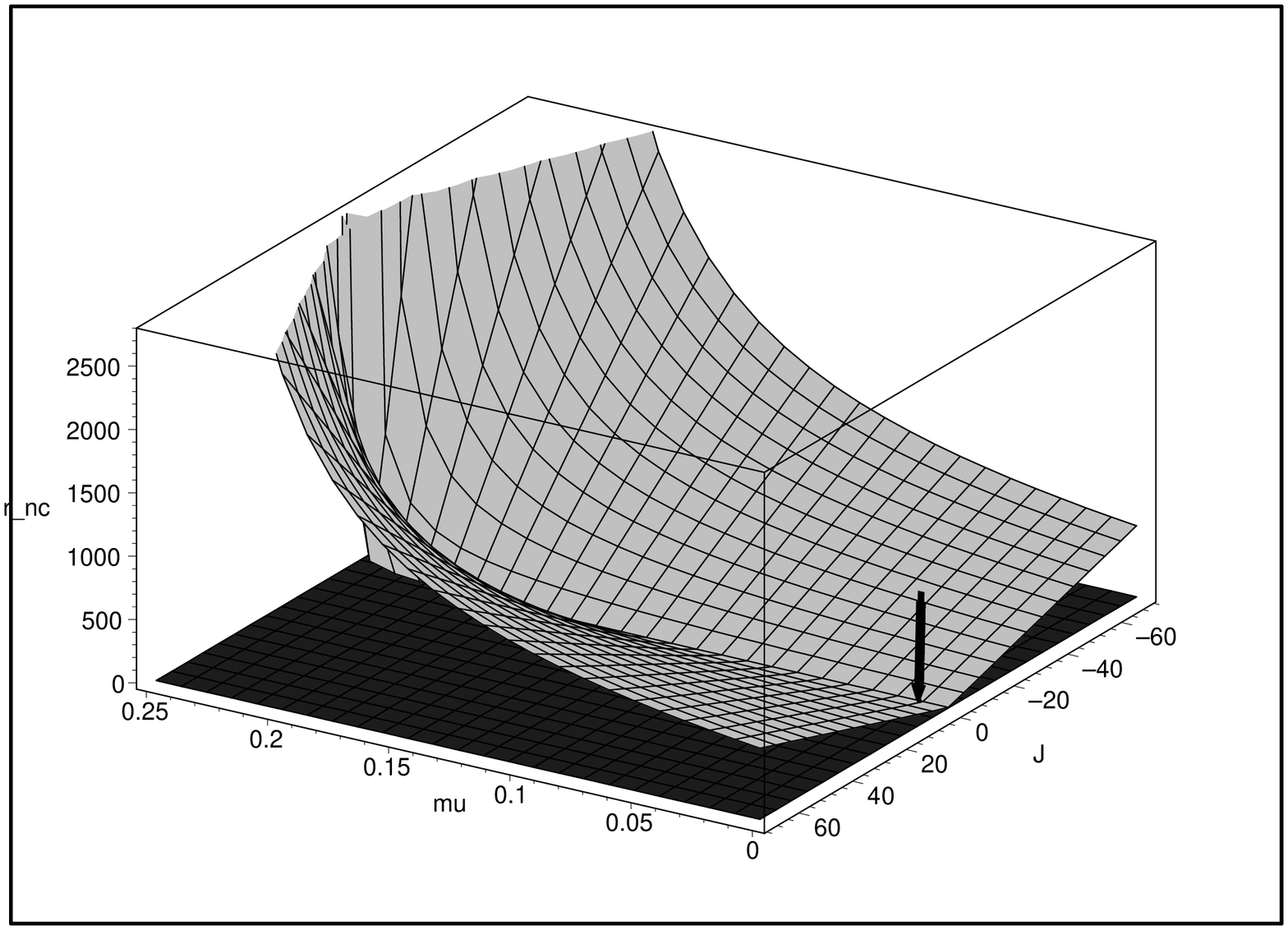,height=7cm,width=7cm, angle=-90}

\epsfig{file =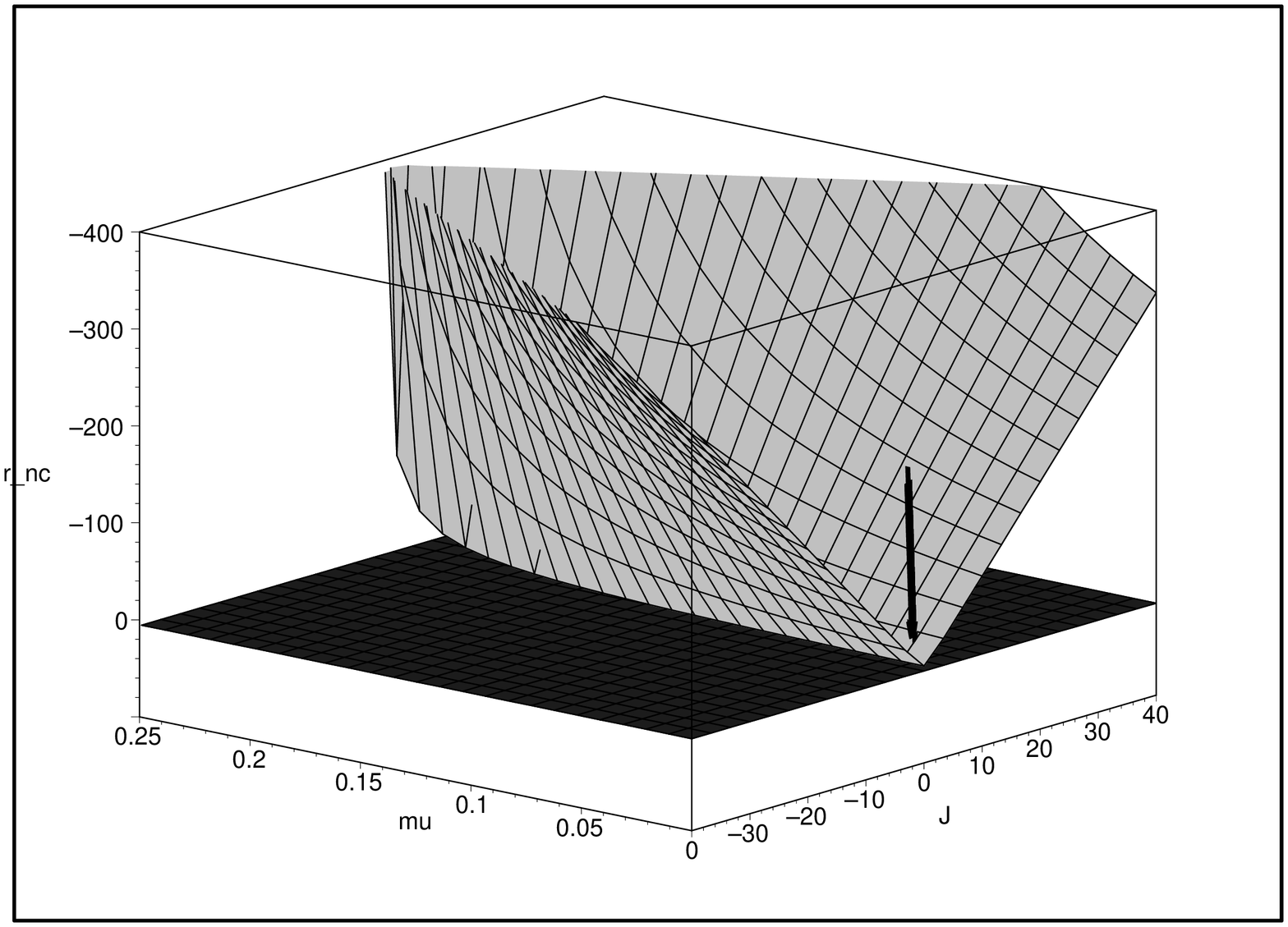,height=7cm,width=7cm, angle=-90}
\epsfig{file =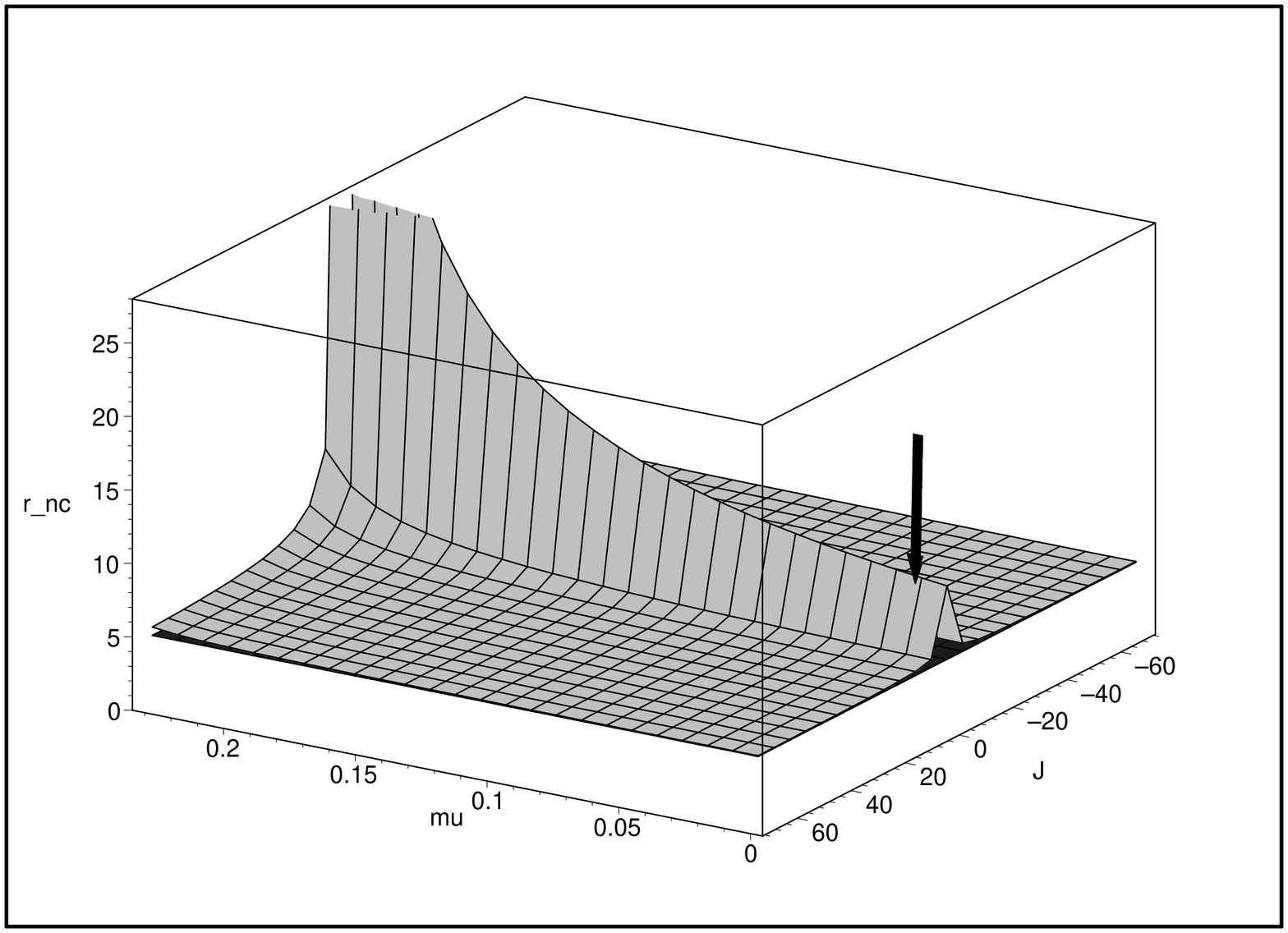,height=7cm,width=7cm, angle=-90}
\caption{{\small Posição das superfícies limite da região não-causal $r_{nc}^{\pm}$ para o cilindro em rotação com interior do tipo ``vaso de flor". O raio $r_{nc}^{+}$ é mostrado nos gráficos acima e $r_{nc}^{-}$ nos gráficos abaixo. Os gráficos da esquerda representam  a escolha de $r_{0}^{+}$ e os da direita $r_{0}^{-}$  na condição (\ref{flcond1}). A superfície escura indica a posição de $r_{s}$ e a superfície cinza indica a posição das superfícies limite da região não-causal. A flecha preta mostra a posição das superfícies limite da região não-causal para $\mu=1.0 \times 10^{-2}$ e $J=1$.}}
\label{paramflowerpot}
\end{figure}

No caso do cilindro com interior tipo ``vaso de flor" a componente $g_{\phi\phi}$ é positiva definida. Portanto, a região interior é livre de curvas fechadas tipo tempo. No exterior deste cilindro, a região não-causal se localiza entre os dois valores limites dados por
\begin{eqnarray}
\label{flcond}
r_{nc}^{-}\ <\ &r&\ <\ r_{nc}^{+}\quad, \nonumber\\
\nonumber\\
\left(-\frac{\sqrt{16J^2}}{(1-4\mu)}-r_{0}\right)\ <\ &r&\ <\ \left(\frac{\sqrt{16J^2}}{(1-4\mu)}-r_{0}\right)\quad,
\end{eqnarray}
onde a constante $r_{0}$ pode ser escolhida entre dois valores
\begin{eqnarray}
\label{flcond1}
r_{0}=r_{0}^{\pm}=\pm\sqrt{\frac{r^2_{s}+16J^2}{(1-4\mu)^2}}-r_{s}\quad.
\end{eqnarray}
A Fig.(\ref{figomega1}) mostra o comportamento da região não-causal para o cilindro cósmico em rotação com interior do tipo ``vaso de flor" para os conjunto de parâmetros $(J=1,\mu=1.0\times10^{-2})$.
Inspecionando a Eq.(\ref{flcond}) vemos que se $|r_{0}|$ for grande o bastante $r_{nc}^{+}<r_{s}$, o que nos permite concluir que a região não-causal é coberta pelo interior do cilindro, ou seja, tal região não existe. Portanto, o valor de $r_{0}$ também é um fator importante para a existência da região não-causal. Para ilustrar a importância de $r_{0}$, no aparecimento da região não-causal, mostramos na Fig.(\ref{paramflowerpot}) a posição das superfícies limites $r_{nc}^{\pm}$ para o cilindro cósmico em rotação com interior do tipo ``vaso de flor" com $r_{s}=5$.

Quando a constante $r_{0}^{+}$ é escolhida em (\ref{flcond1}), observamos que as superfícies limite da região não-causal $r_{nc}^{\pm}$ sempre se localizam antes de $r_{s}$. Para essa escolha, o exterior do cilindro em rotação não admite curvas fechadas tipo tempo, isto é, ele é causalmente bem comportado. Por outro lado, quando escolhemos $r_{0}^{-}$ as superfícies limite da região não-causal sempre se localizam depois da fronteira do cilindro, ou seja, $r_{nc}^{\pm}>r_{s}$ permitindo assim a exis-tência de curvas fechadas tipo tempo no exterior do cilindro. Desta forma, concluimos que a escolha do parâmetro $r_{0}$ dita a presença de curvas fechadas tipo tempo nesse espaço-tempo. Esta conclusão independe do valor de $r_{s}$. Neste caso, nenhuma imposição é necessária para $J$ mas ainda impomos que $\mu<1/4$.

No caso do cilindro em rotação com o interior do tipo ``caneta esfero-gráfica" tanto a região interior quanto a região exterior permitem a existência de curvas fechadas tipo tempo. Por causa do número de parâmetros livres desta solução, uma análise geral é mais complicada. Assim, cada conjunto de parâmetros deve ser analisado independentemente. Contudo, algumas res-trições ao conjunto de parâmetros podem ser impostas de forma a restringir o espaço de parâmetros e tornar nossa análise causal mais geral.

São elas:
\begin{itemize}

\item Preservar a região interior de curvas fechadas tipo tempo verificando se os parâmetros escolhido mantêm a componente $g_{\phi\phi}$ interior positiva definida.

\item A densidade linear de massa deve ser satisfazer a relação $\mu<1/4$ e $\lambda$ deve respeitar $\sqrt{\lambda}r_{s}<\pi/2$.

\end{itemize}

A segunda restrição está relacionada ao comportamento da coordenada radial no exterior solução do cilindro em rotação. Uma restrição semelhante, $\theta_{M}<\pi/2$, foi discutida para a solução do cilindro estático.

\begin{figure}[!h]
\centering
\epsfig{file =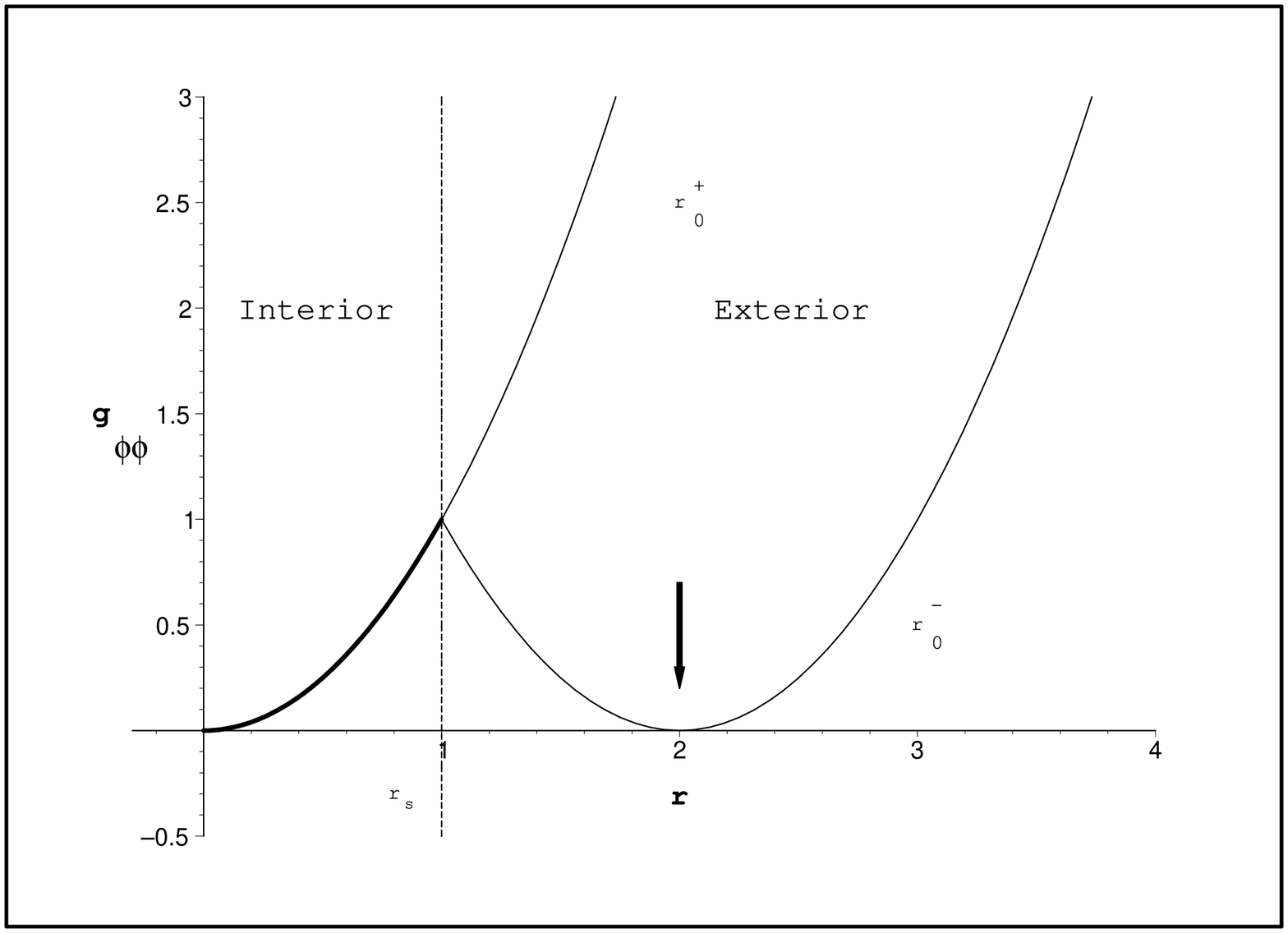, angle=-90, width=0.85 \linewidth}
\epsfig{file =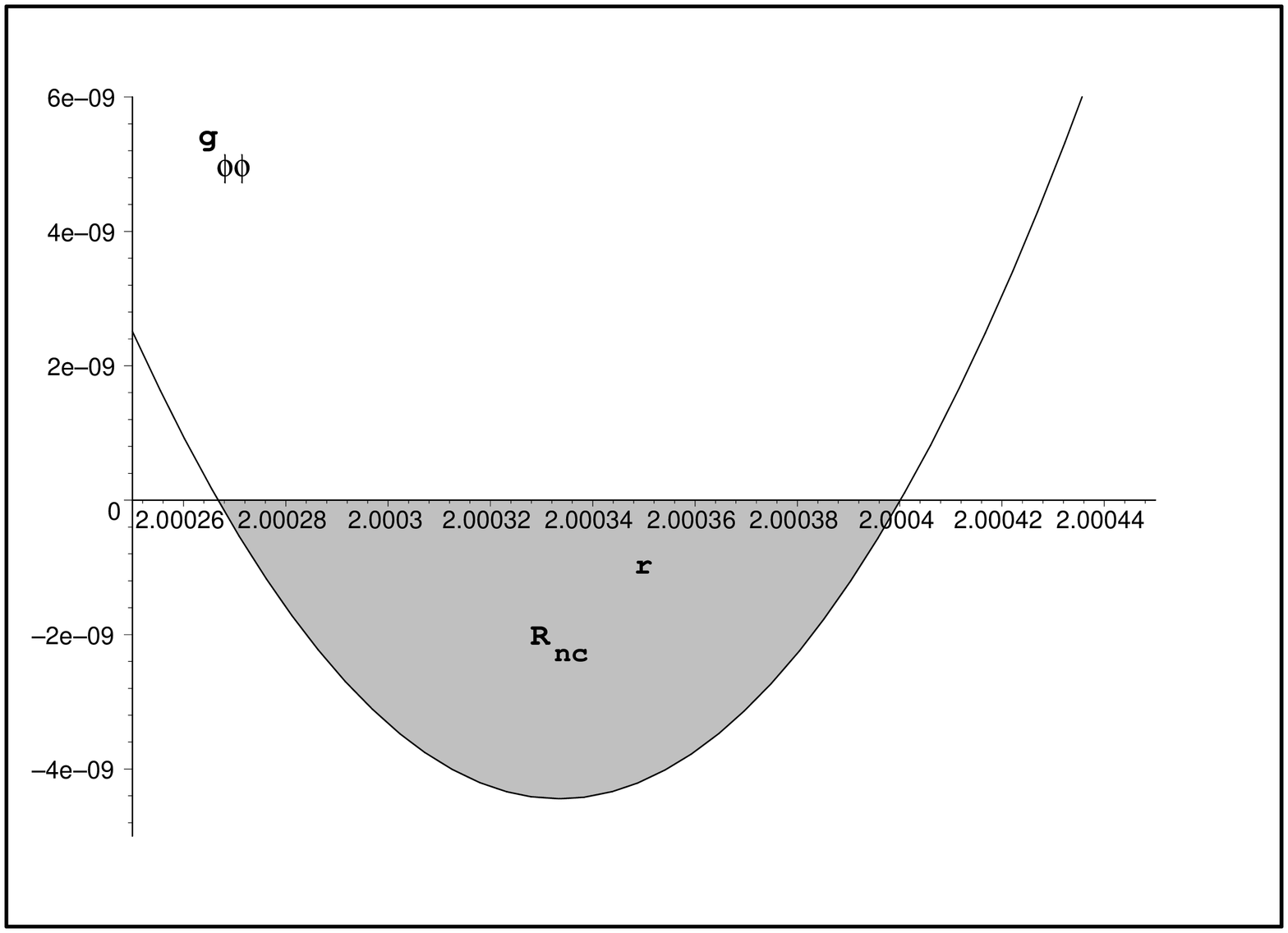, angle=-90, width=0.85 \linewidth}
\caption{Componente $g_{\phi\phi}$ da região interior e exterior do cilindro em rotação com interior do tipo ``caneta esferográfica" para $r_{0}^{+}$ e $r_{0}^{-}$ satisfazendo (\ref{rctcbp}) com $(r_{s}=1, \tilde{\alpha}=0.2, \lambda=1.0\times10^{-3})$ (acima). Detalhes da região não-causal (abaixo).}
\label{figomega1a}
\end{figure}

A estrutura causal desse espaço-tempo dependerá do ajuste das três constantes $r_{s}$, $\lambda$ e $\tilde{\alpha}$ que devem ser escolhidas adequadamente segundo as res-trições declaradas anteriormente. Na Fig.(\ref{figomega1a}) mostramos um exemplo do comportamento da componente $g_{\phi\phi}$ na região interior e exterior do cilindro cósmico em rotação com interior do tipo ``caneta esferográfica".

Estabeleçamos então a condição para existência de curvas fechadas tipo tempo no exterior desse espaço-tempo analisando a componente $g_{\phi\phi}$ do exterior do cilindro com interior do tipo ``caneta esferográfica".

\begin{figure}[!h]
\epsfig{file =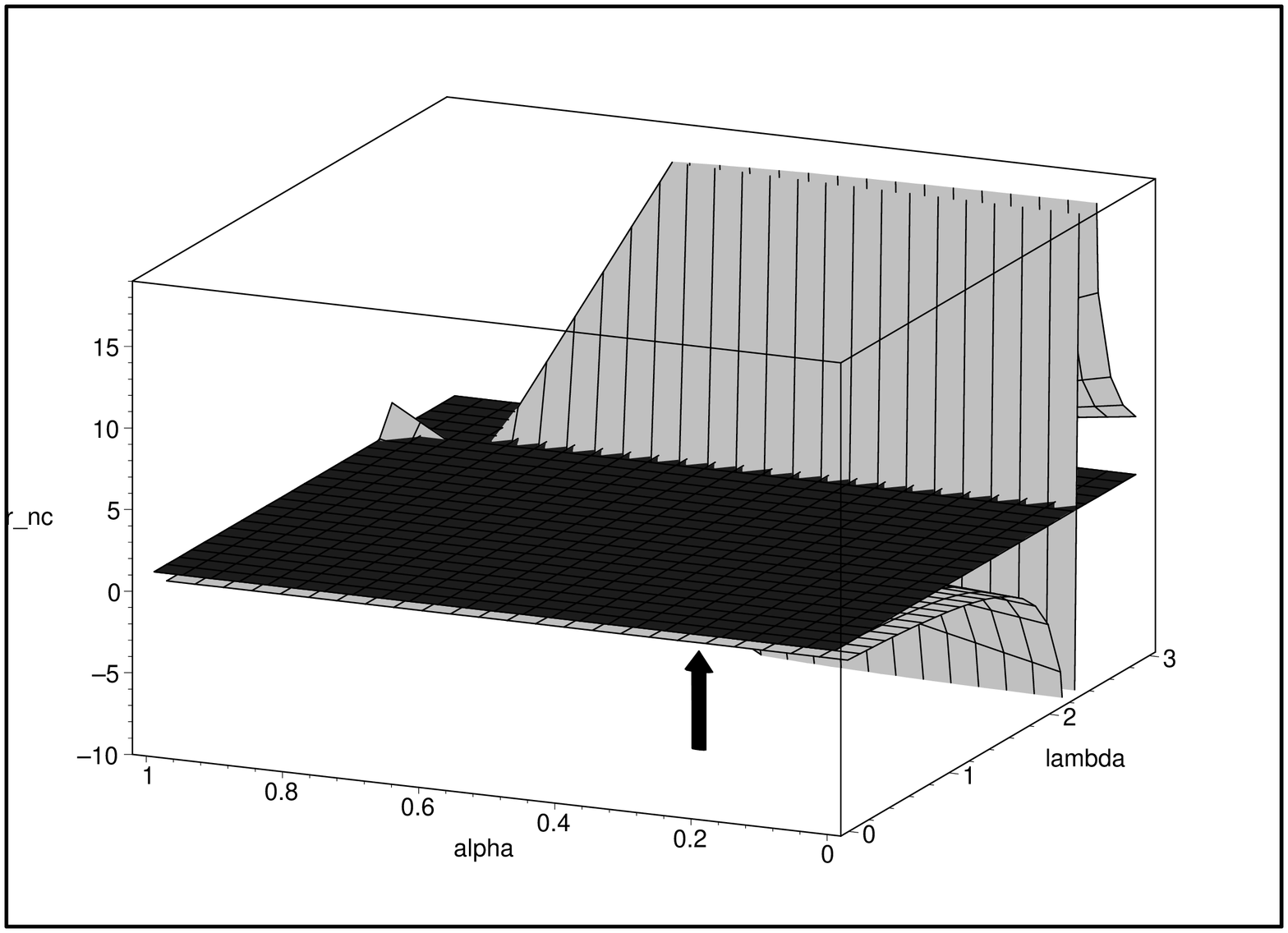,height=7cm,width=7cm, angle=-90}
\epsfig{file =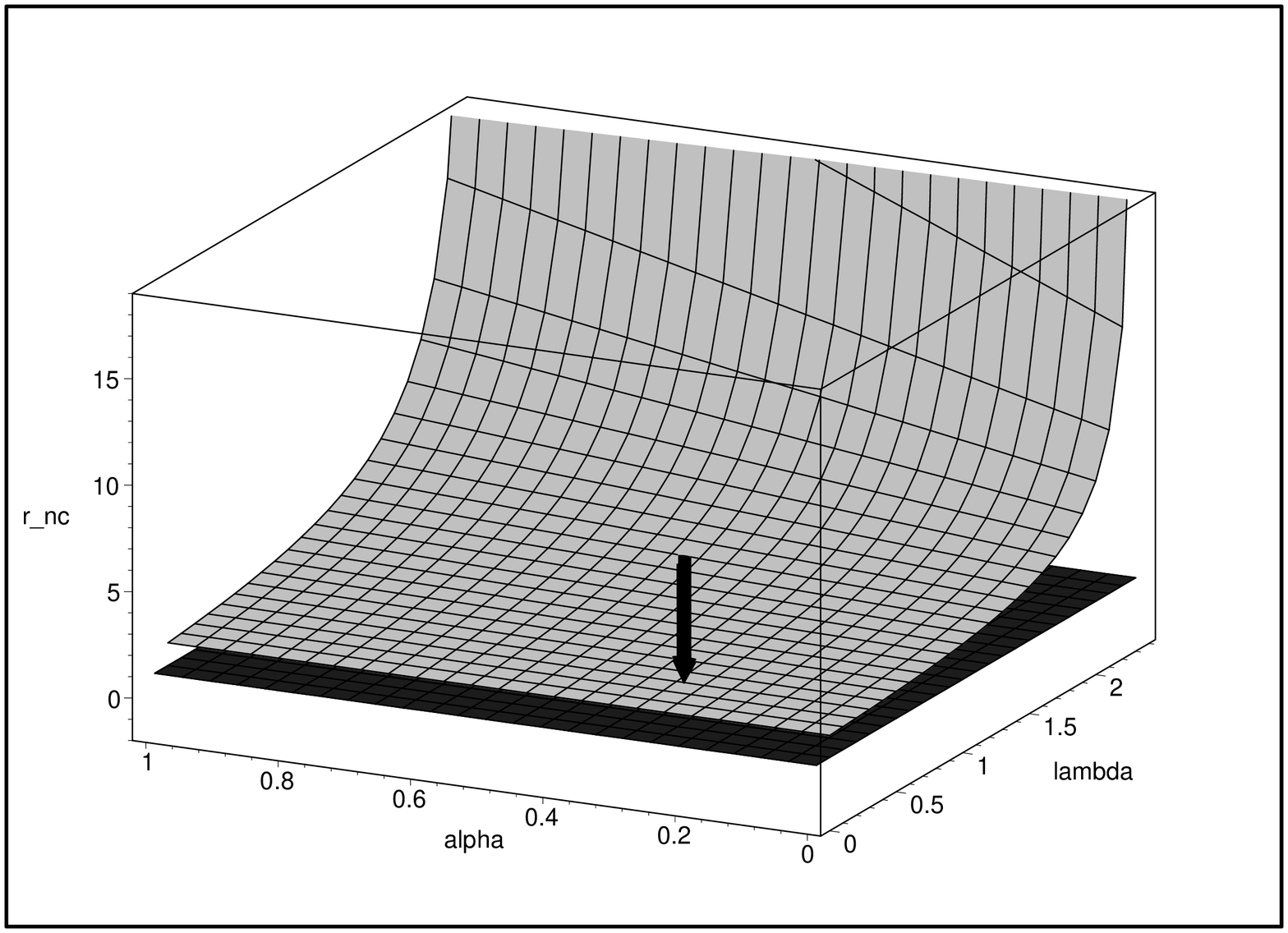,height=7cm,width=7cm,angle=-90}

\epsfig{file =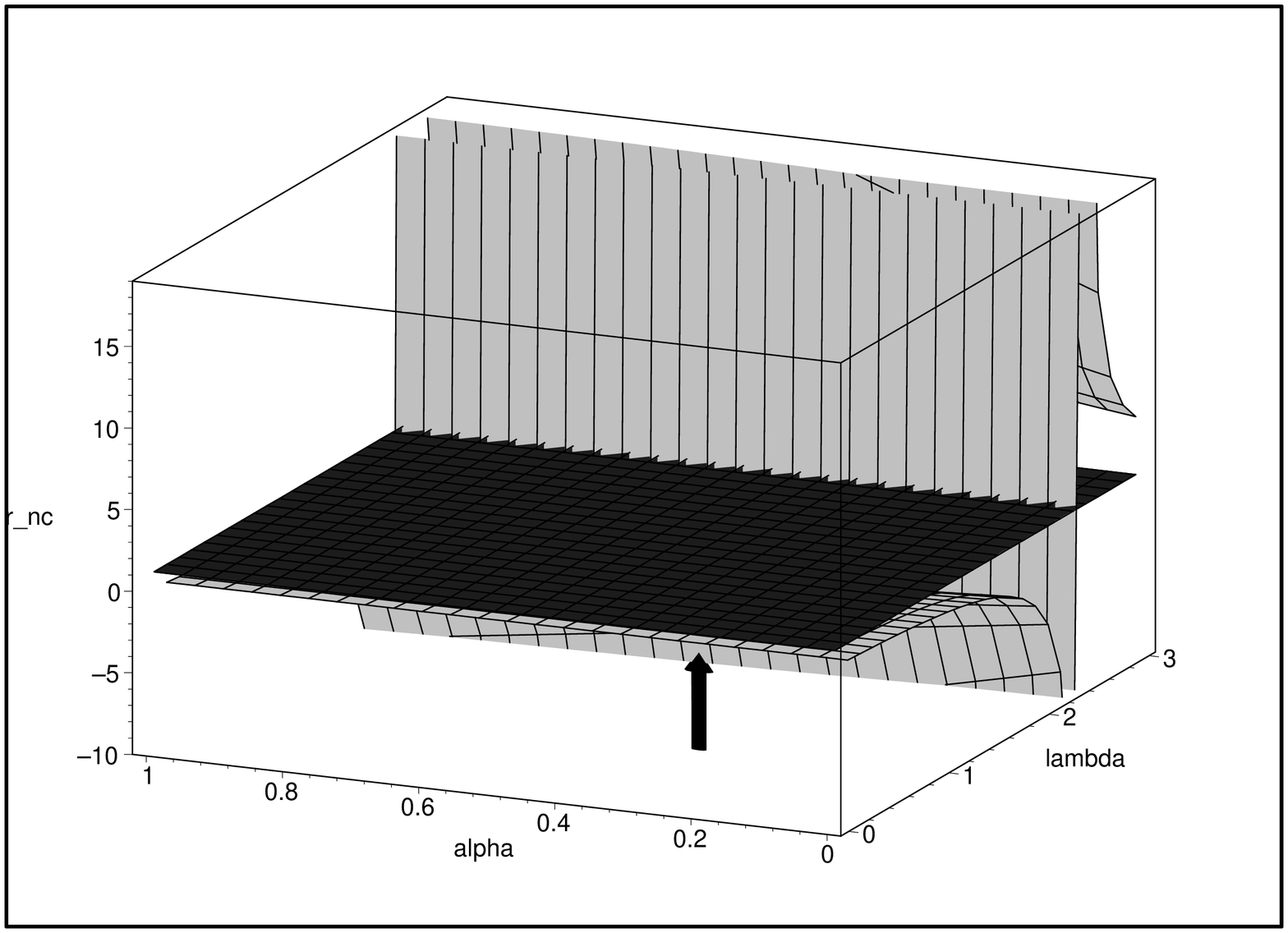,height=7cm,width=7cm, angle=-90}
\epsfig{file =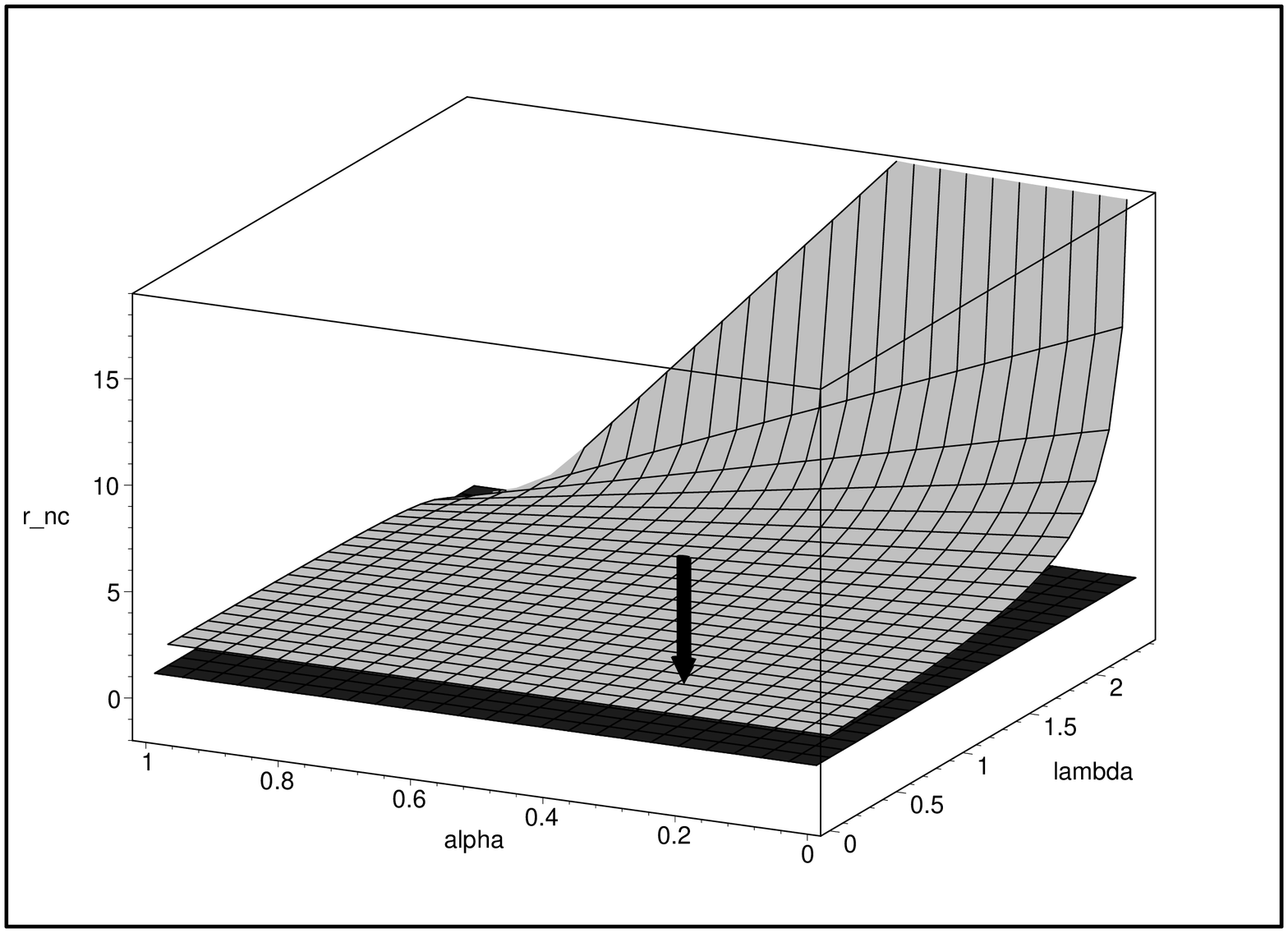,height=7cm,width=7cm,angle=-90}
\caption{{\small Posição das superfícies limite da região não-causal $r_{nc}$ para o cilindro em rotação com interior do tipo ``caneta esferográfica". O raio $r_{nc}^{+}$ é mostrado na parte superior e $r_{nc}^{-}$ na parte inferior. Os gráficos da esquerda representam  a escolha de $r_{0}^{+}$ e os da direita $r_{0}^{-}$  na condição (\ref{rctcbp1}). A superfície escura indica a posição de $r_{s}$ e a superfície cinza indica a posição das superfícies limite da região não-causal. A flecha preta indica a posição das superfícies limite da região não-causal para $( r_{s}=1, \lambda=1.0\times 10^{-3}, \tilde{\alpha}=0.2 )$.}}
\label{rctcbpcausal}
\end{figure}

Neste caso, a condição para a presença de uma região não-causal também dependerá de $r_{0}$. A região não-causal se estenderá por todo o intervalo
\begin{eqnarray}
\label{rctcbp}
r_{nc}^{-}\ <\ &r& \ <\  r_{nc}^{+}\quad, \nonumber\\
\nonumber\\
\left(-\frac{\sqrt{16J^2}}{(1-4\mu)}-r_{0}\right)\ <\ &r& \ <\  \left(-\frac{\sqrt{16J^2}}{(1-4\mu)}-r_{0}\right)\quad,
\end{eqnarray}
onde o parâmetro $r_{0}$ pode assumir dois valores,
\begin{eqnarray}
\label{rctcbp1}
 r_{0}=r_{0}^{\pm}=\pm\frac{\sqrt{1-(1-4\mu)^2}\ r_{s}}{(1-4\mu)\arccos(1-4\mu)}-r_{s}\quad.
\end{eqnarray}
As superfícies limite da região não-causal localizadas em $r_{nc}^{\pm}$ são do tipo luz. Novamente a constante $r_{0}$ assume um papel importante na análise das curvas fechadas tipo tempo. Vejamos a seguir um exemplo que captura a essência dos argumentos desenvolvido aqui. Para tal, escolhemos o conjunto de parâmetros $(\lambda=1.0\times10^{-3}, \tilde{\alpha}=0.2, r_{s}=1)$ que mantêm a componente $g_{\phi\phi}$ interior positiva definida. Na Fig.(\ref{rctcbpcausal}) mostramos a posição das superfícies limite da região não-causal $r_{nc}^{\pm}$ para o cilindro cósmico em rotação com interior do tipo ``caneta esferográfica". Nosso conjunto será representado no gráfico por uma flecha preta.

Quando escolhemos a constante $r_{0}^{+}$ na condição (\ref{rctcbp1}), vemos que a posição da superfície limite depende dos valores de $\lambda$ e $\tilde{\alpha}$. Se $\lambda$ e $\tilde{\alpha}$ forem grande o bastante teremos a presença de curvas fechadas tipo tempo na região exterior. Contudo, neste caso também teremos curvas fechadas tipo tempo na região interior e dessa forma violaremos uma das restrições impostas. Portanto, podemos variar os valores de $\lambda$ e $\tilde{\alpha}$ somente até o ponto onde as curvas fechadas tipo tempo são ausentes no interior da solução. O exemplo apresentado, onde $\lambda=10^{-3}$, $r_{s}=1$ e $\tilde{\alpha}=0.2$ satisfaz as restrições impostas à região interior. Neste caso, as superfícies limite da região não-causal $r_{nc}^{\pm}$ são sempre menores que $r_{s}$, ou seja, ele não apresenta curvas fechadas tipo tempo na região exterior.

Se escolhermos a constante $r_{0}^{-}$ na condição (\ref{rctcbp1}), vemos que a posição da superfície limite da região não-causal na região exterior também depende de $\lambda$ e $\tilde{\alpha}$. Novamente, a região interior será livre de curvas fechadas tipo tempo se $\lambda$ e $\tilde{\alpha}$ não forem muito grande. Todavia, neste caso, as superfícies da região não-causal $r_{nc}^{\pm}$ estão localizadas sempre depois de $r_{s}$. Portanto, escolher a constante $r_{0}^{-}$ implica em permitir a presença de curvas fechadas tipo tempo no exterior do espaço-tempo.

Mais uma vez observamos que os parâmetros $r_{0}^{\pm}$ foram cruciais para o controle da presença de uma região não-causal no espaço-tempo.

\chapter{Geodésicas}
\label{cap8geodesicas}

Neste capítulo exploraremos as órbitas de partículas movimentando-se nos espaços-tempos estudados. Discutiremos as condições para a existência de geodésicas fechadas tipo tempo circulares.
\\
Após a apresentação das principais características dos espaços-tempos que estudaremos devemos nos concentrar em obter informações adicionais sobre a dinâmica de partículas.

Focaremos nossa atenção apenas na dinâmica de partículas no exterior dos espaços-tempos apresentados. Realizando esse estudo de maneira unificada, analisaremos apenas o exterior do cilindro cósmico em rotação, pois todos os outros são casos particulares desse último. Quando necessário, discussões complementares serão apresentadas.
Nos casos dos espaços-tempos apresentados no capítulo \ref{cap6spacetime} essas equações de movimento são resolvidas exatamente em termos de um parâmetro afim $\tau$.
Para demonstrar essa afirmação resolveremos as equações de movimento de partículas apresentadas no capítulo \ref{cap2geo} para a métrica da região exterior do cilindro cósmico em rotação e discutiremos suas respectivas dinâmicas.
Desenvolvendo as Eqs.(\ref{cap2geo2}, \ref{cap2geo3}, \ref{cap2geo4}, \ref{cap2geo5}) para a métrica ($\ref{cr1}$) temos
\begin{eqnarray}
\label{cap8geo2}
\dot{\phi}&=&\frac{\ell+4JE}{\alpha^2(r+r_{0})^2}\quad,\\
\nonumber\\
\label{cap8geo3}
\dot{t}&=&E+\frac{4J\left(\ell+4JE\right)}{\alpha^2(r+r_{0})^2}\quad,\\
\nonumber\\
\label{cap8geo4}
\dot{z}&=&p_{z}\quad,\\
\nonumber\\
\label{cap8geo5}
\dot{r}^2&=&E^2-\epsilon-p_{z}^{2}-\frac{\left(\ell+4JE\right)^2}{\alpha^2(r+r_{0})^2}\quad.
\end{eqnarray}
Aqui, as constantes de integração $\ell$ e $E$ não serão identificadas, em princípio, com momento angular e energia de uma partícula massiva ou de um fóton pois a estrutura assintótica do espaço-tempo não é plana devido a presença do termo $g_{\phi t}$. No caso da partícula massiva devemos normalizar estas constantes pela massa de repouso $m_{p}$ da partícula. Escolhemos $m_{p}=1$ pois tal escolha não reduz a generalidade da análise.

Para encontrarmos as soluções exatas das equações acima enunciadas integramos a Eq.(\ref{cap8geo5}) em função do parâmetro afim $\tau$
\begin{eqnarray}
\label{cap8geo6}
\int \frac{dr}{\sqrt{E^2-\epsilon-p_{z}^{2}-\frac{\left(\ell+4JE\right)^2}{\alpha^2(r+r_{0})^2}}}=\pm \int d\tau\quad,
\end{eqnarray}
resultando em
\begin{eqnarray}
\label{cap8geo7}
\tau = \pm  \sqrt{\left(E^2-\bar{E}^2\right)(r+r_{0})^2-\frac{\left(\ell+4JE\right)^2}{\alpha^2}}+ C_{1}\quad,
\end{eqnarray}
onde fizemos a substituição $\bar{E}^2=\epsilon+p_{z}^2$. Invertendo a Eq.(\ref{cap8geo7}) obtemos a função $r=r(\tau)$
\begin{eqnarray}
\label{cap8geo8}
\left(r+r_{0}\right)^2=\frac{\alpha^2\left(E^2-\bar{E}^2\right)^2\left(\tau-C_{1}\right)^2+\left(\ell+4JE\right)^2}{\alpha^2\left(E^2-\bar{E}^2\right)}\quad,\\
\nonumber\\
\label{cap8geo9}
r(\tau)=\pm\sqrt{\frac{\alpha^2\left(E^2-\bar{E}^2\right)^2\left(\tau-C_{1}\right)^2+\left(\ell+4JE\right)^2}{\alpha^2\left(E^2-\bar{E}^2\right)}}-r_{0}\quad.
\end{eqnarray}
A seguir, substituindo a Eq.(\ref{cap8geo8}) na Eq.(\ref{cap8geo2}) temos
\begin{eqnarray}
\label{cap8geo10}
\frac{d\phi}{d\tau}=\frac{\left(\ell+4JE\right)\left(E^2-\bar{E}^2\right)}{\alpha^2\left(E^2-\bar{E}^2\right)^2\left(\tau-C_{1}\right)^2+\left(\ell+4JE\right)^2}\quad.
\end{eqnarray}
Integrando agora a Eq.(\ref{cap8geo10}) temos a função $\phi=\phi(\tau)$
\begin{eqnarray}
\label{cap8geo11}
\phi(\tau)=\frac{1}{\alpha}\arctan\left(\frac{\alpha\left(E^2-\bar{E}^2\right)(\tau-C_{1})}{\ell+4JE}\right)+C_{2}\quad.
\end{eqnarray}
Repetindo o mesmo processo para a Eq.(\ref{cap8geo3}) teremos a função $t=t(\tau)$
\begin{eqnarray}
\label{cap8geo12}
t(\tau)=\tau E-\frac{4J}{\alpha}\arctan\left(\frac{\alpha\left(E^2-\bar{E}^2\right)(\tau-C_{1})}{\ell+4JE}\right)+C_{3}\quad.
\end{eqnarray}
No caso da Eq.(\ref{cap8geo4}) o resultado é
\begin{eqnarray}
\label{cap8geo12a}
z(\tau)= \tau\ p_{z}+C_{4}\quad.
\end{eqnarray}
As constantes $C_{1}, C_{2}, C_{3}, C_{4}$ são definidas pelas condições iniciais. Essas são as equações que descrevem as coordenadas $(t,\phi,r,z)$ em função do parâmetro afim $\tau$.

Voltando nossa atenção novamente para a Eq.(\ref{cap8geo5}) percebemos que o comportamento da velocidade radial é praticamente o mesmo para partículas massivas e fótons.

Iniciaremos nossa análise do comportamento de fótons e partículas massivas nesse espaço-tempo fazendo uma redefinição das equações movimento. Elas serão reescritas como
\begin{eqnarray}
\label{foton1}
\dot{\phi}&=&E\left(\frac{L+4J}{\alpha^2(r+r_{0})^2}\right)\quad,\\
\nonumber\\
\label{foton2}
\dot{t}&=&E\left(1+\frac{4J\left(L+4J\right)}{\alpha^2(r+r_{0})^2}\right)\quad,\\
\nonumber\\
\label{foton3}
\dot{z}&=&E\ \bar{P}\quad,\\
\nonumber\\
\label{foton4}
\dot{r}^2&=&E^2\left(1-\bar{P}^{2}-\frac{\epsilon}{E^2}-\frac{\left(L+4J\right)^2}{\alpha^2(r+r_{0})^2}\right)\quad,
\end{eqnarray}
onde as constantes $L$ e $\bar{P}$ são definidas como
\begin{eqnarray}
\label{foton5}
L=\frac{\ell}{E}, \qquad \bar{P}=\frac{p_{z}}{E}.
\end{eqnarray}
Inspecionando as Eqs.(\ref{foton1},\ref{foton2},\ref{foton3},\ref{foton4}) vemos que as características das órbitas vão depender do valor da constante $(L+4J)$. Por isso, separaremos os tipos órbitas em três regimes distintos:
\begin{itemize}

\item $(L+4J)=0$ órbitas radiais,

\item $(L+4J)>0$ órbitas diretas,

\item $(L+4J)<0$ órbitas retrógradas.

\end{itemize}

Discutiremos cada caso separadamente prestando especial atenção às geo-désicas nula e sua relação com a estrutura causal do espaço-tempo analisado.

\section{Órbitas radiais $(L+4J)=0$}

Como sabemos, o estudo de geodésicas nulas $(\epsilon=0)$ radiais em um espaço-tempo pode nos revelar muitos aspectos da sua estrutura causal. No caso do cilindro em rotação, a condição necessária para existência de órbitas radiais pode ser obtida através da Eq.(\ref{foton1}) fazendo
\begin{eqnarray}
\label{geod_f1}
(L+4J)=0 \quad \Longrightarrow \quad \dot{\phi}=0\quad.
\end{eqnarray}
Impondo a condição (\ref{geod_f1}) às outras equações de movimento resulta no conjunto
\begin{eqnarray}
\label{geod_f2}
\dot{t}&=&E\quad,\\
\nonumber\\
\label{geod_f3}
\dot{z}&=&E\ \bar{P}\quad,\\
\nonumber\\
\label{geod_f4}
\dot{r}^2&=&E^2\left(1-\bar{P}^{2}\right)\quad.
\end{eqnarray}
Vejamos então quais os tipos de órbitas radiais que os fótons podem realizar. Se escolhermos $\bar{P}=\pm 1$ o resultado será um fóton se propagando paralelamente ao longo do eixo $z$ para cima $(+)$  ou para baixo $(-)$ mantendo um raio $r_{f}$ fixo com energia $E$. Isso pode ser confirmado derivando a Eq.(\ref{geod_f4}) em relação a $\tau$ tendo como resultado $\ddot{r}=\dot{r}=0$, ou seja, não há movimento na direção radial.
\begin{figure}[!h]
\centering
\epsfig{file =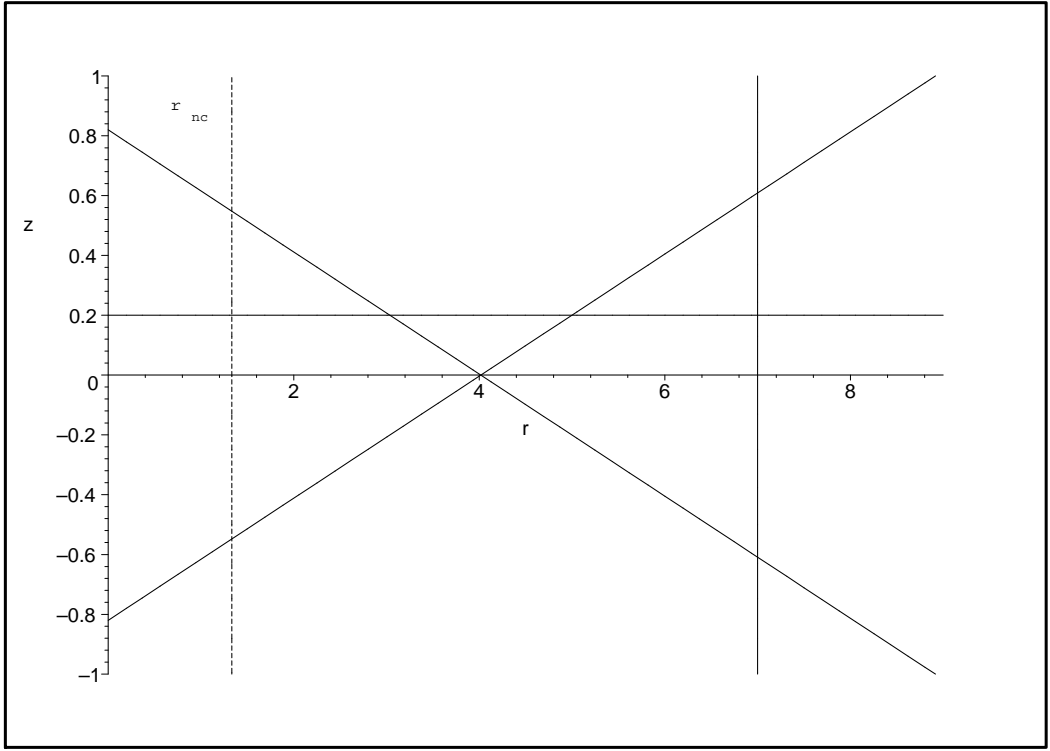, angle=-90, width=1.0 \linewidth}
\caption{Projeção no plano $(z\times r)$ das trajetórias descritas por fótons na corda cósmica em rotação com $J=0.2$ e $\mu=0.1$. Nas trajetórias diagonais $\bar{P}=0.2$ e $E=100$. A região não-causal compreende o intervalo $0<r<r_{nc}$.}
\label{geod1}
\end{figure}
\begin{figure}[!h]
\centering
\epsfig{file =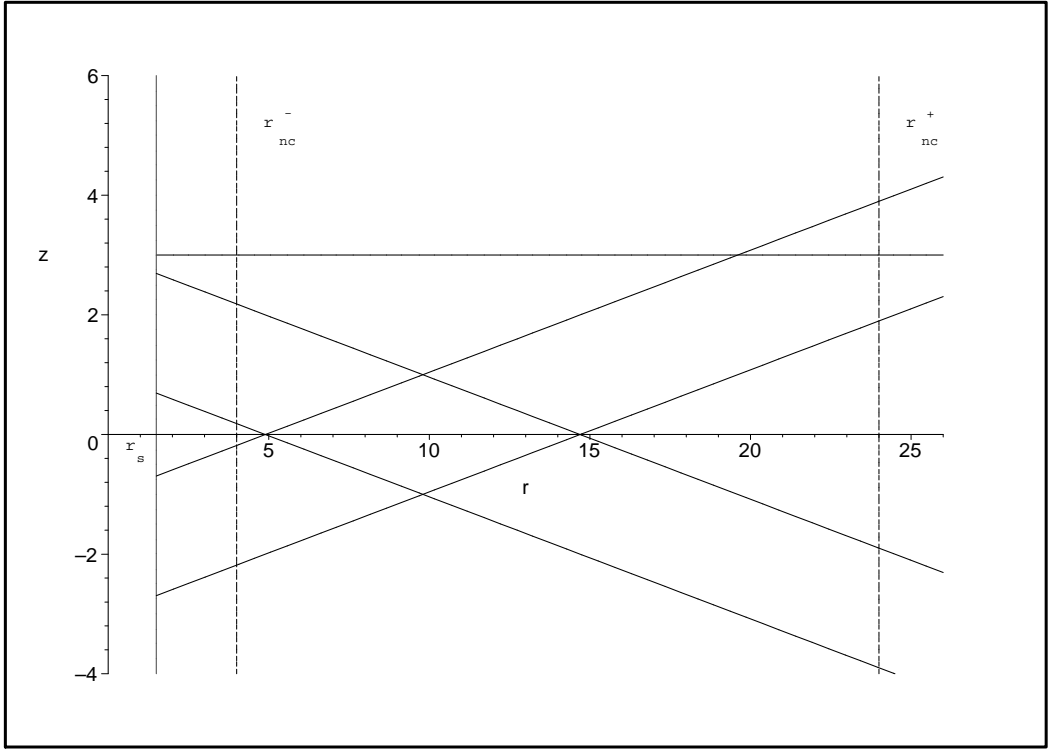, angle=-90, width=1.0 \linewidth}
\caption{\small{Projeção no plano $(z\times r)$ das trajetórias descritas por fótons no cilindro em rotação com interior tipo ``vaso de flor" com $J=0.5$, $\mu=0.2$ e $r_{s}=1.5$. Nas trajetórias diagonais $\bar{P}=0.2$ e $E=100$. A região não-causal compreende o intervalo $r_{nc}^{-}<r<r_{nc}^{+}$.}}
\label{geod1a}
\end{figure}
Se agora escolhermos qualquer valor de $\bar{P}$ tal que $|\bar{P}|<1$ teremos novamente $\ddot{r}=0$, mas agora a velocidade radial será
\begin{eqnarray}
\label{geod_f5}
\dot{r}&=&\pm E\sqrt{\left(1-\bar{P}^{2}\right)}\quad.
\end{eqnarray}
Neste caso, o fóton se afasta $(+)$ do cilindro ou se aproxima $(-)$ dele até alcançar sua fronteira em $r=r_{s}$ ou a origem $r=0$ no caso da corda.
Se $\bar{P}=0$ teremos trajetórias radiais sem movimentos na direção $z$.
As trajetórias acima são apresentadas nas Figs.(\ref{geod1}, \ref{geod1a}).

Vemos que, mesmo no caso do cilindro em rotação, a estrutura das geodésica nulas radiais são bem simples. Curiosamente, o comportamento dos cones de luz radiais também são simples mesmo no caso onde curvas fechadas tipo tempo possam existir. Contudo, isso já era esperado uma vez que o ingrediente essencial para a formação dessas curvas, o movimento na direção $\phi$, é ausente no caso das geodésicas radiais. 

\begin{figure}[!h]
\centering
\epsfig{file =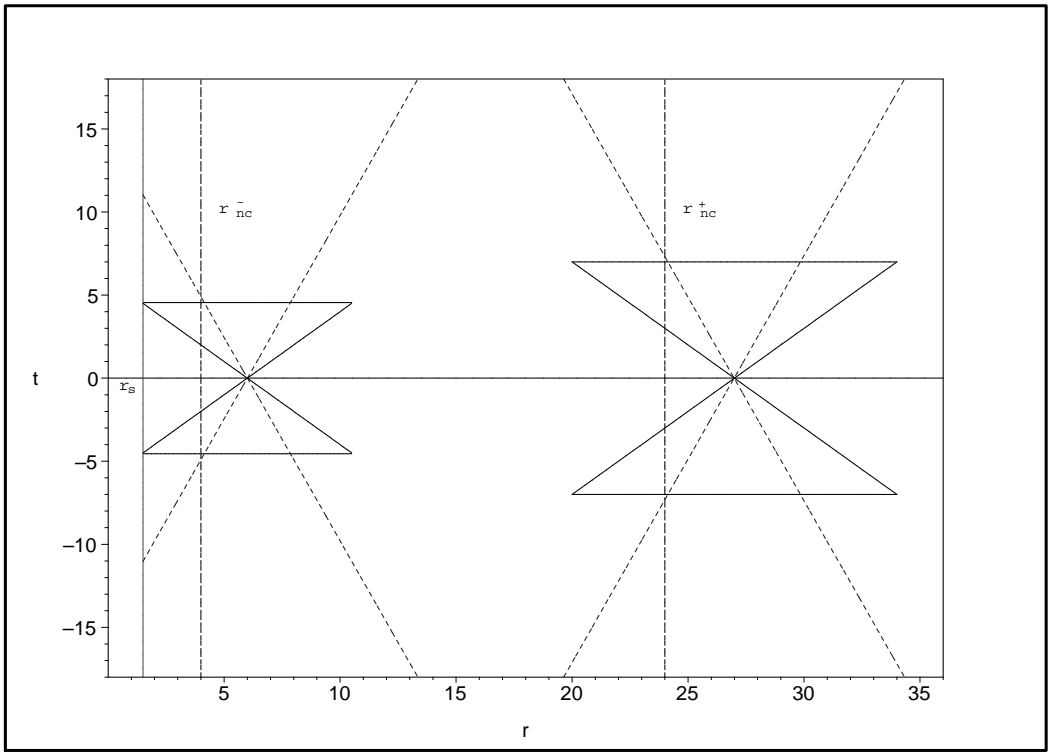, angle=-90, width=1.0 \linewidth}
\caption{\small{Cones de luz (linhas cheias) com $\bar{P}=0$ e $E=100$ e trajetórias radiais (linhas diagonais tracejadas) de partículas com $\bar{P}=0$ e $E=\sqrt{1.2}$ para o cilindro em rotação com interior tipo ``vaso de flor". Os parâmetros da corda são $J=0.5$, $\mu=0.2$ e $r_{s}=1.5$. A região não-causal compreende o intervalo $r_{nc}^{-}<r<r_{nc}^{+}$.}}
\label{geod2}
\end{figure}

O comportamento dos cones de luz pode ser obtido integrando a equação
\begin{eqnarray}
\label{geod_f6}
\frac{\dot{t}}{\dot{r}}=\frac{dt}{dr}=\pm\frac{1}{\sqrt{1-\bar{P}^2}}\quad,
\end{eqnarray}
que tem como resultado
\begin{eqnarray}
\label{geod_f6a}
t=\pm \frac{r}{\sqrt{1-\bar{P}^2}} + C_{1}\quad.
\end{eqnarray}
Vemos na Fig.(\ref{geod2}) que eles se assemelham aos cones de luz do espaço-tempo de Minkowski. As trajetórias radiais de partículas $(\epsilon=1)$ se encontram contidas na região interior dos cones de luz e são descritas pela equação
\begin{eqnarray}
\label{geod_f6b}
t=\pm \frac{r}{\sqrt{1-\frac{1}{E^2}-\bar{P}^2}} + C_{1}\quad,
\end{eqnarray}

Para o caso do cilindro e da corda estática a condição para órbitas radiais é susbstituída por $L=0$ já que $J=0$. Portanto, as trajetórias radiais de fótons e partículas massivas, nestes casos são semelhantes às apresentadas acima.

\section{Órbitas diretas $(L+4J)>0$}

Exploraremos agora órbitas um pouco mais gerais permitindo o movimento em todas as direções, inclusive na direção $\phi$.
O conjunto de equações de movimento para fótons e partículas massivas sob a condição $(L+4J)>0$ resulta nas Eqs.(\ref{foton1}, \ref{foton2}, \ref{foton3}, \ref{foton4}).
Iniciaremos nossa análise calculando os pontos críticos da Eq.(\ref{foton4}). Assim procedendo teremos
\begin{eqnarray}
\label{cap8geo13}
\dot{r}^2&=&0 \quad \Longrightarrow \quad \left(r+r_{0}\right)^2=\frac{\left(L+4J\right)^2}{\alpha^2\left(1-\frac{\epsilon}{E^2}-\bar{P}^2\right)}\quad,\\
\nonumber\\
\nonumber\\
\label{cap8geo13a}
r_{ext}&=&r_{ext}^{(1,2)}=\pm \frac{\left(L+4J\right)}{\alpha\sqrt{\left(1-\frac{\epsilon}{E^2}-\bar{P}^2\right)}}-r_{0}\quad.
\end{eqnarray}
Os pontos $r_{ext}$ são os pontos de retorno da parte radial da trajetória seja para uma partícula massiva ou um fóton. Observa-se que a condição $(E^2-\epsilon-\bar{P}^2)>0$ deve ser satisfeita para que esses pontos existam. Como sabemos, $r_{0}$ pode assumir valores positivos e negativos no caso do cilindro em rotação, ou ainda ser nulo no caso das cordas estática e em rotação e do cilindro estático. Isso implica que a presença de mais de um ponto de retorno estará associada ao valor de $r_{0}$ escolhido em cada caso. Nos casos onde $r_{0}=0$, teremos um único ponto de retorno localizado em
\begin{eqnarray}
\label{cap8geo13b}
r_{ext}&=& \frac{\left(L+4J\right)}{\alpha\sqrt{\left(1-\frac{\epsilon}{E^2}-\bar{P}^2\right)}}\quad.
\end{eqnarray}

\begin{figure}[!h]
\centering
\epsfig{file =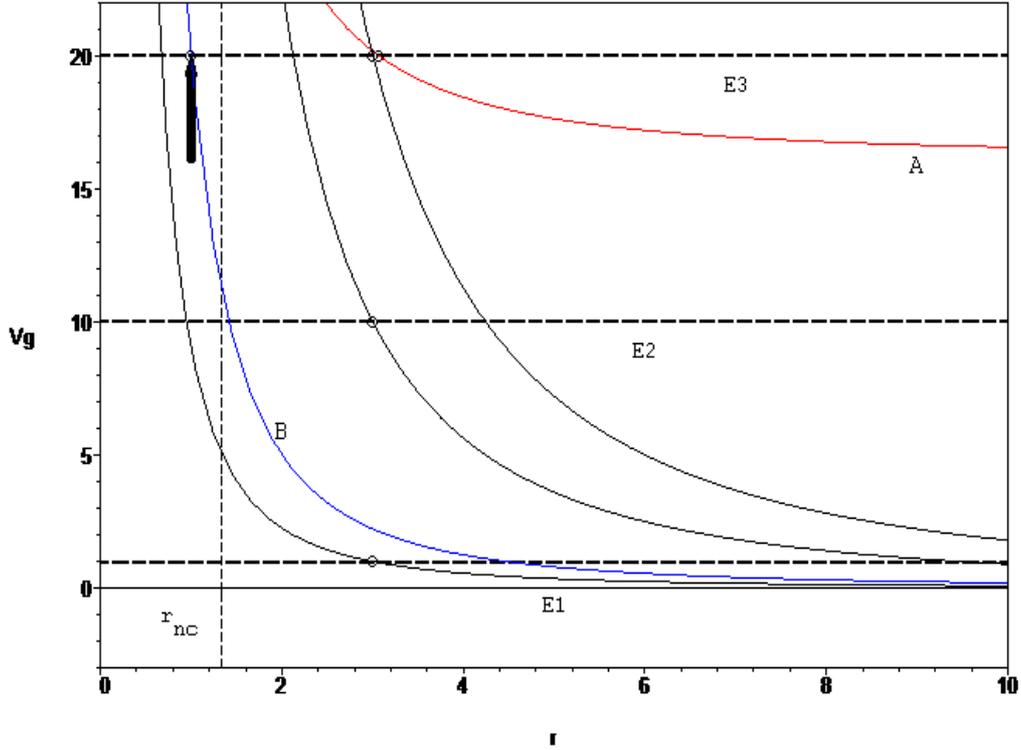, height=10cm,width=13.5cm,angle=0}
\caption{\small{ Potencial efetivo $V_{g}(r)$ para os fotóns se propagando na corda cósmica em rotação com $J=0.2$, $\mu=0.1$, $r_{nc}=1.33$. A linhas tracejadas horizontais representam os diferentes valores de $E^2$ para fótons com $\bar{P}=0$ e $L=1$. São eles $E_{1}=E^2=1, E_{2}=E^2=10, E_{3}=E^2=20$. A região não-causal compreende o intervalo $0<r<r_{nc}^{+}$. Os pontos de retorno são representados pelos círculos pretos.}}
\label{figVglinha}
\end{figure}

\begin{figure}[!h]
\centering
\epsfig{file =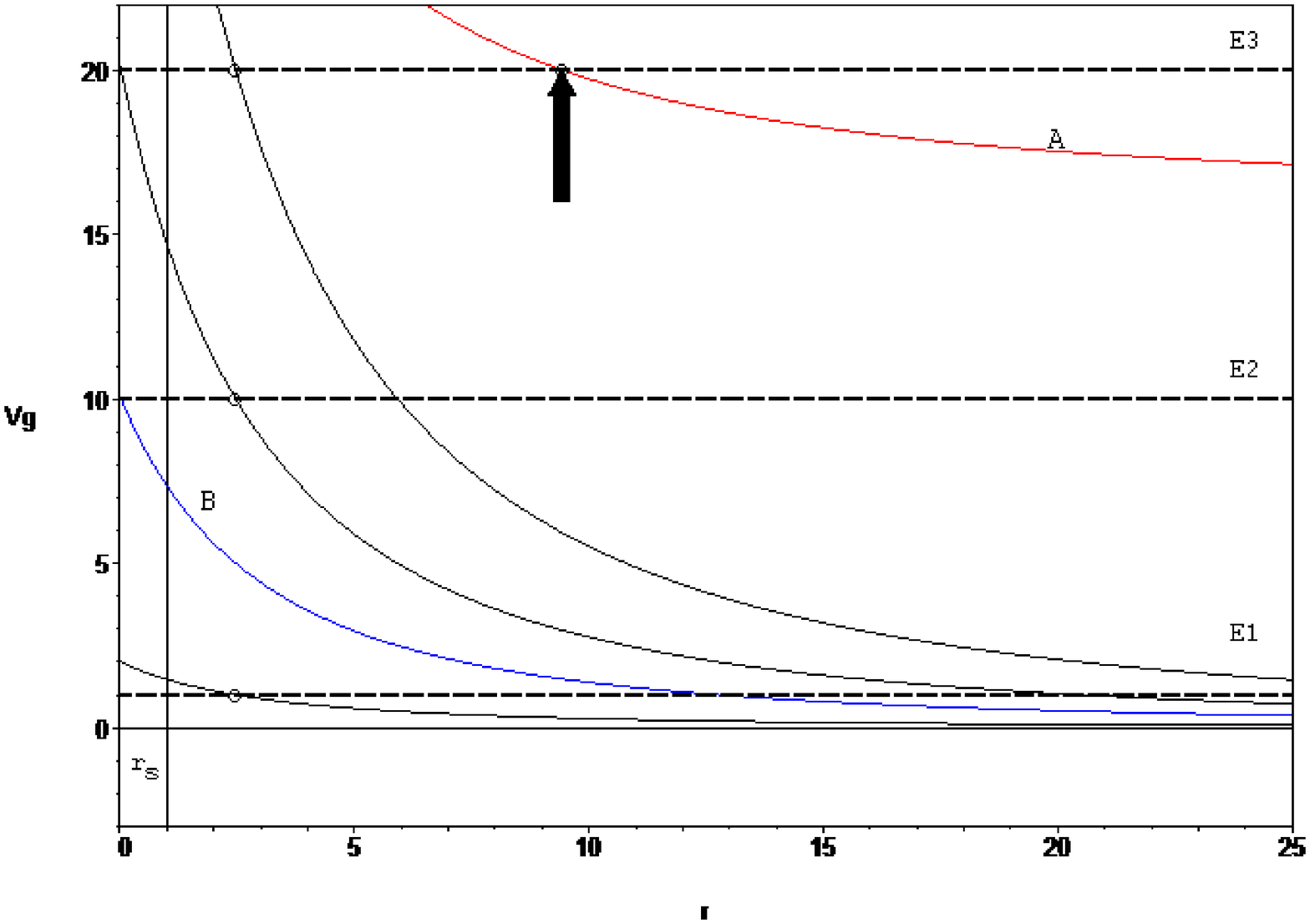, angle=0, width=0.95 \linewidth}
\caption{\small{ Potencial efetivo $V_{g}(r)$ para fotóns se propagando no exterior do cilindro em rotação com interior do tipo ``vaso de flor" com $J=1$, $\mu=0.1$, $r_{s}=1$ e $r_{0}=5.87$. A linhas tracejadas horizontais representam os diferentes valores de $E^2$ para fótons com $\bar{P}=0$ e $L=1$. São eles $E_{1}=E^2=1, E_{2}=E^2=10, E_{3}=E^2=20$.}}
\label{figVgvaso}
\end{figure}

\begin{figure}[!h]
\centering
\epsfig{file =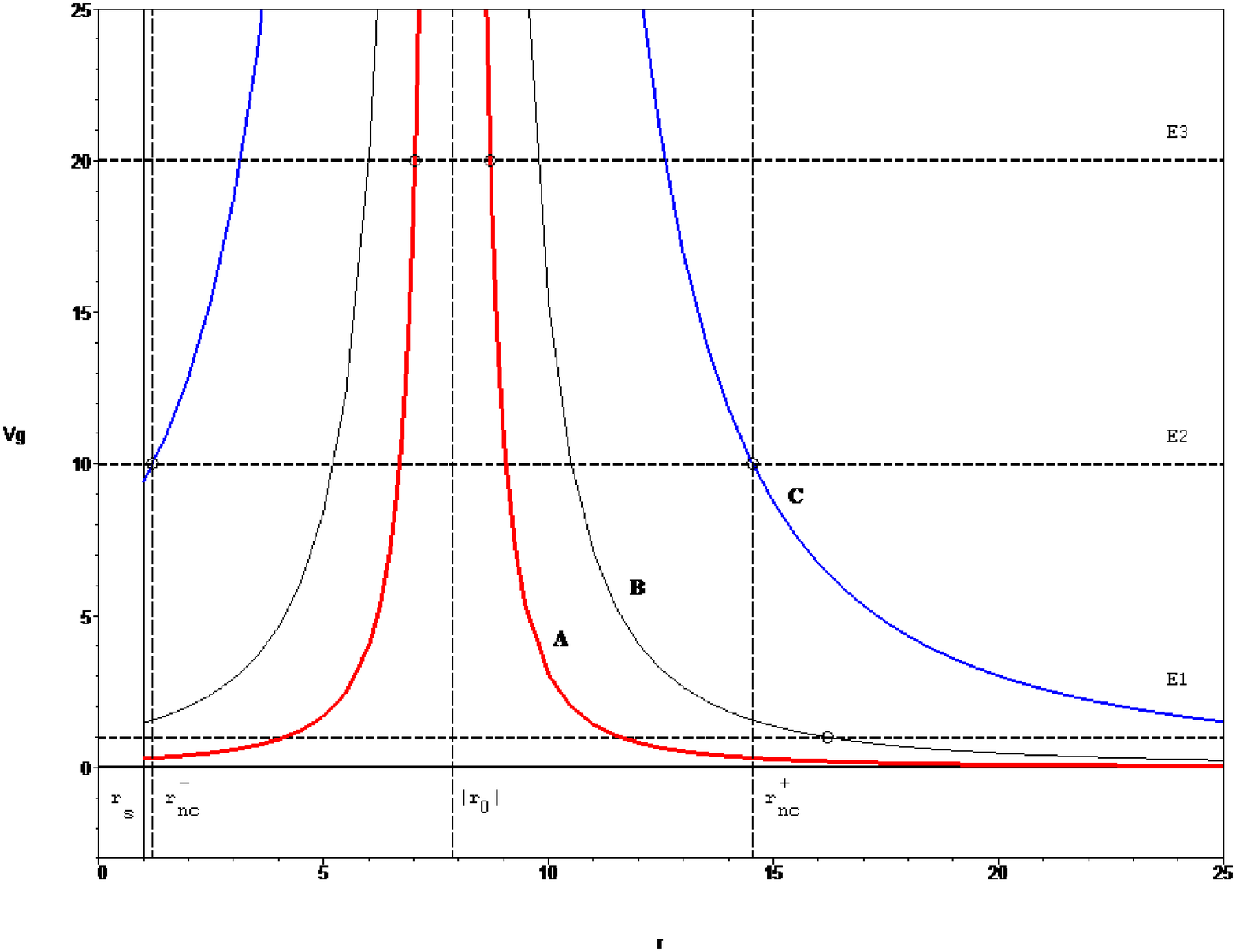,height=12cm,width=14cm, angle=0}
\caption{\small{ Potencial efetivo $V_{g}(r)$ para fotóns se propagando no cilindro em rotação com interior do tipo ``vaso de flor" com $J=1$, $\mu=0.1$, $r_{s}=1$ e $r_{0}=-7.87$. A linhas tracejadas horizontais representam os diferentes valores de $E^2$ para fótons. São eles $E_{1}=E^2=1, E_{2}=E^2=10, E_{3}=E^2=20$.}}
\label{figVgvaso2}
\end{figure}

Se o parâmetro $r_{0}<0$, teremos:

\begin{itemize}

\item dois pontos de retorno se $r_{ext}^{(1,2)}>r_{s}$,

\item um ponto de retorno se $r_{ext}^{(2)}>r_{s}$ ou

\item nenhum ponto de retorno se $r_{ext}^{(1,2)}<r_{s}$

\end{itemize}
Agora, se $r_{0}>0$ temos um ponto de retorno se $r_{ext}^{(2)}>r_{s}$ ou nenhum ponto de retorno se $r_{ext}^{(2)}<r_{s}$. Para tentar fazer cada caso mais claro vamos reescrever a Eq.(\ref{foton4}) em termos de um potencial efetivo geral $V_{g}(r)$ tal que
\begin{eqnarray}
\label{cap8geo13b1}
\dot{r}^2=E^2-V_{g}(r)\quad,
\end{eqnarray}
onde o potencial efetivo $V_{g}(r)$ será
\begin{eqnarray}
\label{cap8geo14}
V_{g}(r)=E^2\left(\bar{P}^{2}+\frac{\epsilon}{E^2}+\frac{\left(L+4J\right)^2}{\alpha^2(r-r_{0})^2}\right)\quad.
\end{eqnarray}
Observemos o comportamento do potencial efetivo geral apresentado nas Figs.(\ref{figVglinha}, \ref{figVgvaso}, \ref{figVgvaso2}) para os casos $r_{0}=0$, $r_{0}>0$ e $r_{0}<0$.
\\
\\
\\
{\bf Caso $r_{0}=0$}\\

Apresentaremos o potencial efetivo para a corda em rotação por ele ser mais rico, mas discutiremos brevemente os casos da corda e do cilindro estáticos.
O valor de $r_{ext}$ depende essencialmente do valor de $L$ e $E$ tanto para fótons quanto para partículas massivas. Focando nossa atenção nos fótons vemos que, se $L\geq 0$ e $E>1$ o ponto de retorno sempre se encontra fora da região não-causal. Por outro lado, se $-4J<L<0$ o ponto de retorno se encontra dentro da região não-causal. Para ilustrar essas situações observemos a Fig.(\ref{figVglinha}). A curva $B$ representa a trajetória de um fóton com os parâmetros $L=-0.2$, $E_{3}=E^2=20$ e $\bar{P}=0$. Seu o ponto de retorno, representado pela flecha preta, é $r_{ext}=1.0$. A curva $A$ representada pela linha tracejada mostra o efeito da mudança de $\bar{P}$, ou seja, o aumento de $\bar{P}$ eleva a cauda do potencial e antecipa o ponto de retorno. Nesta curva $A$ os parâmetros escolhidos são $L=0$, $\bar{P}=0.9$ e $E_{3}=E^2=20$.

De modo geral, $V_{g}$ se comporta como uma barreira de potencial sendo positivo definido. Portanto, não apresenta pontos de mínimo, nem de máximo, implicando na impossibilidade de órbitas circulares ou elípticas seja para fótons ou para partículas massivas.
Podemos confirmar a afirmação sobre geodésicas circulares e elípticas mostrando que $\ddot{r}>0$ nos pontos de retorno. Assim procedendo temos
\begin{eqnarray}
\label{cap8geo15}
\ddot{r}(r)&=&\frac{(L+4J)^2}{\alpha^2r^3}\quad,\\
\nonumber\\
\ddot{r}(r_{ext})&=&\frac{(L+4J)^2}{\alpha^2(r_{ext})^3}\quad.
\end{eqnarray}
Como $r_{ext}$ é sempre positivo $\ddot{r}(r_{ext})$ nunca assume valores negativos, ou seja, existe uma força sempre empurrando tanto fótons quanto partículas massivas para longe da corda. E mais, o único conjunto de parâmetros para o qual $\ddot{r}(r_{ext})=0$ é $L=-4J$. Mas esse valor de $L$ viola a condição $(L+4J)>0$.

Esse aspecto é muito importante pois garante que não teremos geodésicas fechadas tipo-tempo (GFT) circulares nestes espaços-tempos com $r_{0}=0$. No próximo capítulo discutiremos esse ponto. As conclusões são similares para a corda estática e cilindro estático, excetuando-se a discussão sobre a região não-causal. No caso do cilindro estático, se as constantes $E$ ou $L$ forem muito pequenas comparadas com o raio $r_{s}$ do cilindro não existirá ponto de retorno.
\\
\\
\\
{\bf Caso $r_{0}>0$}\\

Para ilustrar o comportamento de $V_{g}$ quando $r_{0}>0$ usaremos o cilindro em rotação com interior ``tipo vaso de flor".

Vemos na Fig. (\ref{figVgvaso}) que o comportamento de $V_{g}$ é muito semelhante ao caso $r_{0}=0$. Como vimos no capítulo anterior, quando $r_{0}>0$ a região não-causal é encoberta pela parte interior da corda. Novamente, os valores de $L$ e $E$ determinam a posição de $r_{ext}$. Como neste caso não temos região não-causal a única discussão de interesse é sobre qual condição $L$ precisa satisfazer para que não exista ponto de retorno. Um exemplo no qual o fóton, com parâmetros $\bar{P}=0$, $E_{3}=20$ e $L=-1.5$, alcança o cilindro e não retorna é dado pela curva $B$. Podemos ver que neste caso o ponto de retorno se localiza em $r_{ext}<0$. Assim, se $L<0$ e a constante $E$ não for muito grande comparada com $r_{s}$ o fóton pode não ter ponto de retorno. A curva $A$ mostra o caso de um fóton onde os parâmetros são $\bar{P}=0.9$, $E_{3}=20$ e $L=0$. Uma análise mais geral sobre todo o espaço de parâmetros seria muito extensa, portanto nos manteremos à condição citada anteriormente.

Neste caso, $V_{g}$ também atua como uma barreira de potencial já que é positivo definido. Isso também acaba impossibilitando a existência de órbitas circulares e elípticas quando $r_{0}>0$. 
\\
\\
\\
{\bf Caso $r_{0}<0$}\\

O caso $r_{0}<0$ é o de maior interesse pois apresenta de uma região não-causal. O cilindro em rotação com interior ``tipo vaso de flor" foi escolhido novamente para ilustrar o comportamento de $V_{g}$. Como esse caso é o mais complexo escolheremos algumas órbitas de interesse para ilustrar a variedade de órbitas possíveis. Neste caso é mais difícil fazermos declarações gerais.

Vemos na Fig.(\ref{figVgvaso2}) que o comportamento de $V_{g}$, apesar de ser positivo definido, é bastante diferente dos outros casos.A região não-causal compreende o intervalo $r_{nc}^{-}<r<r_{nc}^{+}$. Os pontos de retorno para cada um dos casos são indicados pelos círculos pretos sobre as linhas horizontais tracejadas. Essas linhas indicam os valores de $E^2$. As curvas $A$, $B$ e $C$ descrevem fótons com os seguintes parâmetros $(\bar{P}=0, E_{3}=20, L=-3.5)$, $(\bar{P}=0, E_{1}=1, L=1)$ e $(\bar{P}=0, E_{2}=10, L=0)$, respectivamente. 

A diferença marcante apresetada por $V_{g}$ é o ponto de máximo localizado em $r=|r_{0}|$. Na verdade $V_{g}$ diverge neste ponto dividindo o espaço-tempo em duas regiões. Portanto, se $(L+4J)>0$, tanto fótons quanto partículas massivas localizadas entre $r_{s}$ e $|r_{0}|$ não conseguem escapar para o infinito espacial seguindo uma geodésica. Pela mesma razão, se eles estiverem localizados entre $|r_{0}|$ e $r=\infty$ não podem alcançar o cilindro.

Embora o ponto $|r_{0}|$ separe o espaço-tempo criando um poço de potencial infinito de um lado e uma barreira infita do outro não é possível a existência de órbitas circulares ou elípticas. 

\subsection{Cones de luz}

\begin{figure}[!h]
\centering
\epsfig{file =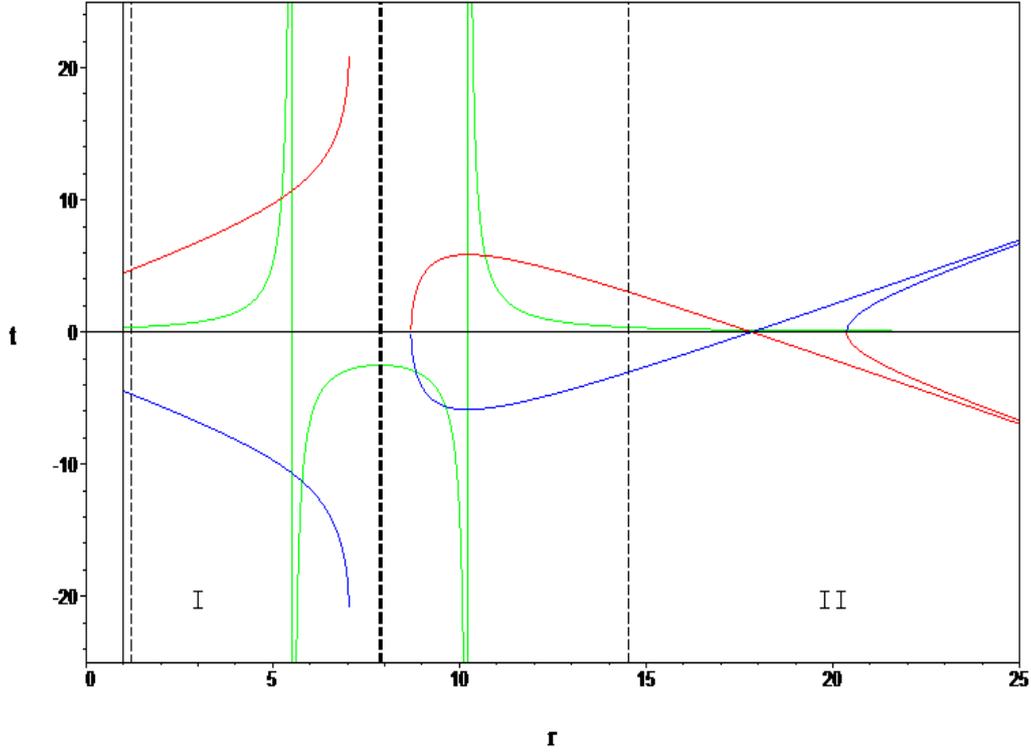, height=10cm,width=14cm,angle=0}
\caption{\small{Estrutura dos cones de luz com $\bar{P}=0$ e $L=-3.5$ (descontínua) e $L=3.5$ (contínua) para o cilindro em rotação com interior do tipo ``vaso de flor" com $J=1$, $\mu=0.1$, $r_{s}=1$ e $r_{0}=-7.87$. As linhas tracejadas verticais posicionadas da esquerda para direita representam $r_{nc}^{-}$, $|r_{0}|$ e $r_{nc}^{+} $ respectivamente.}}
\label{figcone}
\end{figure}

\begin{figure}[!h]
\centering
\epsfig{file =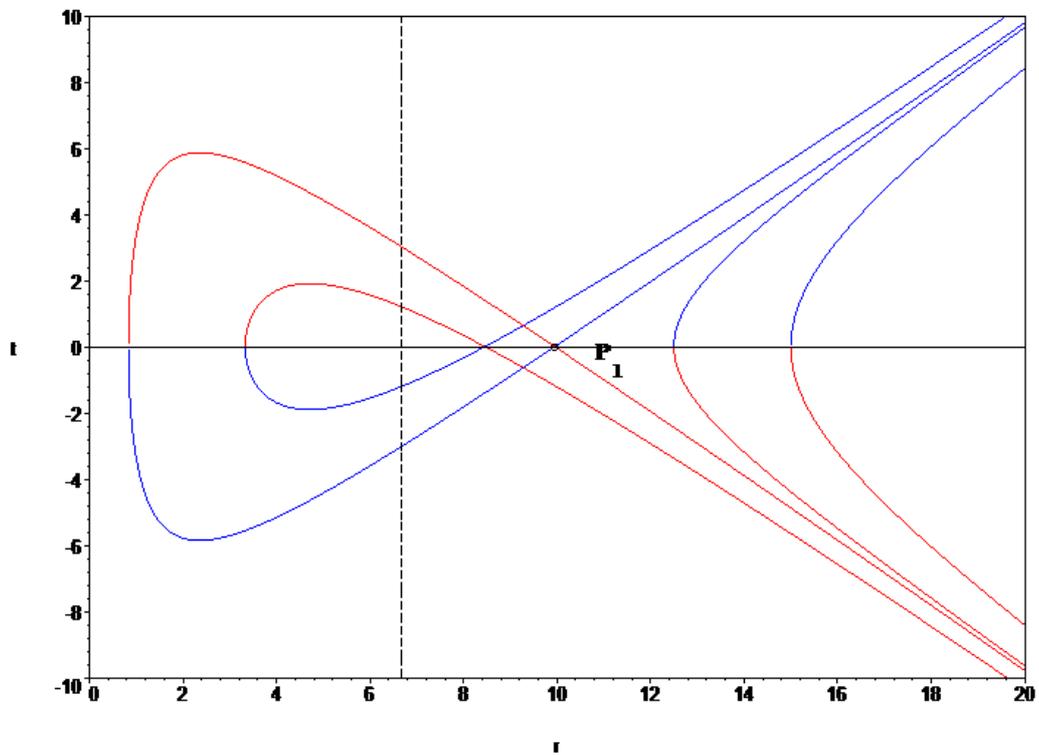, height=10cm,width=14cm,angle=0}
\caption{\small{Estrutura dos cones de luz com $\bar{P}=0$ e $L=(-3.5, -2, 3.5, 5)$ (esquerda para direita) para a corda em rotação com $J=1$ e $\mu=0.1$. A linha vertical tracejada representa $r_{nc}^{+} $. Os fótons escapando para o infinito e movimentando-se em direção a corda são representados pelas linhas azuis e vermelhas, respectivamente.}}
\label{figcone0}
\end{figure}

Continuando nossa análise veremos que, quando $(L+4J)>0$, a estrutura dos cones de luz se apresenta muito mais rica devido ao movimento na direção $\phi$. Vejamos então o comportamento dos cones de luz integrando a equação
\begin{eqnarray}
\label{cap8geo16}
\frac{dt}{dr}=\pm \left[\frac{\alpha(r+r_{0})}{\sqrt{\alpha^2(r+r_{0})^2\left(1-\frac{\epsilon}{E^2}-\bar{P}^2\right)-(L+4J)^2}}-\right.\nonumber\\
-\left.\frac{4J(D+4J)}{\alpha(r+r_{0})\sqrt{\alpha^2(r+r_{0})^2\left(1-\frac{\epsilon}{E^2}-\bar{P}^2\right)-(L+4J)^2}}\right]\quad,
\end{eqnarray}
obtida das Eqs.(\ref{foton4},\ref{foton3}). O resultado da integração é
\begin{eqnarray}
\label{cap8geo17}
t=\pm \frac{\sqrt{\alpha^2(r+r_{0})^2\left(1-\frac{\epsilon}{E^2}-\bar{P}^2\right)-(L+4J)^2}}{\alpha\left(1-\frac{\epsilon}{E^2}-\bar{P}^2\right)}-\nonumber\\
-\frac{4J}{\alpha}\ \textrm{arcsec}\left((r+r_{0})\frac{\sqrt{1-\frac{\epsilon}{E^2}-\bar{P}^2}}{(L+4J)}\right)+C_{3}\quad.
\end{eqnarray}
Nas Figs.(\ref{figcone}, \ref{figcone0}) apresentamos o comportamento dos cones de luz $(\epsilon=0)$ para $r_{0}<0$ e para $r_{0}=0$ quando existe uma região não-causal. Não discutiremos os casos em que curvas fechadas tipo tempo não são possíveis pois a estrutura dos cones de luz é semelhante ao espaço-temo de Minkowski.
A influência da região não-causal no comportamento dos cones de luz do cilindro em rotação é evidente. As linhas vermelhas representam o fóton se movimentando em direção ao cilindro e as linhas azuis representa o fóton escapando do cilindro. Vemos na Fig.(\ref{figcone}) que o comportamento da coordenada $t$, quando $L=-3.5$, na região I é distinto da região II. Na região I, o fóton se movimenta apenas na direção negativa de $t$, isto é, viaja para o passado e seu movimento é descontínuo saltando de $t=-20$ para $t=20$ quando alcança o ponto de retorno. Na região II ele se movimenta nos dois sentidos dependendo da posição radial, podendo influenciar o próprio passado. Nessa região a coordenada $t$ é contínua tanto para $L=-3.5$ (esquerda) quanto para $L=3.5$ (direita).

Podemos observar que o ponto onde os cones de luz para $L=-3.5$ mudam de direção coincide, exatamente com a divergência da sua velocidade angular, representada pela linha verde na região II. A velocidade angular foi amplificada em 10 vezes. Essa mudança na direção é a responsável pela volta ao passado do fóton. Na região I a dinâmica do fóton é bastante estranha já que ele inicia seu movimento escapando do cilindro, voltando ao passado, saltando para o futuro e voltando ao cilindro.

Vemos ainda que, todos os fótons com $\dot{\phi}\neq0$ que tentam escapar ou alcançar o cilindro acabam confinados ao seu local de origem. A única maneira de se escapar ou alcançar o cilindro é realizando um movimento puramente radial.

Observemos na Fig.(\ref{figcone0}) o comportamento dos cones de luz no caso da corda em rotação .

Neste caso, o ponto $|r_{0}|$ se localiza sobre a corda em rotação de modo que agora temos apenas a região II. Como na região II do cilindro em rotação, a evolução do fóton na coordenada $t$ se comporta de maneira contínua permi-tindo-lhes o movimento tanto na direção positiva quanto na direção negativa de $t$. Tomemos como exemplo, o fóton localizado no ponto $P_{1}$. Se acompanharmos a evolução da geodésica descrita por esse fóton veremos que seu passado e futuro irão se interceptar exatamente em $P_{1}$.

\subsection{Órbitas de fótons e partículas massivas}

\begin{figure}[!h]
\centering
\epsfig{file =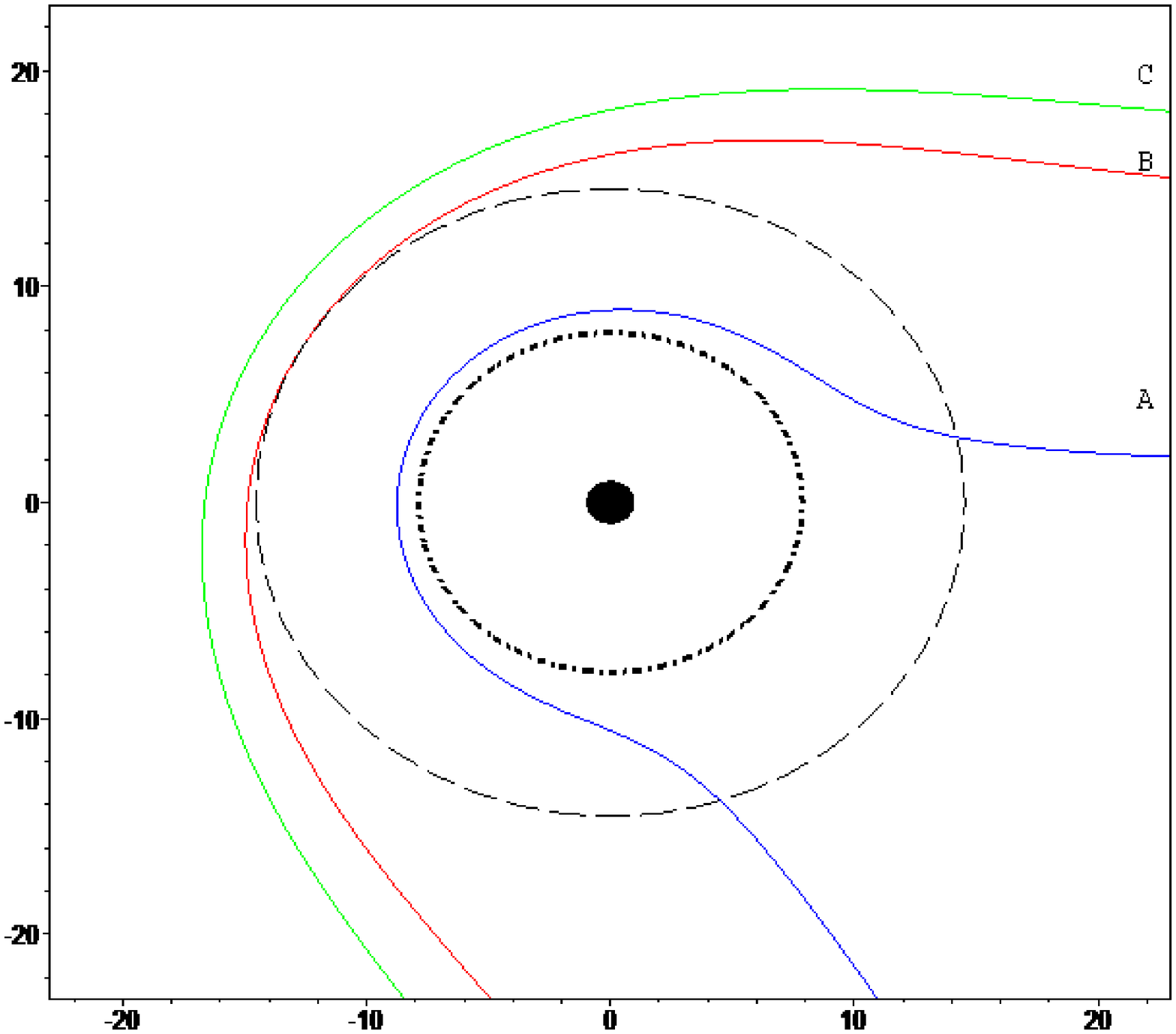, angle=0, width=0.67 \linewidth}
\epsfig{file =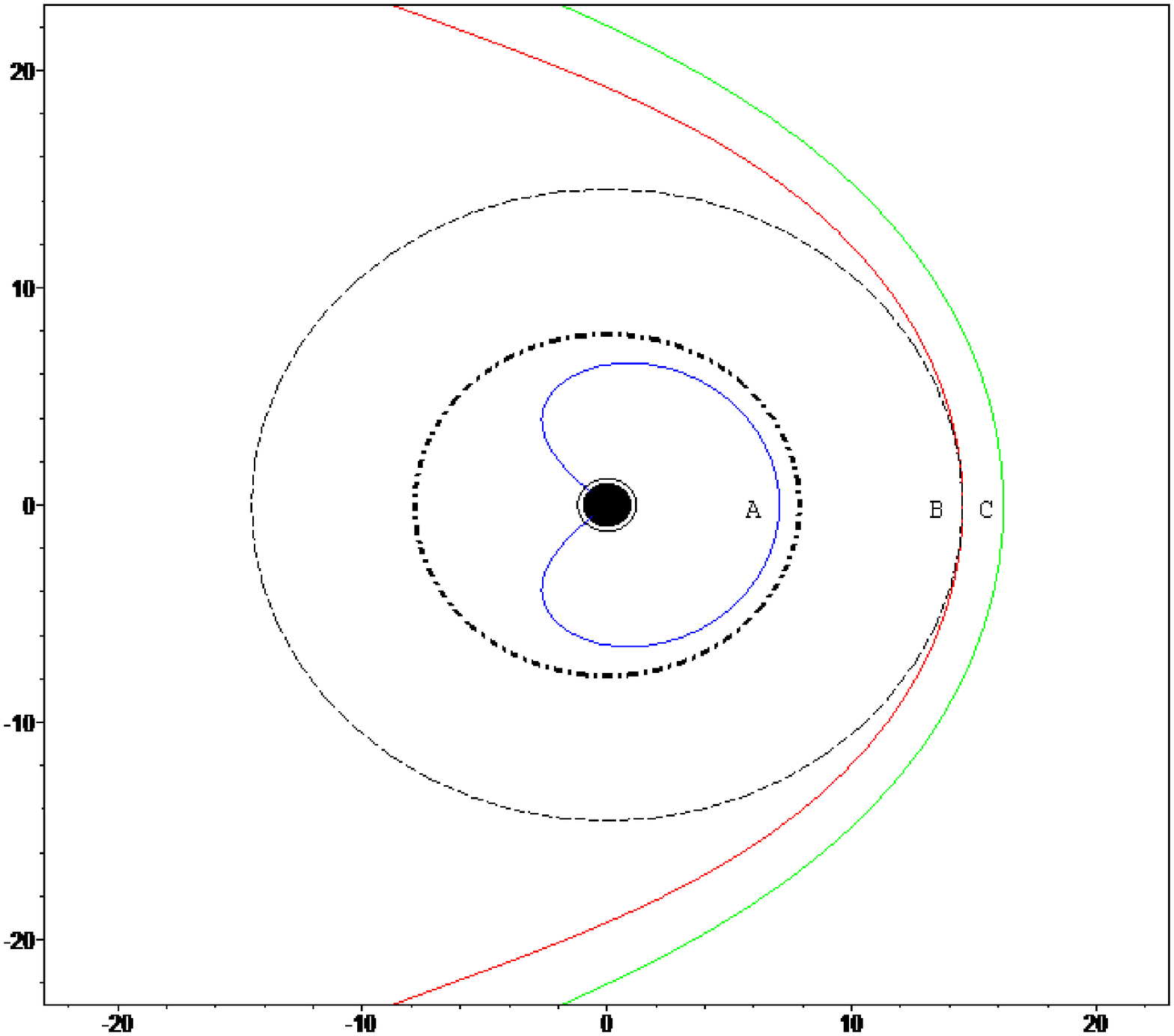, angle=0, width=0.67 \linewidth}
\caption{\small{Trajetória dos fótons no plano $(r\times\phi)$ para o cilindro em rotação com interior do tipo ``vaso de flor" com $J=1$, $\mu=0.1$, $r_{s}=1$ e $r_{0}=-7.87$. As linhas tracejadas localizadas do centro para fora representam $r=r_{nc}^{-}$, $r=|r_{0}|$ e $r=r_{nc}^{+} $ respectivamente.}}
\label{figorbita_I}
\end{figure}

\begin{figure}[!h]
\centering
\epsfig{file =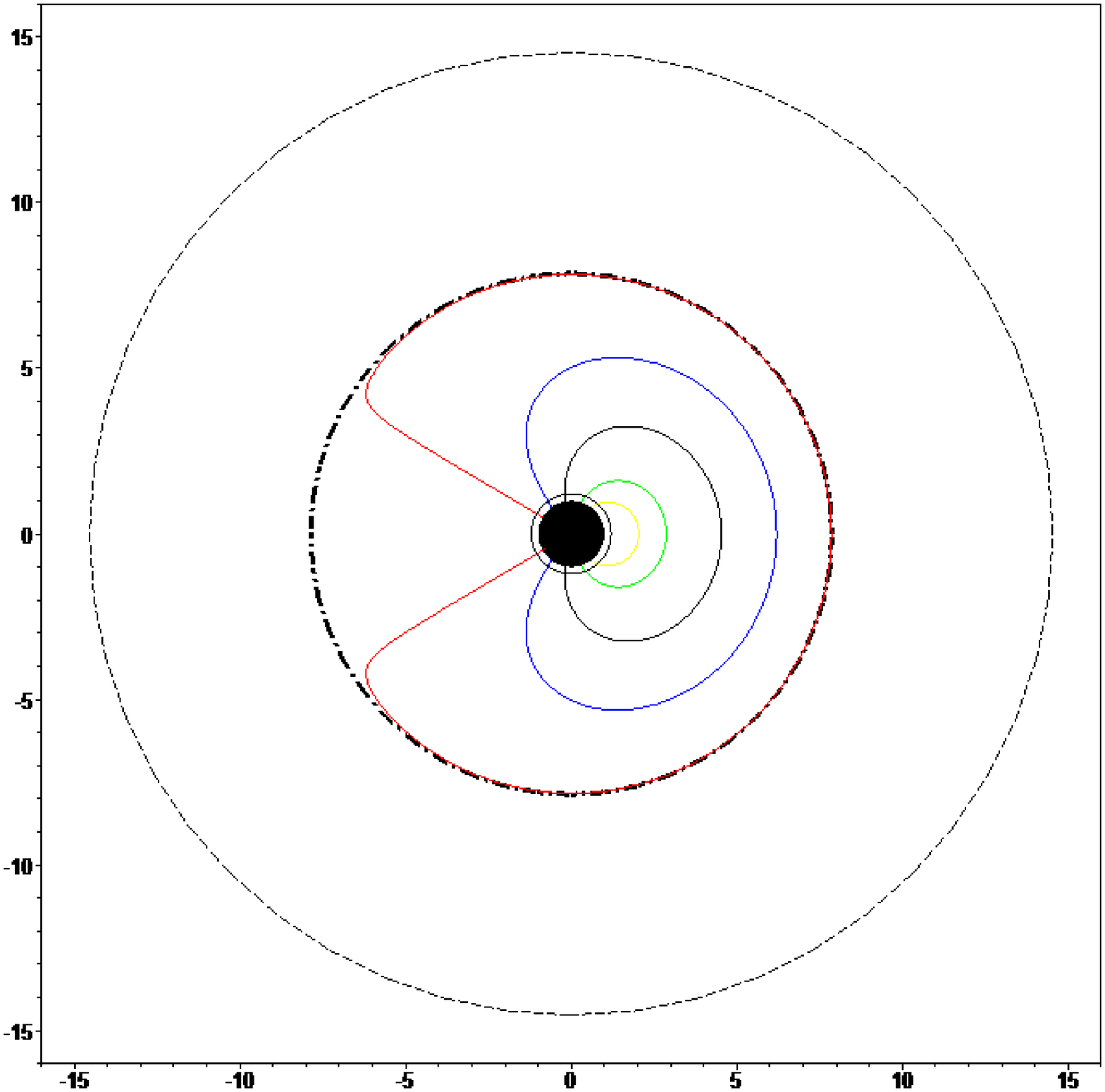, angle=0, width=0.6 \linewidth}
\epsfig{file =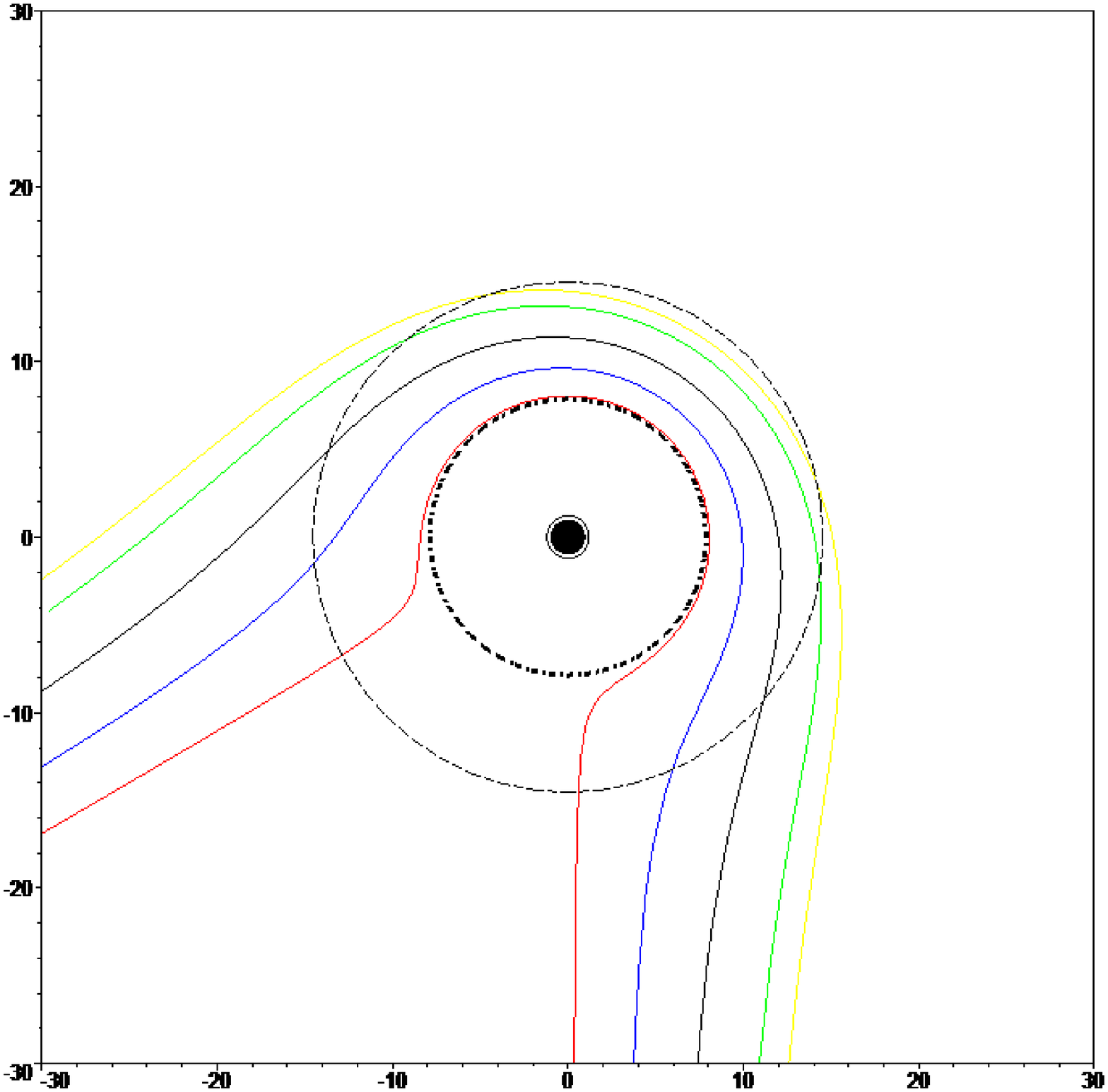, angle=0, width=0.6 \linewidth}
\caption{\small{Trajetória dos fótons no plano $(r\times\phi)$ para o cilindro em rotação com interior do tipo ``vaso de flor" com $J=1$, $\mu=0.1$, $r_{s}=1$ e $r_{0}=-7.87$. As linhas tracejadas localizadas do centro para fora representam $r=r_{nc}^{-}$, $r=|r_{0}|$ e $r=r_{nc}^{+} $ respectivamente.}}
\label{figorbita_II}
\end{figure}

\begin{figure}[!h]
\centering
\epsfig{file =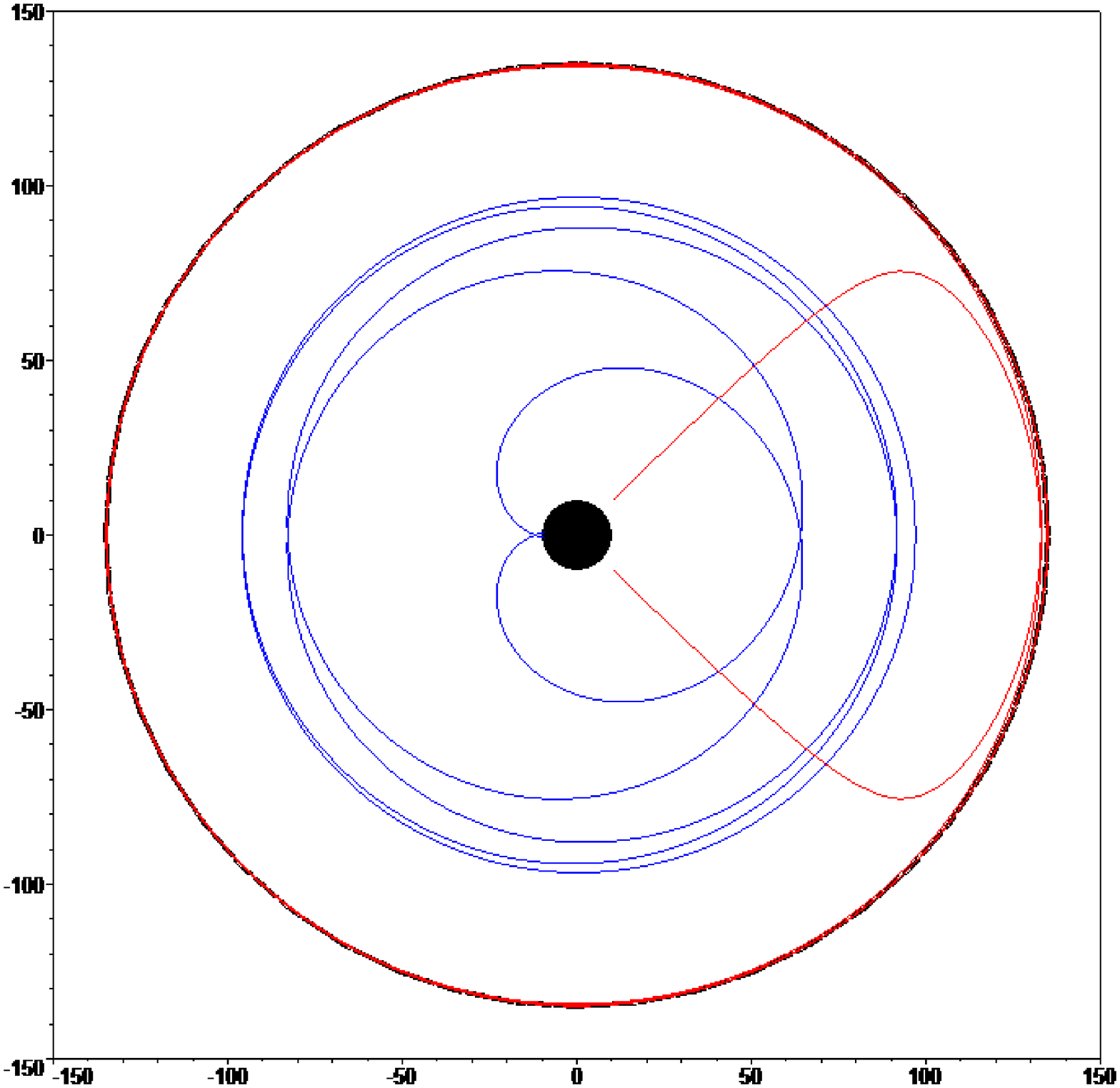, angle=0, width=0.60 \linewidth}
\epsfig{file =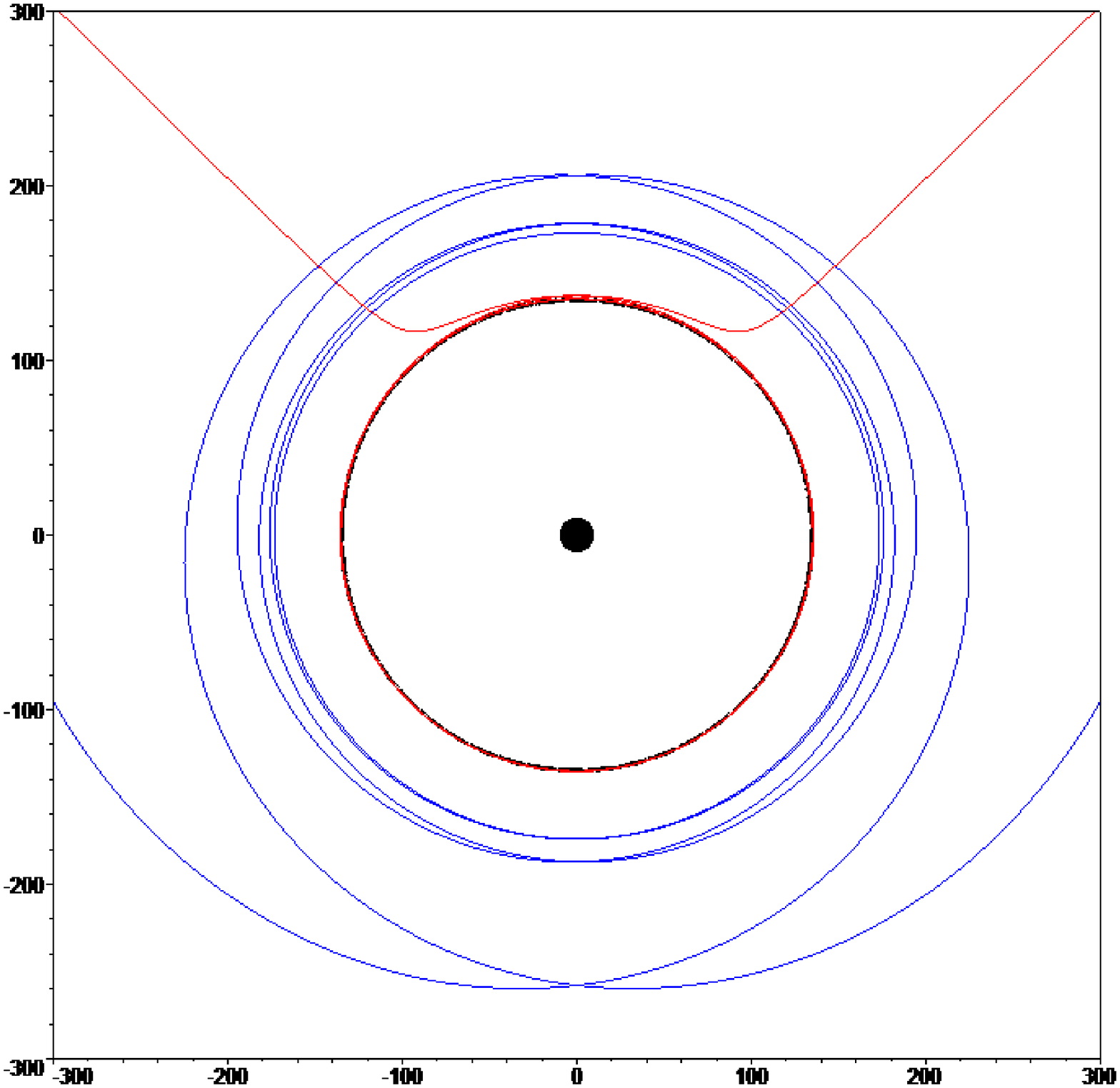, angle=0, width=0.60 \linewidth}
\caption{\small{Trajetória dos fótons no plano $(r\times\phi)$ para o cilindro em rotação com interior do tipo ``vaso de flor" com $J=1.0\times10^{-2}$, $\mu=0.23$, $r_{s}=10$ e $r_{0}=-135$. As linhas tracejadas representam $r=r_{nc}^{-}$, $r=|r_{0}|$ e $r=r_{nc}^{+} $ respectivamente.}}
\label{figorbita_III}
\end{figure}

O próximo passo nessa análise é descobrir quais tipos de trajetórias fótons e partículas massivas podem realizar no plano $(r,\phi)$. Para tal, dividiremos a Eq.(\ref{foton1}) pela Eq.(\ref{foton4}) obtendo
\begin{eqnarray}
\label{cap8geo18}
\frac{d\phi}{dr}=\pm \frac{(L+4J)}{\alpha(r+r_{0})\sqrt{\alpha^2(r+r_{0})^2\left(1-\frac{\epsilon}{E^2}-\bar{P}^2\right)-(L+4J)^2}}\quad.
\end{eqnarray}
Integrando (\ref{cap8geo18}) temos como resultado
\begin{eqnarray}
\label{cap8geo19}
\phi=\pm \frac{1}{\alpha}\textrm{arcsec}\left(\frac{(r+r_{0})\alpha\sqrt{1-\frac{\epsilon}{E^2}-\bar{P}^2}}{(L+4J)}\right) +C_{1}\quad.
\end{eqnarray}
Invertendo a Eq.(\ref{cap8geo19}) teremos a coordenada $r$ em função da coordenada angular $\phi$. Isso conduz à equação
\begin{eqnarray}
\label{cap8geo20}
r(\phi)=\frac{(L+4J)}{\alpha\sqrt{1-\frac{\epsilon}{E^2}-\bar{P}^2}}\sec\left[\pm \frac{}{}\alpha(\phi-\phi_{0})\right]-r_{0}\quad.
\end{eqnarray}
Nas Figs.(\ref{figorbita_I}, \ref{figorbita_II}, \ref{figorbita_III}) apresentamos algumas órbitas descritas fótons nos espaços-tempos dos cilindros cósmicos em rotação. Como observaremos, tanto partículas massivas quanto fótons descrevem, essencialmente, dois tipos de trajetórias. O primeiro tipo chamaremos de espalhamento. Neste caso, partículas massivas e fótons vindos de um região distante são espalhados pelo cilindro cósmico e escapam para o infinito com um ângulo que depende dos parâmetros escolhidos. O segundo tipo chamaremos de trajetória confinada. Este tipo de trajetória acontece apenas quando $r_{0}<0$ devido ao comportamento do potencial efetivo $V_{g}$ entre $r_{s}\leq r\leq |r_{0}|$. Na Fig.(\ref{figVgvaso2}) as curvas $A$, $B$ e $C$ ilustram a presença do poço de potencial inifnito no intervalo comentado.

Na Fig.(\ref{figorbita_I}) podemos observar tanto as trajetórias do tipo espalhamento (acima) quanto do tipo confinada (abaixo). As curvas $A$, $B$ e $C$ apresentadas têm os seguintes os parâmetros:
\begin{itemize}
\item Curva $A$: fóton $\epsilon=0$ com $\bar{P}=0$ e $L=-3.5$,
\item Curva $B$: fóton $\epsilon=0$ com $\bar{P}=0$ e $L=0$,
\item Curva $C$: fóton $\epsilon=0$ com $\bar{P}=0$ e $L=1$.
\end{itemize}
As diferenças observadas entre as trajetórias apresentadas na parte superior e na parte inferior da figura são devidas às diferentes condições iniciais. Para um mesmo conjunto de parâmetros podemos ter tanto trajetórias espalhadas quanto confinadas. A curva $A$ é um destes exemplos. Também é evidente a influência da barreira de potencial em $|r_{0}|$ divindo as regiões onde temos trajetórias espalhadas e confinadas.

A Fig.(\ref{figorbita_II}) apresenta a mudança nas órbitas confinadas e espalhadas quando diminuímos o parâmetro $L$. As órbitas são obtidas mantendo o parâmetro $\bar{P}=0$ fixo e variando os valores de $L$. Na Fig.(\ref{figorbita_II}) são mostradas as trajetórias para fótons com $L=(-0.5,-1, -2,-3,-3.99)$. Na figura à esquerda (direita), quanto menor o valor de $L$ mais externa (interna) é a curva.

Como argumentamos anteriormente e pudemos observar nas trajetórias acima apresentadas, órbitas do tipo circular e elípticas não são permitidas no cilindro em rotação. Contudo, mudando alguns parâmetros do cilindro observamos que órbitas fechadas são possíveis. Essas órbitas podem ser observadas na Fig.(\ref{figorbita_III}).

Na parte superior da figura vemos duas órbitas descrevendo fótons confinados entre $r_{s}$ e $|r_{0}|$, enquanto na parte inferior observamos órbitas descrevendo fótons espalhados pelo cilindro. Neste caso os parâmetros escolhidos para os fótons foram $L=-0.03$ (linha vermelha) e $L=3$ (linha azul). Observa-se também que a região não-causal praticamente não existe. Contudo, o ponto $|r_{0}|$ ainda produz uma barreira infinita que divide os tipos de trajetórias. Assim, apesar da existência de órbitas fechadas elas não são tipo tempo, confirmando a ausência de geodésicas fechadas tipo tempo nesses espaços-tempos.

No caso da corda em rotação a situação é um pouco diferente. Aqui, a órbita se fecha dentro da região não-causal possibilitando a existência de geodésicas fechadas tipo tempo.

\begin{figure}[!h]
\centering
\epsfig{file =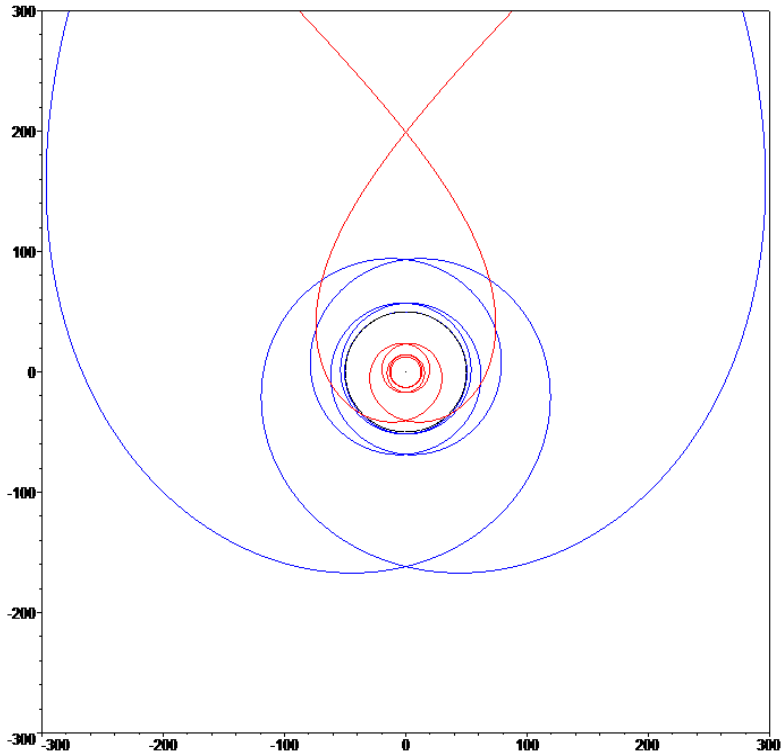, angle=0, width=0.75 \linewidth}
\caption{\small{Trajetória dos fótons no plano $(r\times\phi)$ para a corda em rotação com $J=1.0$, $\mu=0.23$. A linha tracejada representa $r_{nc}^{+} $.}}
\label{figorbita_IV}
\end{figure}

Apesar das órbitas apresentadas serem tipo luz é bem simples encontrar órbitas tipo tempo que apresentam quase os mesmos comportamentos que os fótons descritos na Fig.(\ref{figorbita_IV}). As órbitas apresentadas nesta figura descrevem fótons cujos parâmetros são $L=0$ (linha azul) e $L=-3$ (linha vermelha). Portanto, embora não existam geodésicas fechadas tipo tempo circulares ou elípticas, ainda temos trajetórias como as apresentadas acima que permitem uma viagem ao passado no caso da corda em rotação.

\newpage

\section{Órbitas retrógradas $(L+4J)<0$}

Finalmente, analisaremos que mudanças acontecem na dinâmica das geo-désicas quando $(L+4J)<0$.

Inspecionando a Eq.(\ref{cap8geo14}) vemos que o comportamento do potencial efetivo é independente do sinal da constante $(L+4J)$. Deste modo, as condições necessárias para a existência de pontos de retorno são as mesmas discutidas no caso $(L+4J)>0$. Resta-nos, então, analisar o comportamento dos cones de luz e os tipos de órbitas que fótons e partículas realizam sob essa condição.

\subsection{Cones de luz}

\begin{figure}[!h]
\centering
\epsfig{file =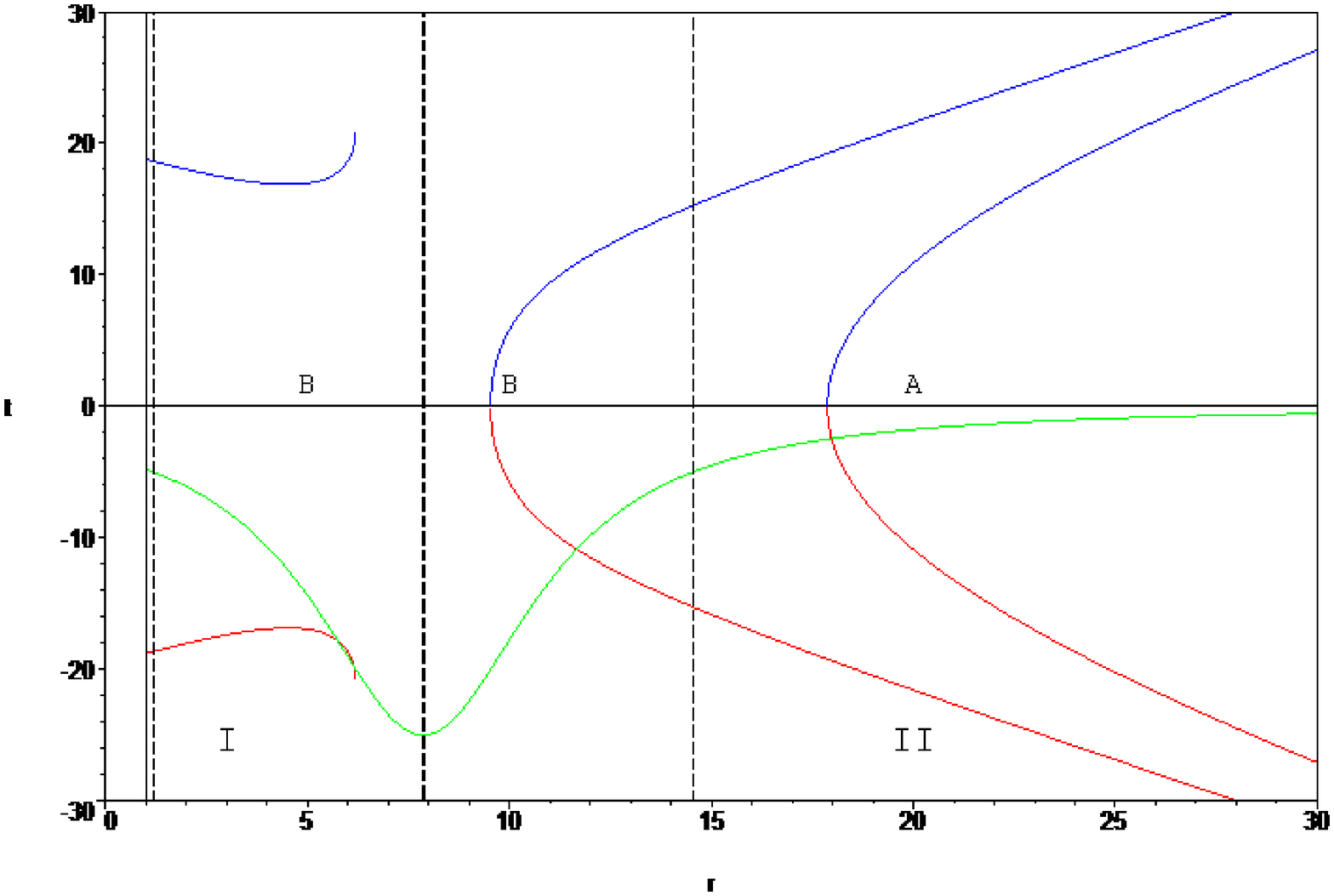,height=10cm,width=14cm, angle=0}
\caption{\small{Estrutura dos cones de luz para o cilindro em rotação com interior do tipo ``vaso de flor" com $J=1$, $\mu=0.1$, $r_{s}=1$ e $r_{0}=-7.87$. As linhas tracejadas verticais, localizadas da esquerda para direita representam $r=r_{nc}^{-}$, $r=|r_{0}|$ e $r=r_{nc}^{+} $, respectivamente.}}
\label{figconeD}
\end{figure}

\begin{figure}[!h]
\centering
\epsfig{file =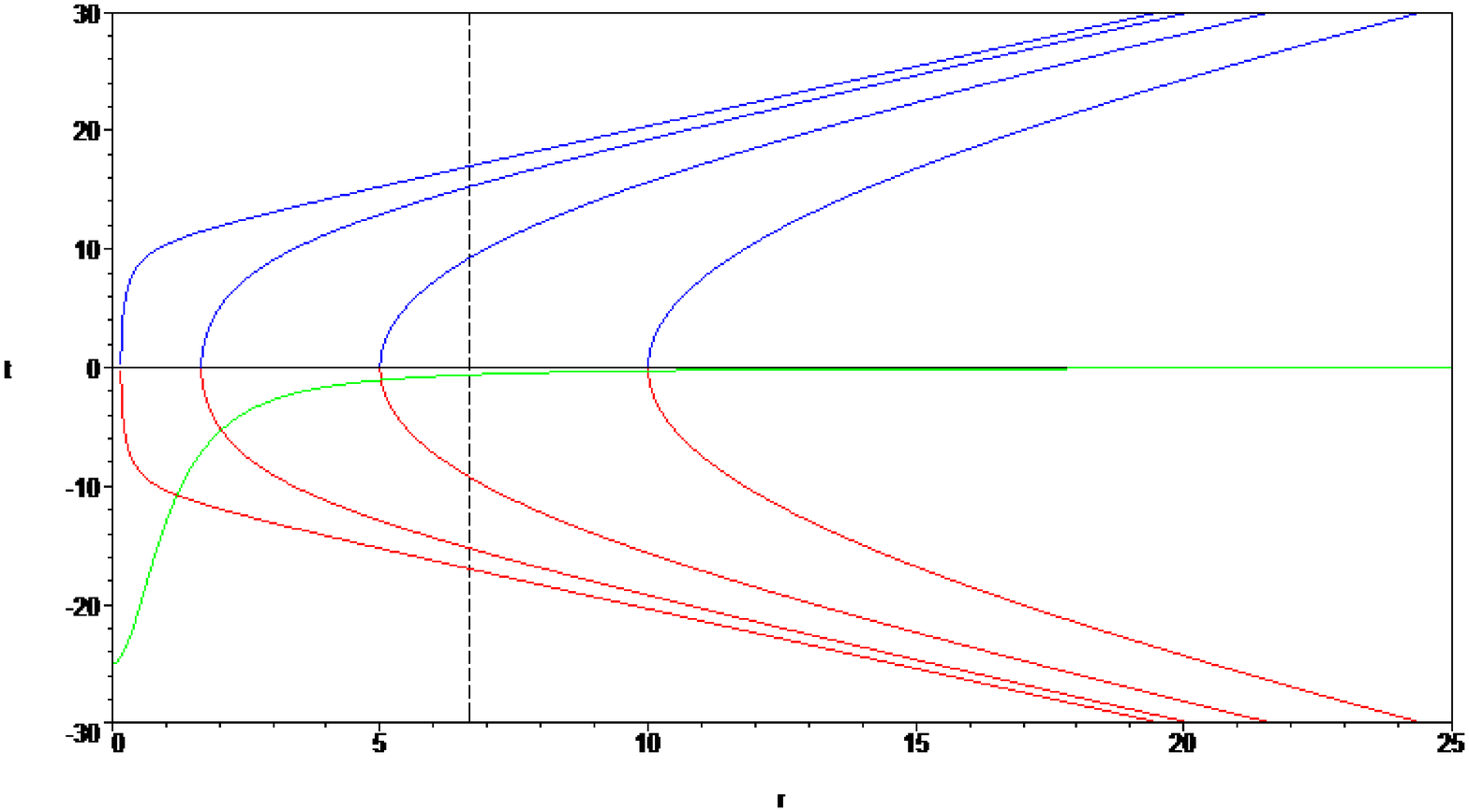,height=10cm,width=14cm, angle=0}
\caption{\small{Estrutura dos cones de luz para a corda em rotação com $J=1$ e $\mu=0.1$. A linha vertical tracejada representa $r_{nc}^{+} $. Os fótons representados acima possuem os seguintes parâmetros: (da esquerda para a direita) $L=(-4.1,-5,-7,-10)$.}}
\label{figconeD0}
\end{figure}

O comportamento dos cones de luz pode ser obtido integrando a equação
\begin{eqnarray}
\label{cap8geoD1}
\frac{dt}{dr}=\pm \left[\frac{\alpha(r+r_{0})}{\sqrt{\alpha^2(r+r_{0})^2\left(1-\frac{\epsilon}{E^2}-\bar{P}^2\right)-(L+4J)^2}}\right.+\nonumber\\
+\left.\frac{4J|L+4J|}{\alpha(r+r_{0})\sqrt{\alpha^2(r+r_{0})^2\left(1-\frac{\epsilon}{E^2}-\bar{P}^2\right)-(L+4J)^2}}\right]\quad,
\end{eqnarray}
e o resultado da integração é
\begin{eqnarray}
\label{cap8geoD2}
t=\pm \left[\frac{\sqrt{\alpha^2(r+r_{0})^2\left(1-\frac{\epsilon}{E^2}-\bar{P}^2\right)-(L+4J)^2}}{\alpha\left(1-\frac{\epsilon}{E^2}-\bar{P}^2\right)}+\right.\nonumber\\
+\left.\frac{4J}{\alpha}\ \textrm{arcsec}\left((r+r_{0})\frac{\sqrt{1-\frac{\epsilon}{E^2}-\bar{P}^2}}{|L+4J|}\right)\right]+C_{3}\quad.
\end{eqnarray}
Vejamos, a seguir, como se comportam alguns cones de luz quando, $r_{0}<0$ e $r_{0}=0$, nas Figs.(\ref{figconeD},\ref{figconeD0}). Novamente, os casos sem curvas fechadas tipo tempo não serão discutidos.

Observemos na Fig.(\ref{figconeD}) a mudança no comportamento dos cones de luz do cilindro em rotação. Neste caso, a velocidade angular do fóton quando $L=-5$, representada pela linha verde, não apresenta divergência. Para uma melhor visualização do seu comportamento, $\frac{d\phi}{dt}$ foi amplificada em 100 vezes. As linhas vermelhas representam o fóton se movimentando em direção ao cilindro e as linhas azuis representam o fóton escapando do cilindro. O ponto $|r_{0}|$ divide o espaço-tempo em duas regiões de interesse.
A curva $A$ descreve a trajetória de um fóton com $\bar{P}=0$ e $L=-10$ espalhado pelo cilindro. Já a curva $B$ descreve a trajetória de um fóton com $\bar{P}=0$ e $L=-5$ que é espalhado se estiver na região II ou é confinado se estiver na região I. Vemos que, diferentemente da curva $A$, a curva $B$ tem dois pontos de retorno. Na região I a evolução do fóton na coordenada $t$ é novamente descontínua, saltando de $t=20$ para $t=-20$. Entretanto, na região II, $t$ é contínua e causalmente bem comportada. Esse comportamento difere daquele apresentado no caso $(L+4J)>0$.

No caso da corda em rotação os cones de luz se comportam de maneira um tanto surpreendente. Este comportamento é apresentado na Fig.(\ref{figconeD0}).

Neste caso, a evolução do fóton na coordenada $t$ é sempre contínua e causalmente bem comportada, mesmo dentro da região não-causal. Aparentemente o fóton é espalhado pela corda, entrando e saindo da região não-causal, sem retornar ao próprio passado.
Aqui também mostramos o comportamento da velocidade angular do fóton quando $L=-4.1$, que foi amplificada em 100 vezes e é representada pela linha verde. Ela é sempre negativa e finita.

\subsection{Órbitas de fótons e partículas massivas}

\begin{figure}[!ht]
\centering
\epsfig{file =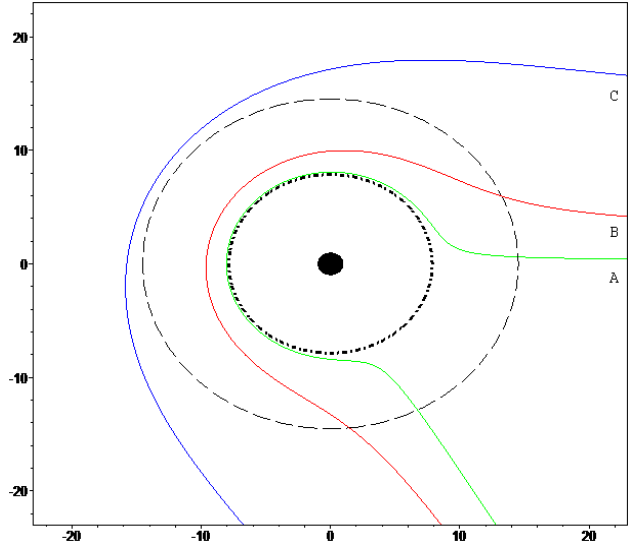, angle=0, width=0.6 \linewidth}
\caption{\small{Trajetória dos fótons com $L=-8.5$ (linha azul), $L=-5$ (linha vermelha) e $L=-4.1$ (linha verde) no plano $(r\times\phi)$ para o cilindro em rotação com interior do tipo ``vaso de flor" com $J=1.0$, $\mu=0.1$, $r_{s}=1$ e $r_{0}=-7.87$ . As linhas tracejadas localizadas da esquerda para a direita representam $r_{nc}^{-}$, $|r_{0}|$ e $r_{nc}^{+} $, respectivamente.}}
\label{figorbita_D0}
\end{figure}

\begin{figure}[!ht]
\centering
\epsfig{file =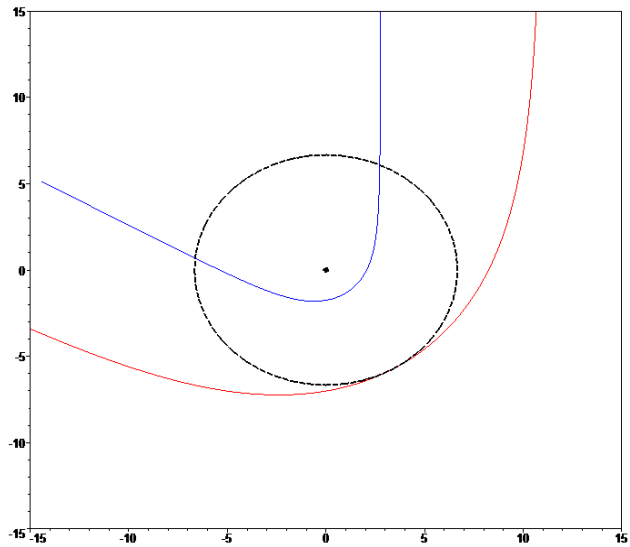, angle=0, width=0.6 \linewidth}
\caption{\small{Trajetória dos fótons com $L=-5$ (linha azul) e $L=-8$ (linha vermelha) no plano $(r\times\phi)$ para a corda em rotação com $J=1.0$, $\mu=0.1$. A linha tracejada representa $r_{nc}^{+} $.}}
\label{figorbita_D01}
\end{figure}

Vejamos agora quais tipos de trajetórias podem ser realizadas quando $(L+4J)<0$. Dividindo a Eq.(\ref{foton1}) pela Eq.(\ref{foton4}) obteremos a equação
\begin{eqnarray}
\label{cap8geoD3}
\frac{d\phi}{dr}=\pm \frac{|L+4J|}{\alpha(r+r_{0})\sqrt{\alpha^2(r+r_{0})^2\left(1-\frac{\epsilon}{E^2}-\bar{P}^2\right)-(L+4J)^2}}\quad.
\end{eqnarray}
Integrando a Eq.(\ref{cap8geoD3}) temos como resultado
\begin{eqnarray}
\label{cap8geoD4}
\phi=\pm \frac{1}{\alpha}\textrm{arcsec}\left(\frac{(r+r_{0})\alpha\sqrt{1-\frac{\epsilon}{E^2}-\bar{P}^2}}{|L+4J|}\right) +C_{1}\quad.
\end{eqnarray}
Invertendo a Eq.(\ref{cap8geoD4}) temos
\begin{eqnarray}
\label{cap8geoD5}
r(\phi)=\frac{|L+4J|}{\alpha\sqrt{1-\frac{\epsilon}{E^2}-\bar{P}^2}}\sec\left[\pm \frac{}{}\alpha(\phi-\phi_{0})\right]-r_{0}\quad.
\end{eqnarray}
Sob essa condição, os tipos de trajetórias descritas tanto por fótons quanto por partículas massivas são essencialmente os mesmos apresentados no caso $(L+4J)>0$.
Escolhemos alguns valores do parâmetro $L$ para ilustrar nas Figs.(\ref{figorbita_D0}, \ref{figorbita_D01}) algumas trajetórias de fótons tanto no cilindro em rotação como na corda em rotação.


\chapter{Curvas e Geodésicas fechadas tipo tempo}
\label{cap9CTC}

Quando pensamos em curvas fechadas tipo-tempo, imaginamos que para obtê-las é necessário que a matéria ou o espaço-tempo se comporte de maneira exótica. Entretanto como vimos nos capítulos anteriores não é este o caso. Objetos em rotação com simetria cilíndrica geralmente admitem curvas fechadas tipo-tempo. Parece-nos que a rotação dos objetos pode fornecer as condições necessárias para a formação dessas curvas, como vimos nos exemplos citados acima. 

Agora, curvas fechadas tipo-tempo parece-nos tão acessíveis que existem soluções onde tais curvas são geodésicas. A existência de geodésicas fechadas tipo-tempo (GFT) são mais um indício de que tal aspecto das soluções não podem ser ignorados. A vantagem de possuirmos GFT's é que não precisamos de nenhuma força externa para gerar tais curvas e que podemos, conseqüentemente, criar máquinas do tempo sem fornecer energia externa para a manutenção das GFT's.

Neste capítulo nos devotaremos a encontrar novos tipos de curvas fechadas tipo tempo e se possível, GFT's. Discutiremos também algumas de suas consequências.

Seja $S$ uma curva arbitrária parametrizada pelas coordenadas $x^{\mu}(\tau)$, onde $\tau$ é o parâmetro afim. O vetor tangente a curva $S$ será dado por
\begin{eqnarray}
\label{CFT1}
k^{\mu}=\frac{dS}{d\tau}\quad,
\end{eqnarray}
de modo que a curva $S$ será tipo tempo, tipo luz ou tipo espaço se a norma quadrática do vetor tangente $<k,k>=g_{\mu\nu}k^{\mu}k^{\nu}$ for negativa, nula ou positiva respectivamente.

Para uma geometria arbitrária a mesma curva genérica $S(x^{\mu})$ é solução da equação
\begin{eqnarray}
\label{CFT2}
\frac{d^2x^{\mu}}{d\tau^2}+\Gamma^{\mu}_{\alpha\beta}\frac{dx^{\alpha}}{d\tau}\frac{dx^{\beta}}{d\tau}=F^{\mu}\quad,
\end{eqnarray}
onde $F^{\mu}$ é o campo vetorial de forças externas que geram a curva $S$. Quando $S$ for uma geodésica, $F^{\mu}=0$ e a Eq.(\ref{CFT2}) se reduz a
\begin{eqnarray}
\label{CFT3}
\frac{d^2x^{\mu}}{d\tau^2}+\Gamma^{\mu}_{\alpha\beta}\frac{dx^{\alpha}}{d\tau}\frac{dx^{\beta}}{d\tau}=0\quad.
\end{eqnarray}
Para obtermos uma curva fechada tipo tempo queremos uma solução da Eq.(\ref{CFT2}) que possua a seguinte propriedade
\begin{eqnarray}
\label{CFT4}
x^{\mu}(\tau_{0})=x^{\mu}(\tau_{1})\quad,
\end{eqnarray}
onde $x^{\mu}$ faz o papel da coordenada temporal do espaço-tempo. Por exemplo, se $x^{\mu}$ for a coordenada $\phi$, a condição (\ref{CFT4}) é satisfeita devido à identifição entre $0$ e $2\pi$. Por outro lado, se $x^{\mu}$ for a coordenada $t$ sua evolução deve ter uma região onde $t(\tau_{0})=t(\tau_{1})$.

A curva fechada tipo tempo $S_{c}$ deve ter portanto uma característica periódica e tem o formato de círculo. Isso significa que ela sempre retorna ao ponto inicial. Essa característica é que nos permite voltar ao nosso passado. Sob essa curva voltamos ao nosso passado um número infinito de vezes.

Pode também ocorrer o caso de termos uma curva fechada tipo tempo $S_{lc}$ do tipo laço. Nesse tipo de curva a condição (\ref{CFT4}) também é sastisfeita infinitas vezes, mas diferentemente do caso da curva $S_{c}$ circular nem sempre voltamos ao ponto inicial da trajetória. Sob esses laços geralmente voltamos ao ponto inicial uma única vez.

Uma expressão geral para a obtenção de $S_{c}$ e $S_{lc}$ pode ser calculada, para espaços-tempos cilindricamente simétricos e estacionários que dependem apenas da coordenada radial, como se segue. Dada uma curva $S$ qualquer, ela satisfaz a seguinte equação
\begin{eqnarray}
\label{CFT5}
-\epsilon =g_{\phi\phi}\dot{\phi}^2+g_{rr}\dot{r}^2+g_{zz}\dot{z}^2+g_{tt}\dot{t}^2+2g_{t\phi}\dot{t}\dot{\phi}\quad,
\end{eqnarray}
onde $\ \dot{}\ $ é a derivada comum com relação a $\tau$ e $\epsilon$ define se a curva é tipo tempo $(\epsilon=1)$, tipo luz $(\epsilon=0)$ ou tipo espaço $(\epsilon=-1)$.
Uma curva do tipo $S_{c}$ circular pode ser obtida se escolhermos as condições
\begin{eqnarray}
\label{CFT6}
\epsilon=1\quad,\quad t(\tau)&=&t_{0}\quad,\quad r(\tau)=r_{c}\quad,\nonumber\\
\nonumber\\
z(\tau)=z_{0}&,& \phi(\tau)=A\tau\quad,
\end{eqnarray}
onde $A=\frac{1}{\sqrt{g_{\phi\phi}(r_{c})}}$ e a condição $g_{\phi\phi}(r_{c})<0$ deve ser satisfeita juntamente com a condição (\ref{CFT6}). Essas curvas são círculos de raio $r_{c}$ e a coordenada $\phi$ faz o papel de um tempo periódico satisfazendo a condição (\ref{CFT4}). Essa curva é o tipo de curvas fechada tipo tempo mais simples e mais estudado.

Curvas do tipo laço $S_{lc}$ e outras mais gerais, podem ser obtidas, dependendo da complexidade dos termos da métrica, se relaxarmos algumas imposições feitas na condição (\ref{CFT6}). Tomemos um novo e mais geral conjunto de condições dado por
\begin{eqnarray}
\label{CFT7}
t(\tau)&=&t_{0}\quad,\qquad g_{\phi\phi}(r(\tau))<0\quad,
\end{eqnarray}
de modo que a Eq.(\ref{CFT5}) se tornará
\begin{eqnarray}
\label{CFT8}
\dot{\phi}^2=\frac{-\epsilon - \left(g_{rr}\dot{r}^2+g_{zz}\dot{z}^2\right)}{g_{\phi\phi}}\quad.
\end{eqnarray}
A Eq.(\ref{CFT8}) acima tem como soluções curvas do tipo $S_{c}$, $S_{lc}$ e também curvas fechadas tipo luz. Para o caso do cilindro em rotação ou da corda em rotação temos os três tipos de soluções. Podemos checar os tipos de curvas calculando a norma quadrática do vetor tangente e observando seu sinal pois pode acontecer de uma curva mudar de tipo.

Após obtermos uma solução de curva fechada tipo tempo podemos tentar descobrir se existe alguma geodésica fechada tipo-tempo relacionada a esta curva. O procedimento é o seguinte:
\begin{itemize}
\item Primeiro substituímos as funções que definem a curva fechada tipo tempo na Eq.(\ref{CFT2}) e calculamos quais são as componentes da força $F^{\mu}$ que geram essa curva.

\item Procuramos pontos onde $F^{\mu}(x^{\mu}_{GFT})=0$. A geodésica fechada tipo tempo (GFT) será dada pela restrição imposta pela condição anterior. Portanto na região onde temos curvas fechadas tipo tempo geralmente podemos encontrar GFT's seguindo esse procedimento.
\end{itemize}

Letelier e Rosa aplicaram esse procedimento em diversos trabalhos onde analizaram espaços-tempos que admitem curvas fechadas tipo tempo \cite{Letelier,Letelier2}. Contudo, devemos ficar atentos ao realizar tal procedimento. Precisamos sempre checar se essa GFT implica na imposição de valores reais para as constantes de movimento associadas a ela. Algumas vezes é possível que a solução exista, apenas quando impomos valores complexos para essas constantes de movimento. Neste caso a ``GFT" encontrada deve ser descartada.

Usaremos esse procedimento para mostrar que não existem GFT's circulares associadas à curva fechada tipo tempo $S_{c}$ tanto no caso do cilindro quanto da corda em rotação.

Aplicando as Eqs.(\ref{cap2geo6}, \ref{cap2geo7}, \ref{cap2geo8}, \ref{cap2geo9}) obtidas no capítulo \ref{cap2geo} à métrica exterior do cilindro em rotação dada pela Eq.(\ref{cr1}) resulta no conjunto de equações
\begin{eqnarray}
\label{CFT9}
\ddot{r}-(1-4\mu)^2(r+r_{0})\dot{\phi}^2&=&0\quad,\\
\label{CFT9a}
\nonumber\\
\ddot{t}-\frac{8J\dot{\phi}\dot{r}}{(r+r_{0})}&=&0\quad,\\
\label{CFT9b}
\nonumber\\
\ddot{\phi}+\frac{2\dot{\phi}\dot{r}}{(r+r_{0})}&=&0\quad,\\
\label{CFT9c}
\nonumber\\
\ddot{z}&=&0\quad.
\end{eqnarray}
Escolhendo $r_{0}<0$ para garantir a existência da curva fechada tipo tempo e substituindo as funções (\ref{CFT6}) que definem a curva $S_{c}$ nas Eqs.(\ref{CFT9}) teremos
\begin{eqnarray}
\label{CFT10}
-(1-4\mu)^2(r_{c}-|r_{0}|)A^2&=&0\quad.
\end{eqnarray}
A única solução possível para a Eq.(\ref{CFT10}) é $r_{c}=|r_{0}|$. Portanto, a princípio, existe uma GFT associada à curva $S_{c}$ dada por
\begin{eqnarray}
\label{CFT11}
\epsilon=1,\quad t(\tau)&=&t_{0},\quad r(\tau)=|r_{0}|,\nonumber\\
\nonumber\\
z(\tau)=z_{0}&,& \phi(\tau)=A\tau
\end{eqnarray}
Entretanto, investigando as Eqs.(\ref{foton1},\ref{foton2},\ref{foton4}) observaremos que nenhum valor das constantes $L$ e $E$ é capaz de reproduzir as condições impostas pela ``GFT" circular dada pela Eq.(\ref{CFT11}). Além disso, como pode ser observado no capítulo anterior, quando $r_{0}<0$, $\dot{\phi}$ e $V_{g}$ divergem exatamente neste ponto. Consequentemente, essa solução deve ser descartada, mostrando a inexistência de GFT circulares.

No caso da corda em rotação, como $r_{0}=0$ a solução da Eq.(\ref{CFT10}) será $r_{c}=0$. Portanto, a possível ``GFT" circular é na verdade um ponto e não uma curva.  Assim, devemos descarta-lá neste caso também.

A seguir, apresentaremos algumas soluções de curvas fechadas tipo tempo obtidas nos espaço-tempos do cilindro e da corda em rotação.





Devido à simplicidade dos coeficientes de suas métricas exteriores tanto o cilindro quanto a corda em rotação possuem várias curvas fechadas tipo-tempo. Vejamos alguns exemplos.

A primeira curva $S_{c}$ é do tipo circular, sendo descrita pelas seguintes condições
\begin{eqnarray}
\label{scccr1}
z(\tau)=z_{0}, \quad t(\tau)=t_{0}, \quad r(\tau)=r_{c}, \quad \phi(\tau)=\frac{\tau}{\sqrt{g_{\phi\phi}}}
\end{eqnarray}
onde $t_{0}$, $r_{c}$ e $z_{0}$ são constantes e $g_{\phi\phi}(r_{c})<0$. Essas curvas são círculos sobre o plano $(r,\phi)$. Sob essa curva, a partícula fica parada.

A segunda curva, $S^{2}_{lc}$, é tipo laço, sendo descrita pelas expressões
\begin{eqnarray}
\label{slcccr1}
z(\tau)=\frac{\cosh(\tau)}{\sqrt{g_{zz}}},\quad t(\tau)=t_{0}, \quad r(\tau)=r_{c}, \quad \phi(\tau)=\frac{\sinh(\tau)}{\sqrt{g_{\phi\phi}}}
\end{eqnarray}
onde $t_{0}$ e $r_{c}$ são constantes e $g_{\phi\phi}(r_{c})<0$ e $g_{zz}(r_{c})>0$. Neste caso a partícula se movimenta apenas na direção $z$. O movimento na direção $\phi$ descreve a evolução temporal da partícula.

A terceira curva $S^{3}_{lc}$ é do tipo laço também e é dada por
\begin{eqnarray}
\label{slcccr2}
z(\tau)=\frac{(\tau-C_{1})^2}{2\sqrt{g_{zz}}},\quad t(\tau)=t_{0}, \quad r(\tau)=r_{0},\nonumber\\
\\
\phi(\tau)=\frac{\left[(\tau-C_{1})\sqrt{1+(\tau-C_{1})^2}+\textrm{arcsinh}(\tau-C_{1})\right]}{2\sqrt{g_{\phi\phi}}}\nonumber
\end{eqnarray}
onde $t_{0}$ e $r_{c}$ são constantes e $g_{\phi\phi}(r_{c})<0$ e $g_{zz}(r_{c})>0$. Neste caso também teremos uma partícula se movimentando apenas na direção $z$.

\begin{figure}[!h]
\centering
\epsfig{file =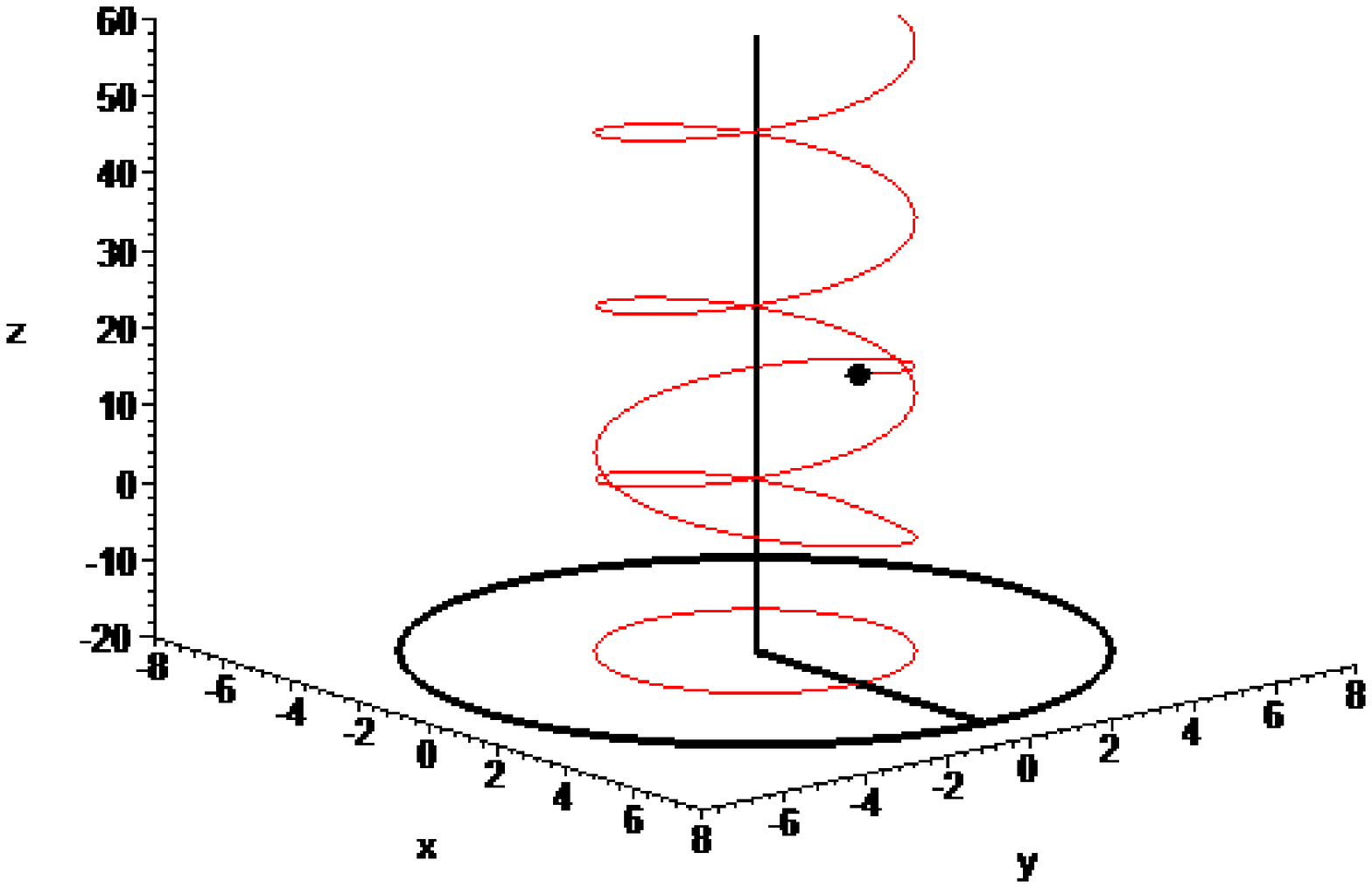,height=8cm,width=9.5cm, angle=0}

\epsfig{file =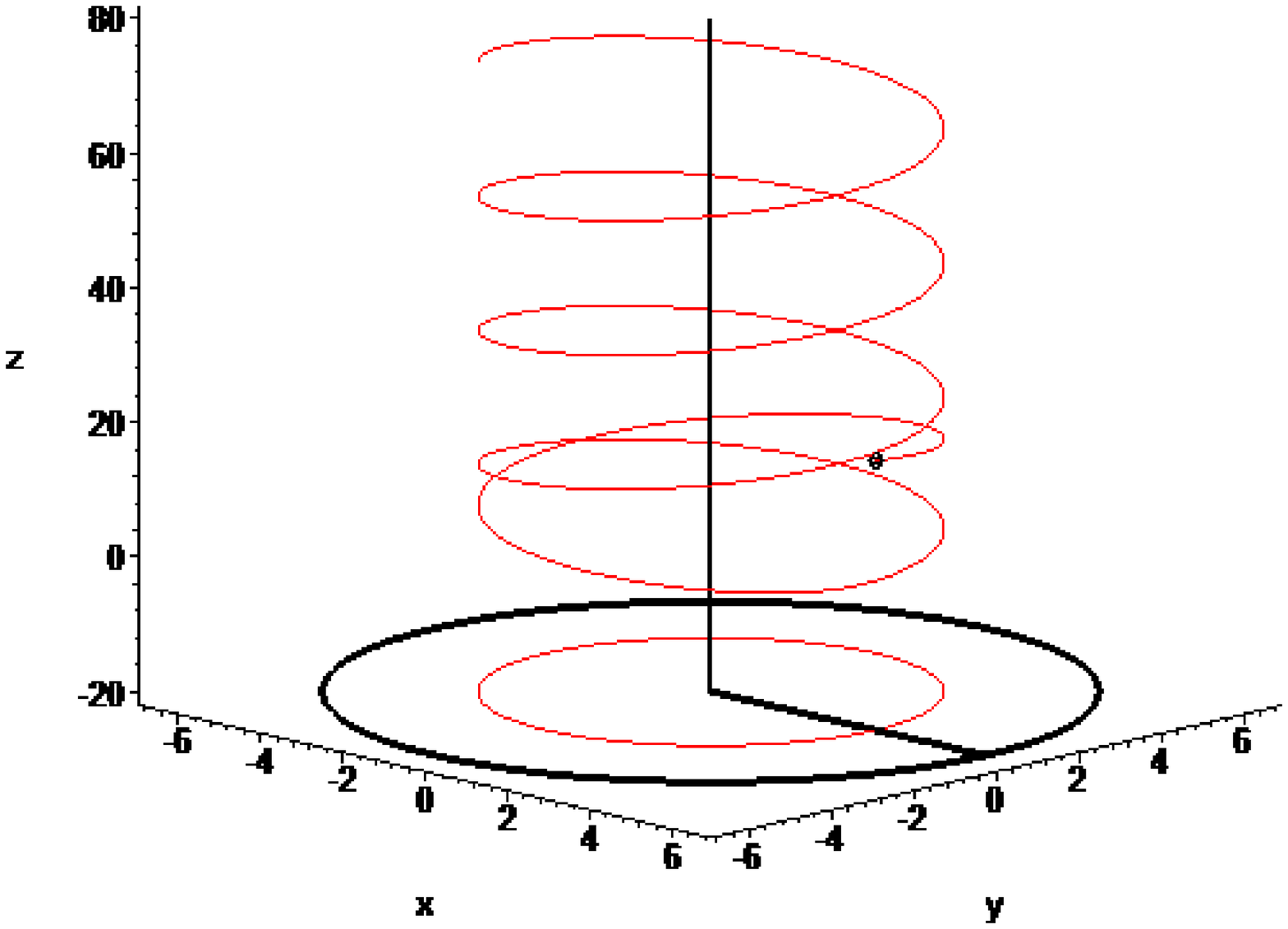,height=8cm,width=9.5cm, angle=0}
\caption{\small{ Comportamento das curvas $S^{2}_{lc}$ com $r_{c}=3$ (acima) e $S^{3}_{lc}$ com $r_{c}=4$ (abaixo) na corda em rotação com $J=1$ e $\mu=0.1$. O ponto escuro marca a posição inicial da partícula. O círculo preto delimita a região não-causal.}}
\label{ctclinha}
\end{figure}

\begin{figure}[!h]
\centering
\epsfig{file =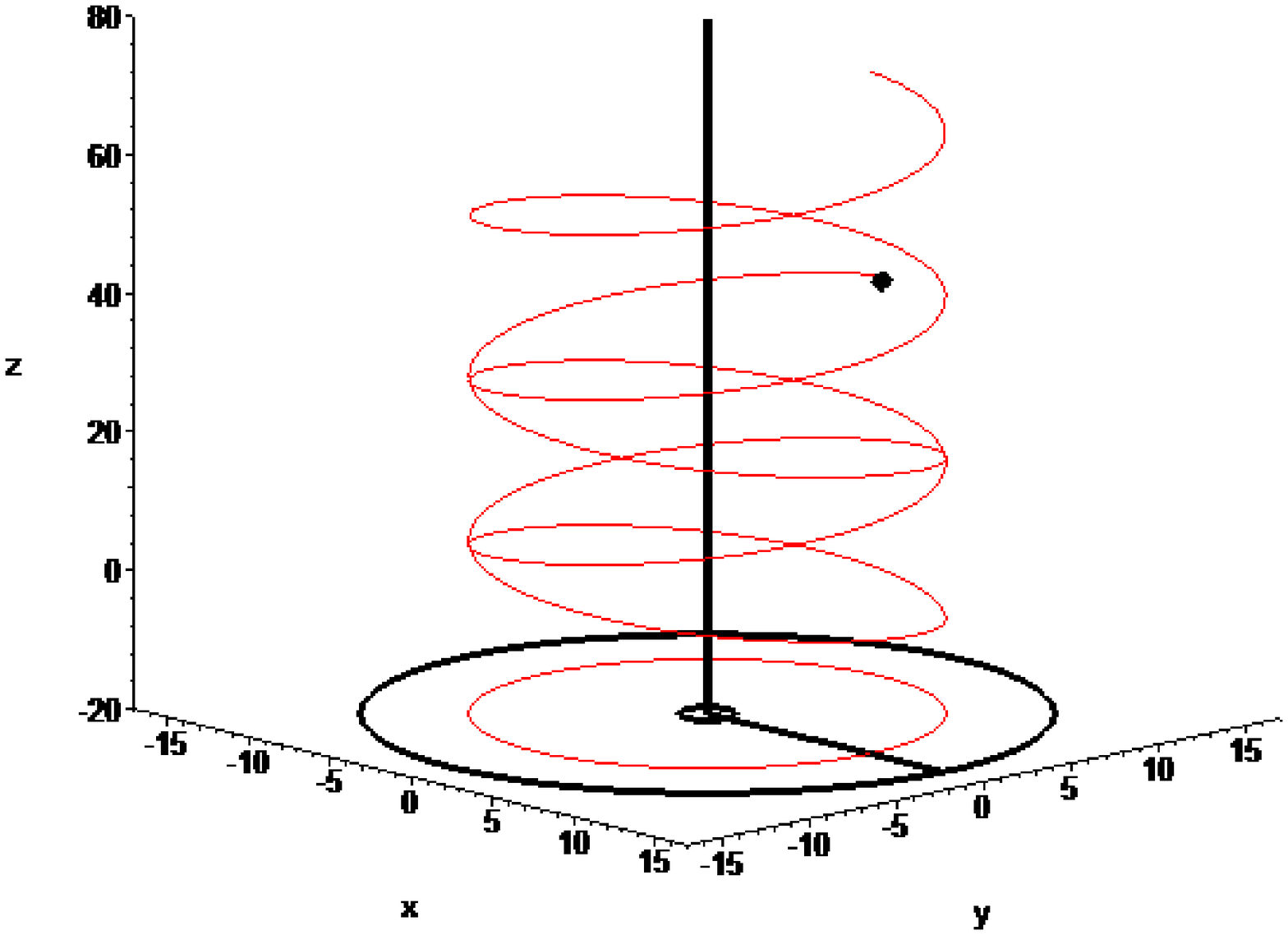,height=8cm,width=9.5cm, angle=0}

\epsfig{file =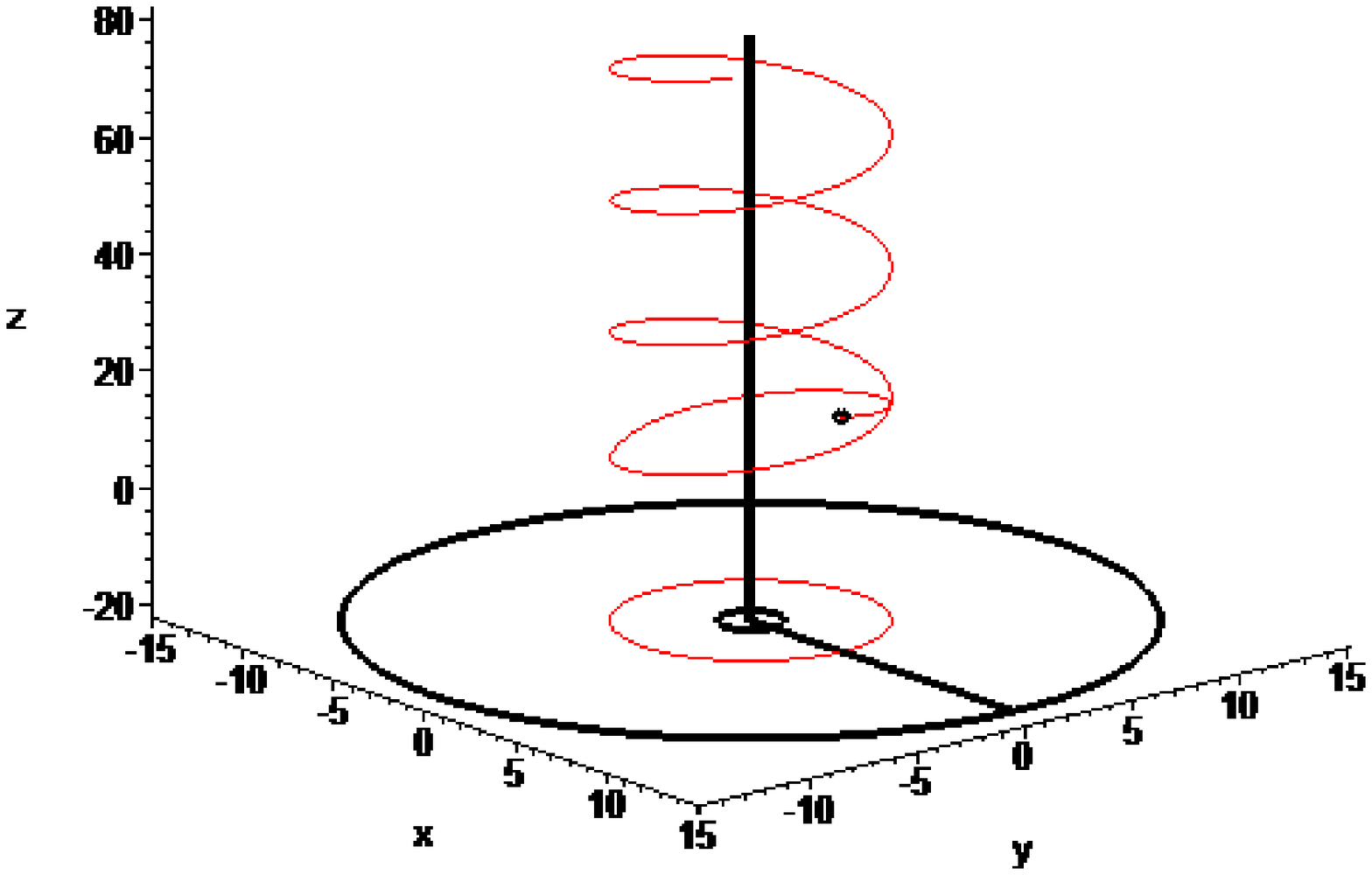,height=8cm,width=9.5cm, angle=0}
\caption{\small{ Comportamento das curvas $S^{2}_{lc}$ com $r_{c}=10$ (acima) e $S^{3}_{lc}$ com $r_{c}=5$ (abaixo) no cilindro em rotação com interior tipo ``vaso de flor" com $J=1$, $\mu=0.1$, $r_{s}=1$ e $r_{0}=-7.87$. O ponto escuro marca a posição inicial da partícula. Os círculos pretos delimitam a região não-causal.}}
\label{ctcvaso}
\end{figure}

Vejamos o comportamento das curvas citadas acima nas Figs.(\ref{ctclinha},\ref{ctcvaso}).

A evolução temporal das curvas $S^{2}_{lc}$ e $S^{3}_{lc}$ é praticamente a mesma tanto no cilindro quanto na corda em rotação. Entretanto, o movimento das curvas na direção $z$ é distinto. Diferentemente da curva $S_{c}$, nas curvas $S^{2}_{lc}$ e $S^{3}_{lc}$ a partícula não retorna ao passado no mesmo ponto inicial. Após a coordenada $\phi$ realizar um período de $2\pi$ a partícula geralmente encontra-se em outra posição no eixo $z$. No caso da curva $S^{3}_{lc}$ a partícula cruza o ponto inicial enquanto a curva $S^{2}_{lc}$ não.

A quantidade de pontos de intersecção das curvas depende geralmente das condições iniciais. Portanto, elas podem ser ajustadas para termos quantos pontos de intersecção quisermos. O fato interessante desses pontos é que eles marcam o encontro da partícula que viaja para o passado com a partícula que vem do futuro, ou seja, ele marca o ponto crucial do paradoxo do avô.

Vejamos agora alguns exemplos de curvas fechadas tipo luz. Usaremos o espaço-tempo da corda em rotação para ilustrá-los.

\begin{figure}[!h]
\centering
\epsfig{file =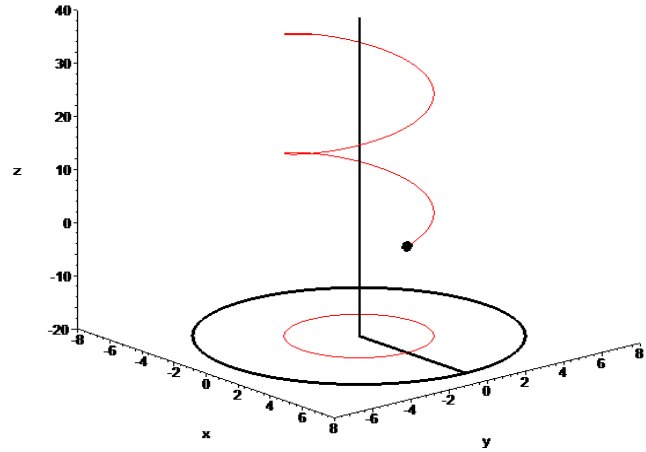,height=9cm,width=10.5cm, angle=0}

\epsfig{file =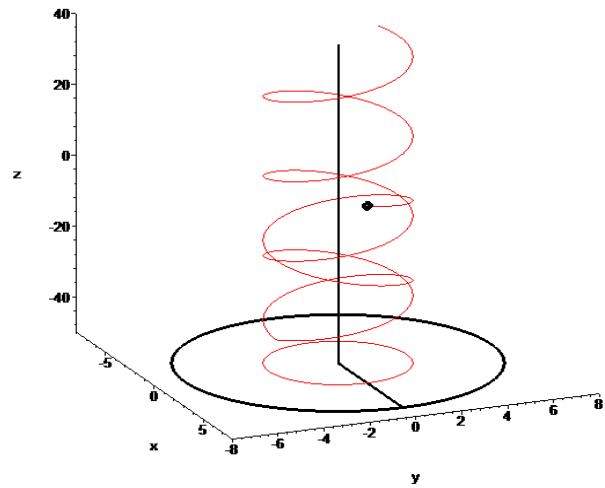,height=9cm,width=10.5cm, angle=0}
\caption{\small{ Comportamento das curvas fechadas tipo luz $S^{1}_{luz}$ com $r_{c}=3$ (acima) e $S^{2}_{luz}$ com $r_{c}=3$ (abaixo) na corda em rotação com $J=1$ e $\mu=0.1$. O ponto escuro marca a posição inicial da partícula. O círculo preto delimita a região não-causal.}}
\label{cfnlinha}
\end{figure}

A primeira curva chamaremos $S^{1}_{luz}$ e é descrita pelas condições
\begin{eqnarray}
\label{S1luz}
z(\tau)=\tau, \quad t(\tau)=t_{0}, \quad r(\tau)=r_{c}, \quad \phi(\tau)=\frac{\tau}{\sqrt{g_{\phi\phi}}}
\end{eqnarray}
onde $t_{0}$, $r_{c}$ são constantes e $g_{\phi\phi}(r_{c})<0$. Essas curvas descrevem espirais em torno do eixo $z$. Sob essa curva, o fóton segue na direção positiva de $z$. A cada volta que o fóton completa ele tem a possibilidade de influenciar seu passado. Contudo, como seu movimento sempre o afasta do ponto inicial ele não pode alterar seu passado. A curva $S^{1}_{luz}$ pode ser observada na Fig.(\ref{cfnlinha} (acima)).

A segunda curva será chamada de $S^{2}_{luz}$ e satisfaz
\begin{eqnarray}
\label{S2luz}
z(\tau)=\frac{\tau^2}{B}, \quad t(\tau)=t_{0}, \quad r(\tau)=r_{c}, \quad \phi(\tau)=\tau^2\sqrt{\frac{1}{B^2 g_{\phi\phi}}}
\end{eqnarray}
onde $t_{0}$, $r_{c}$ são constantes e $g_{\phi\phi}(r_{c})<0$. Essas curvas descrevem um comportamento oscilatório inicialmente e depois seguem em espiral na direção positiva do eixo $z$. Sob essa curva, o fóton é capaz de voltar ao próprio passado e interferir em sua trajetória pois a curva apresenta 3 pontos de intersecção. A curva $S^{2}_{luz}$ pode ser observada na Fig.(\ref{cfnlinha} (abaixo)).

\chapter{Evolução do campo escalar nos espaços-tempos das cordas e dos cilindros}
\label{cap10Evoescalar_et}

Neste capítulo apresentaremos as soluções das equações que governam a dinâmica de campos escalares nos espaços-tempos apresentados no capítulo \ref{cap6spacetime}.

\section{Cilindro cósmico estático}
\label{evocordaestatica}
Como mostrado na Seção (\ref{cordaestatica}) o cilindro cósmico estático possui uma métrica interior e uma métrica exterior. O elemento de linha que caracteriza a primeira é
\begin{eqnarray}
\label{pcc1}
ds^2=-dt^2+dr^2+dz^2+r^2_{0}\sin(r/r_{0})^{2} d\phi^2\quad,
\end{eqnarray}
e o elemento de linha da segunda é
\begin{eqnarray}
\label{pcc2}
ds^2=-dt^2+dr^2+dz^2+(1-4\mu)^2r^2d\phi^2\quad,
\end{eqnarray}
onde nos manteremos no intervalo de parâmetro $\left(\mu<\frac{1}{4}\right)$, pelo motivos declarados anteriormente.
A equação mestra da evolução do campo e o próprio campo escalar $\Psi$ são dados pelas Eqs.(\ref{cap3a20}, \ref{cap3a21}, \ref{cap3a22},).
Desta forma, podemos calcular a evolução do campo nas duas métricas e conectar as duas soluções de modo suave e contínuo.
Para a métrica exterior (\ref{pcc2}) a equação do campo se torna
\begin{eqnarray}
\label{pcc3a}
\frac{(-g)_{,r}}{2}\ \frac{dR}{dr}+\alpha^2r^2\frac{d^2R}{dr^2}&-&k^2\alpha^2r^2R=\nonumber\\
&=&R\left[\frac{}{}\alpha^2r^2\beta^2+m^2-\alpha^2r^2\omega^2\right]\quad.
\end{eqnarray}
Reescrevendo a Eq.(\ref{pcc3a}) de forma mais compacta temos
\begin{eqnarray}
\label{pcc4}
r^2\frac{d^2R}{dr^2}&+&r\ \frac{dR}{dr} +R\left[\left(\omega^2-k^2-\beta^2\right)r^2-\frac{m^2}{\alpha^2}\right]=0\quad,
\end{eqnarray}
onde usamos as relações $\alpha=(1-4\mu)$ e $\mathcal{X}=-g=\alpha^2r^2$ para realizar os cálculos. Inspecionando a Eq.(\ref{pcc4}) vemos que se assumirmos que $\omega$ é um número real e $\omega^2>k^2+\beta^2$ então a Eq.(\ref{pcc4}) é a equação de Bessel. A solução geral da Eq.(\ref{pcc4}) é dada em termos das funções de Bessel e Neumann sendo escrita como
\begin{eqnarray}
\label{pcc4a}
R(r)=C_{1}\ J_{\nu}\left(\sqrt{\omega^2-k^2-\beta^2}\ r\right) + C_{2}\ N_{\nu}\left(\sqrt{\omega^2-k^2-\beta^2}\ r\right)\quad.
\end{eqnarray}
O índice $\nu=\frac{m}{\alpha}$ será um número real discreto já que $m$ é um inteiro. Outras combinações lineares dessas funções ainda podem ser realizadas de modo a gerar outras soluções com características distintas para a Eq.(\ref{pcc4}). Uma delas gera como solução as funções de Hankel que são de grande interesse uma vez que elas se comportam assintoticamente como ondas viajantes.

Por outro lado se $\omega$ for real mas $\omega^2<k^2+\beta^2$ então a Eq.(\ref{pcc4}) é a equação de Bessel modificada e possui dois tipos de soluções independentes,
\begin{eqnarray}
\label{pcc4b}
R(r)=C_{1}\ I_{\nu}\left(\sqrt{k^2+\beta^2-\omega^2}\ r\right) + C_{2}\ K_{\nu}\left(\sqrt{k^2+\beta^2-\omega^2}\ r\right)\quad.
\end{eqnarray}
Ainda temos o caso onde $\omega^2=k^2+\beta^2$ de modo que a Eq.(\ref{pcc4}) terá por solução geral
\begin{eqnarray}
\label{pcc4c}
R(r)=C_{1}\ r^{\frac{m}{\alpha}} + C_{2}\ r^{-\frac{m}{\alpha}}\quad.
\end{eqnarray}
Se quisermos uma solução menos restrita podemos supor que $\omega$ seja um número complexo. Esse será o caso quando calcularmos os modos quasinormais deste espaço-tempo. Neste caso a Eq.(\ref{pcc4}) terá como solução funções de Bessel como as apresentadas na Eq.(\ref{pcc4a}), mas o argumento das funções poderá assumir valores complexos. 

Para a métrica interior (\ref{pcc1}) a equação do campo pode ser escrita como
\begin{eqnarray}
\label{pcc5}
\frac{(-g)_{,r}}{2}\ \frac{dR}{dr}+\mathcal{X}\frac{d^2R}{dr^2}-k^2\mathcal{X}R-
R\left[\frac{}{}\mathcal{X}\beta^2+m^2-L\omega^2\right]=0\quad.
\end{eqnarray}
Substituindo as funções $L$, $g$ e $\mathcal{X}$ na Eq.(\ref{pcc5}) temos
\begin{eqnarray}
\label{pcc6}
\frac{d^2R}{dr^2}+\frac{1}{r_{0}\tan(\frac{r}{r_{0}})} \frac{dR}{dr} +R\left[\frac{}{}\left(\omega^2-k^2-\beta^2\right)-\frac{m^2}{(r_{0})^2\sin(\frac{r}{r_{0}})^2}\right]=0\quad,
\end{eqnarray}
onde usamos as relações $\mathcal{X}=-g =r^2_{0}\sin(r/r_{0})^2$. Se fizermos uma transformação de coordenadas onde $r/r_{0}=x$ então temos
\begin{eqnarray}
\label{pcc7}
\frac{d^2R}{dx^2}+\frac{1}{\tan(x)} \frac{dR}{dx} +R\left[r^2_{0}\left(\omega^2-k^2-\beta^2\right)-\frac{m^2}{\sin(x)^2}\right]=0.
\end{eqnarray}
As possíveis soluções da Eq.(\ref{pcc7}) podem ser classificadas através das ca-racterísticas adotadas para $\omega$ da mesma maneira como foi realizada para a métrica exterior. Tomando o caso de maior interesse, assumiremos que $\omega$ é um número complexo. Neste caso a solução da geral da Eq.(\ref{pcc7}) será uma combinação linear de funções de Legendre Associadas dadas por
\begin{eqnarray}
\label{pcc7a}
R(r)=C_{1}\ P_{n}^{m}\left(\cos\left(x\right)\right) + C_{2}\ Q_{n}^{m}\left(\cos\left(x\right)\right)\quad,
\end{eqnarray}
onde o índice $n$ é dado por
\begin{eqnarray}
\label{pcc7b}
n=-\frac{1}{2}+\frac{\sqrt{4r_{0}^{2}\left(\omega^2-k^2-\beta^2\right)+1}}{2}\quad.
\end{eqnarray}
A escolha adequada de combinações de soluções das Eqs.(\ref{pcc4}, \ref{pcc6}) está estritamente relacionada ao tipo de problema físico que se deseja estudar e as condições de contorno que esse problema devem satisfazer. O problema de interesse a ser tratado nesta tese é o cálculo dos modos quasinormais. Discutiremos as condições de contorno para os modos quasinormais e suas implicações nos próximos capítulos.

\section{Corda cósmica estática}

A propagação do campo escalar no espaço-tempo da corda cósmica estática pode ser pensada como um caso particular dessa propagação no cilindro estático. Isso por que a métrica da corda, dada pela Eq.(\ref{pcc2}) é a mesma daquela que descreve o exterior do cilindro. A diferença agora é que, as soluções obtidas para o campo escalar, são válidas em todo lugar exceto em $r=0$.

A equação que governa a dinâmica do campo assim como as soluções possíveis dessa equação são aquelas apresentadas nas Eqs.( \ref{pcc4}, \ref{pcc4a}, \ref{pcc4b}, \ref{pcc4c},) que foram obtidas para a região exterior do cilindro. A decomposição do campo é dada pela Eq.(\ref{cap3a20}) e a discussão realizada anteriormente sobre as características de $\omega$ ainda são válidas para este caso.

\section{Corda cósmica em rotação}
\label{evolinharotacao}

Como vimos anteriormente, a corda cósmica em rotação não possui métrica interior, de modo que podemos usar as Eqs.(\ref{cap3a20}, \ref{cap3a21}, \ref{cap3a22}) diretamente para a métrica apresentada pela Eq.(\ref{lr1}). As relações
\begin{eqnarray}
\label{plcr1}
\mathcal{X}=\sigma^2r^2=-g\quad,\qquad \sigma=(1-4\mu)\quad,
\end{eqnarray}
são utilizadas no cálculo das equações do campo de modo que teremos uma equação diferencial ordinária de segunda ordem unidimensional dada por
\begin{eqnarray}
\label{plcr2}
\frac{(-g)_{,r}}{2}\ \frac{dR}{dr}&+&\mathcal{X}\frac{d^2R}{dr^2}-R\left[\mathcal{X}(\beta^2+k^2)+\right.\nonumber\\
\nonumber\\
&&\qquad \qquad \left.+ m^2-L\ \omega^2-2Mm\omega\right]=0\quad.\ \ \
\end{eqnarray}
A substituição dos valores das funções $F$, $L$, $M$ e $\mathcal{X}$ na Eq.(\ref{plcr2}) resulta em
\begin{eqnarray}
\label{plcr3}
\frac{(-g)_{,r}}{2}\ \frac{dR}{dr}&+&\sigma^2r^2\frac{d^2R}{dr^2}-R\left\{\frac{}{}\sigma^2r^2(\beta^2+k^2)+m^2-\right.\nonumber\\
\nonumber\\
&&\qquad \qquad  -\left. \frac{}{}\left[\sigma^2r^2-16J^2\right] \omega^2+8Jm\omega\right\}=0\quad.\ \
\end{eqnarray}
Reescrevendo a Eq.(\ref{plcr3}) de maneira mais conveniente temos
\begin{eqnarray}
\label{plcr4}
r^2\frac{d^2R}{dr^2}&+&r \frac{dR}{dr}+R\left[\left(\omega^2-k^2-\beta^2\right)r^2-\frac{(m+4J\omega)^2}{\sigma^2}\right]=0\quad.\ \ \ \ \ \ \ \ \
\end{eqnarray}
Observando atentamente a Eq.(\ref{plcr4}) vemos que ela é muito semelhante à Eq.(\ref{pcc4}), ou seja, ela também é uma equação de Bessel. Na verdade, se fizermos $J=0$ recuperamos exatamente a Eq.(\ref{pcc4}) descrevendo o caso da corda estática. Embora a corda cósmica em rotação apresente termos não diagonais na métrica eles podem ser agrupados de modo que a única diferença entre as equações mencionadas aparecerá na dependência não-trivial em $\omega$ da ordem das funções de Bessel. Novamente, a análise a respeito do comportamento das soluções em função das características de $\omega$ podem ser realizadas de maneira semelhante. 

Escrevemos a solução geral da Eq.(\ref{plcr4}) como (\ref{pcc4a})
onde o índice $\nu$  que indica a ordem da função de Bessel é
\begin{eqnarray}
\label{plcr6}
\nu=\pm\sqrt{\frac{(m+4J\omega)^2}{\sigma^2}}\quad.
\end{eqnarray}
Sendo $\omega$ um número complexo vemos que a ordem $\nu$ das funções de Bessel também será complexa.

\section{Cilindro cósmico em rotação}
\label{evocordarotacao}

No caso da cilindro em rotação devemos tratar da propagação do campo escalar tanto na região exterior do espaço-tempo como na região interna. Para a região interna as soluções não são únicas, portanto analisaremos as duas soluções internas descritas anteriormente:
\begin{itemize}
\item Solução tipo ``vaso de flor",

\item Solução tipo ``caneta esferográfica".
\end{itemize}
A solução que descreve a região externa é a mesma para ambas as soluções internas.

\subsection*{ Solução exterior}

Como sabemos, a métrica que descreve a região exterior é dada pela Eq.(\ref{cr1}). Aplicaremos as Eqs.(\ref{cap3a20}, \ref{cap3a21}, \ref{cap3a22}) a essa métrica de modo direto também. As relações
\begin{eqnarray}
\label{pccr1}
\mathcal{X}=\sigma^2(r+r_{0})^2=-g,\qquad \sigma=(1-4\mu)\quad,
\end{eqnarray}
são utilizadas no cálculo das equações do campo de modo que
\begin{eqnarray}
\label{pccr2}
&&\frac{(-g)_{,r}}{2}\ \frac{dR}{dr}+\mathcal{X}\frac{d^2R}{dr^2}+(-k^2)\mathcal{X}R-R\left[\frac{}{}\mathcal{X}\beta^2+\right.\nonumber\\
\nonumber\\
&&\qquad \qquad \qquad \qquad \left. \frac{}{}+m^2-L\ \omega^2-2Mm\omega\right]=0\quad.
\end{eqnarray}
Substituindo as funções $F$, $M$, $L$, $\mathcal{X}$ e $g$ na Eq.(\ref{pccr2}) temos
\begin{eqnarray}
\label{pccr3}
&&\sigma^2(r+r_{0}) \frac{dR}{dr}+\sigma^2(r+r_{0})^2\frac{d^2R}{dr^2}-R\left[\frac{}{}\sigma^2(r+r_{0})^2(\beta^2+k^2)+\right.\nonumber\\
\nonumber\\
&& \qquad \qquad +\left. \frac{}{}m^2-\left[\sigma^2(r+r_{0})^2-16J^2\right] \omega^2+8Jm\omega\right]=0\quad,
\end{eqnarray}
que escrito de forma compacta resulta em
\begin{eqnarray}
\label{pccr4}
&&x^2\frac{d^2R}{dx^2}+ x\frac{dR}{dx} + R\left[\left(\omega^2-k^2-\beta^2\right)x^2-\frac{(m+4J\omega)^2}{\sigma^2}\right]=0\quad.\ \qquad
\end{eqnarray}
No cálculo da equação fizemos a mudança de coordenadas $x=(r+r_{0})$. As soluções da Eq.(\ref{pccr4}), no caso mais geral onde $\omega$ é um número complexo, são novamente as funções de Bessel descritas pelas Eqs.(\ref{pcc4a}, \ref{plcr6}) apenas com a dependência radial trocada de $r\rightarrow r+r_{0}$. Vemos que quando $J=0$ e $r_{0}=0$ recuperamos a mesma solução obtida para o cilindro estático e quando fazemos apenas $r_{0}=0$ recuperamos a solução obtida para a corda cósmica em rotação.
Essa solução exterior é muito importante pois descreve a evolução do campo escalar em um espaço-tempo que possui curvas fechadas tipo tempo. Apesar desta característica exótica ainda podemos calcular os MQN's deste espaço-tempo se tivermos a solução da parte interior. Este é o nosso próximo passo.

\subsection*{ Solução vaso de flor }

A primeira solução interna a ser analisada é a do tipo ``vaso de flor" com a métrica sendo dada pela Eq.(\ref{cr3}). Assim como foi feito para a região exterior, seguiremos os mesmos passos usando as relações
\begin{eqnarray}
\label{pccri11}
\mathcal{X}=r^2\sigma_{1}^2=-g\quad,   \qquad \sigma_{1}^2=\left(\frac{r_{s}^2+16J^2}{r_{s}^2}\right)\quad,
\end{eqnarray}
e as Eqs.(\ref{cap3a20}, \ref{cap3a21}, \ref{cap3a22}) obtemos a equação para a evolução do campo que será
\begin{eqnarray}
\label{pccri12}
r^2\frac{d^2R}{dr^2}+ r\frac{dR}{dr} + R \left\{\left[\frac{1}{\sigma_{1}^2}\left(\omega-\frac{4Jm}{r^2_{s}}\right)^2-k^2-\beta^2\right]r^2-m^2 \right\}=0\ .
\end{eqnarray}
A solução mais geral da Eq.(\ref{pccri12}) quando $\omega$ é complexo são as funções de Bessel, de modo que
\begin{eqnarray}
\label{pccri13}
R(r)=C_{1}\ J_{m}\left(p\ r\right) + C_{2}\ N_{m}\left(p\ r\right)\quad,
\end{eqnarray}
onde a constante $p$ é dada por
\begin{eqnarray}
\label{pccri14}
p=\sqrt{\left[\frac{1}{\sigma_{1}^2}\left(\omega-\frac{4Jm}{r^2_{s}}\right)^2-k^2-\beta^2\right]}\quad.
\end{eqnarray}
O efeito da rotação do espaço-tempo pode ser observado na mudança da dependência de $p$. A frequência $\omega$ sofre um deslocamento devido ao momento angular $J$ do cilindro, o índice $m$ e o raio $r_{s}$ da casca cilíndrica. Interessante é o fato que este deslocamento é discreto uma vez que $m$ é um número inteiro. O momento angular e $r_{s}$ afeta também a amplitude de $p$. Usaremos essa solução juntamente com a solução para região exterior para calcular os modos quasinormais deste espaço-tempo.

\subsection*{ Solução caneta esferográfica}

A segunda solução a ser analisada é a do tipo ``caneta esferográfica" cuja métrica é dada pela Eq.(\ref{cr7}). Assim como foi feito para a outra solução interior, seguiremos os mesmos passos usando agora as relações
\begin{eqnarray}
\label{pccri21}
&&\mathcal{X}=\frac{\sin(\sqrt{\lambda}r)^2}{\lambda}=-g\quad,  \qquad -g_{,r}=\frac{\sin(\sqrt{\lambda}r)\cos(\sqrt{\lambda}r)}{\sqrt{\lambda}}\quad,\nonumber\\
\nonumber\\
&&M=-2\tilde{\alpha} f\quad, \qquad f=\left[(r-r_{s})\cos(\sqrt{\lambda}r)-\frac{\sin(\sqrt{\lambda}r)}{\sqrt{\lambda}}+r_{s}\right]\ ,\ \ \ \ \
\end{eqnarray}
e as Eqs.(\ref{cap3a20}, \ref{cap3a21}, \ref{cap3a22}) obtendo a equação para a evolução do campo que será
\begin{eqnarray}
\label{pccri22}
&&\sin(x)^2\frac{d^2R}{dx^2}+ \sin(x)\cos(x)\frac{dR}{dx} + \qquad \qquad \nonumber\\
\nonumber\\
&&\qquad + R\left\{\frac{\left(\omega^2-k^2-\beta^2\right)}{\lambda}\sin(x)^2-[m+2\tilde{\alpha}\omega f]^2\right\}=0\quad.\ \ \
\end{eqnarray}
No cálculo da Eq.(\ref{pccri22}) fizemos a mudança de coordenadas $x=\sqrt{\lambda}r$. Neste caso a Eq.(\ref{pccri22}) não possui solução exata conhecida exceto para o caso muito particular onde impomos $\tilde{\alpha}=0$ e $\sqrt{\lambda}=\frac{1}{r_{0}}$. Neste caso, são recuperadas a Eq.(\ref{pcc7}) e sua solução dada pelas Eqs.(\ref{pcc7a}, \ref{pcc7b}) para o cilindro estático.
Apesar desse fato, pudemos obter duas soluções aproximadas dessa equação nos seguintes regimes
\begin{itemize}
\item limite quasi-estático quando $\sqrt{\lambda}r_{s}\ll 1$
\item limite de altos valores de $m$
\end{itemize}
Para estes dois limites fomos capazes de calcular os modos quasinormais desse espaço-tempo. Eles serão apresentados nos próximos capítulos.
\newpage

\chapter{Modos quasinormais  }
\label{MQN}

Em 1957, as discussões sobre qual seria o melhor entendimento de certas soluções singulares das equações de Einstein, que depois receberiam o nome de ``buracos negros", ainda eram bastante controversas. Neste mesmo ano, Regge e Wheeler \cite{Regge} lançaram alguma luz sobre esse asunto iniciando um novo ramo de pesquisa em Relatividade Geral. Eles estudaram a singularidade de Schwarzschild analisando o espectro de frequências complexas de perturbações lineares na métrica e concluiram que tal singularidade era estável.

Partindo dessa iniciativa uma série de outros trabalhos relacionando a propagação de campos de matéria e a estabilidade de geometrias foi realizada \cite{Abdalla,Molina}. Dentre as muitas opções de estudo escolhemos trabalhar com a análise dos \emph{modos quasinormais}.

No contexto da Relatividade Geral, os modos quasinormais (MQN's) podem ser entendidos como o conjunto de frequências complexas de oscilação, tanto de campos de matéria evoluindo em uma geometria quanto da própria geometria perturbada gravitacionalmente. A componente complexa desses modos pode gerar um amortecimento ou uma amplificação exponencial das oscilações indicando a estabilidade ou a instabilidade do campo ou da geometria. Um outro aspecto importante desses modos é que em geral eles guardam estreita relação com os parâmetros que definem o espaço-tempo analisado. Por essa razão eles são considerados impressões digitais dos objetos gravitacionais que os produziram \cite{Nollert}.

No caso de buracos negros os modos quasinormais são intimamente ligados aos parâmetros que os caracterizam tais como a massa, a carga elétrica e o momento angular. Ainda hoje o estudo desses modos de oscilação em buracos negros é um ramo de pesquisa que desperta grande interesse \cite{Chandra,Cardoso0,Kokkotas}.

Pensando nestas razões investigaremos a existência de tais modos nos espaços-tempos estudados, com interesse especial naqueles que possuem curvas fechadas tipo-tempo. Nossa ``ferramenta de trabalho" será o campo escalar que será inserido na métrica fundo como uma perturbação, de tal sorte que o tensor energia-momentum do campo é desprezível. Desta forma precisamos resolver apenas a equação de Klein-Gordon, que governa a evolução do campo escalar. A evolução do campo será afetada pela curvatura da métrica e assim relacionaremos a evolução do campo com os parâmetros da métrica.

Condições de contorno bem específicas devem ser impostas nas soluções das equações de perturbação para a obtenção dos MQN's.
\begin{figure}[!ht]
\begin{center}
\epsfig{file =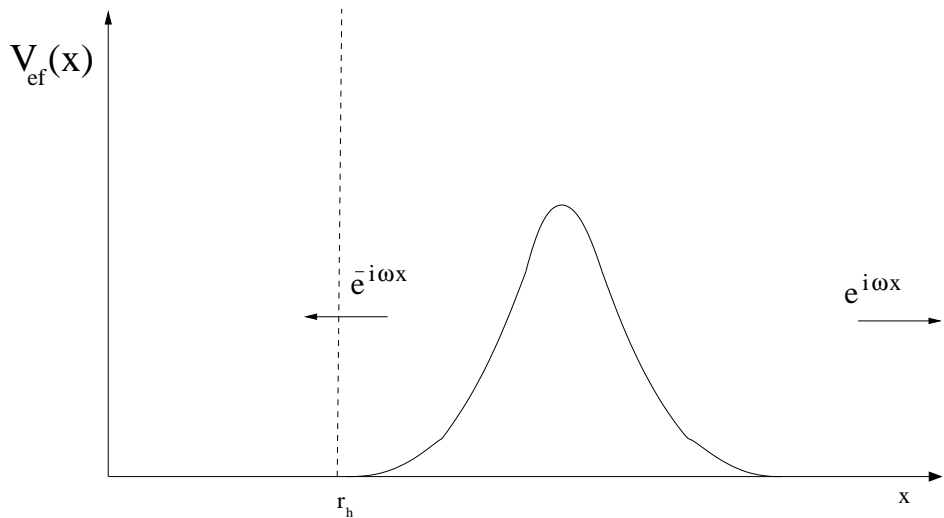, width=0.9 \linewidth, clip=}
\end{center}
\caption{{\small Condições de contorno para MQN em buracos negros: ondas entrando no horizonte de eventos e ondas escapando para o infinito.}}
\label{figmqn_bn}
\end{figure}

No caso de buracos negros assintóticamente planos essas condições \cite{Chandra} pedem que
\begin{eqnarray}
\label{mq1}
\Psi_{hor}^{in}\  \sim \ e^{-i\omega x}, \qquad x\rightarrow r_{h}\\
\nonumber\\
\label{mq2}
\Psi_{\infty}^{out}\ \sim\ e^{i\omega x}, \qquad x\rightarrow \infty
\end{eqnarray}
ou seja, próximo ao horizonte de eventos queremos apenas ondas entrando $\Psi^{in}$ e nenhuma saindo e no infinito queremos apenas ondas escapando para o infinito $\Psi^{out}$ e nenhuma vinda dele. A Fig.(\ref{figmqn_bn}) ilustra as condições acima.


\begin{figure}[!ht]
\begin{center}
\epsfig{file =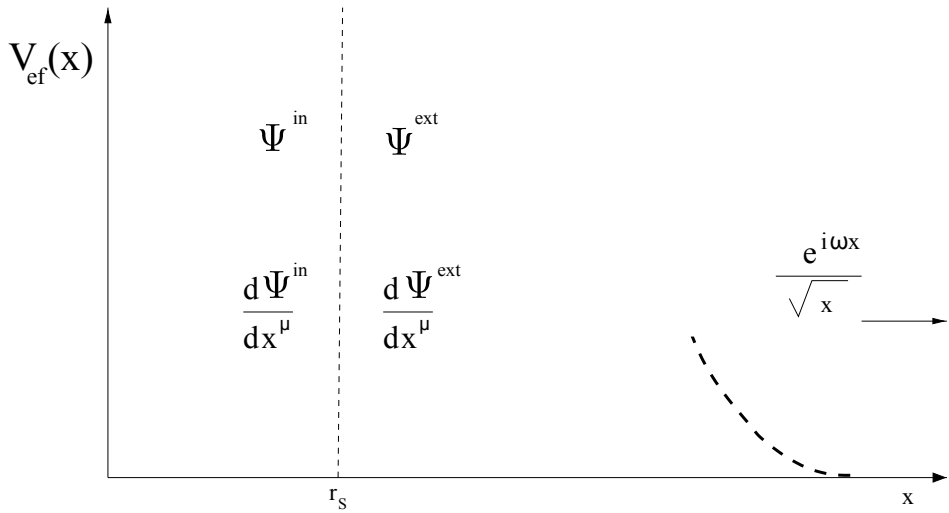, width=0.9 \linewidth, clip=}
\end{center}
\caption{{\small Condições de contorno para MQN em espaços-tempos com solução interna: Continuidade na fronteira e ondas escapando para o infinito}}
\label{figmqn_sol_int}
\end{figure}

No caso de objetos que possuem uma solução interna com um conteúdo de matéria específico essas condições de contorno devem ser alteradas. A maioria dos casos estudados neste trabalho possuem soluções internas e são cilindricamente simétricos. Para esses casos os MQN's serão definidos através das seguintes condições de contorno
\begin{eqnarray}
\label{mq2a}
|\Psi^{in}|&<&\infty \hspace{3,5cm} \textrm{quando}\quad x\rightarrow 0\quad,\\
\nonumber\\
\label{mq3}
\Psi^{in}&=&\Psi^{ext}\nonumber\\
&& \hspace{4cm}\textrm{quando}\quad x = r_{s}\quad,\\
\frac{\partial\Psi^{in}}{\partial x^{\nu}}&=&\frac{\partial\Psi^{ext}}{\partial x^{\nu}} \nonumber\\
\nonumber\\
\label{mq4}
\Psi_{\infty}^{ext}\ &\sim&\ \frac{e^{i\omega x}}{\sqrt{x}}, \hspace{3cm}\textrm{quando} \quad x\rightarrow \infty\quad,
\end{eqnarray}
ou seja, impomos continuidade para as soluções internas e externas dos campos e suas derivadas, na fronteira $r_{s}$ e mantemos a condição de que apenas ondas cilíndricas escapando para o infinito são permitidas. Veja a Fig.(\ref{figmqn_sol_int}).

Entretanto, devemos nos lembrar que alguns espaços-tempos estudados neste trabalho admitem a presença de curvas fechadas tipo tempo. Isso poderia, a princípio, inviabilizar o estudo da evolução do campo escalar nesses espaços-tempos. Contudo, esse não é um problema insuperável. Na verdade, para tornar esse problema bem posto precisamos apenas impor condições de contorno extras.

Seguindo a idéia de Nokivov e colaboradores \cite{Novikov}, utilizaremos o princípio de auto-consistência para contornar esse problema. O princípio de auto-consistência declara que:
\begin{quote}
\emph{As únicas soluções das leis físicas que podem ocorrer localmente no Universo real são aquelas que são globalmente auto-consistentes.}
\end{quote}
Isso implica que todos os eventos sobre uma curvas fechadas tipo tempo influencia outros eventos na sua vizinhaça apenas de uma maneira auto-ajustada. Nesta situação uma partícula pode voltar ao passado mas nenhuma alteração pode ser efetuada. Ela apenas repete a mesma história indefinidamente. Portanto, devemos garantir que a propagação do campo escalar seja periódica na região onde a coordenada temporal é identificada.

Da imposição desse princípio resulta a condição de contorno adicional dada por
\begin{eqnarray}
\label{CTC}
\Psi(t,0,r,z)=\Psi(t,2\pi,r,z)\qquad\Longrightarrow\qquad \Phi(0)=\Phi(2\pi)\quad.
\end{eqnarray}
Inspecionando os espaços-tempos aqui estudados vemos que a condição (\ref{CTC}) é satisfeita tanto na região causal quanto na região não-causal. Assim, tornamos o problema do cálculo dos modos quasinormais de um campo escalar bem posto.

\chapter{Modos quasinormais - Campo Escalar}

Neste capítulo apresentaremos o cálculo dos modos quasinormais referentes à propagação de um campo escalar nos espaços-tempos apresentados.

\section{Modos quasinormais do cilindro cósmico estático}
\label{mqncordaestatica}

Iniciaremos o cálculo dos MQN's com o espaço-tempo do cilindro estático. Primeiramente definimos os parâmetros do nosso problema. Neste caso os parâmetros livres são a densidade de energia $\rho=\frac{1}{8\pi r_{0}}$ do cilindro e o raio da fronteira $r_{s}^{int}$ que limita o cilindro. Essas duas constantes se relacionam da seguinte maneira,
\begin{eqnarray}
\label{mqncc1}
r_{s}^{int}=r_{0}\theta_{M}\quad,\qquad (1-4\mu)=\cos(\theta_{M})\quad, \qquad r_{b}^{ext}=\frac{r_{0}\sin(\theta_{M})}{(1-4\mu)}\quad.
\end{eqnarray}
Devemos observar que $\mu$ possui um valor crítico $\mu_{c}=1/4$. Portanto, trabalharemos apenas com valores de $\mu<\mu_{c}$ pelos motivos declarados na Seção (\ref{cordaestatica}) onde descrevemos esse espaço-tempo.

Para este espaço-tempo temos soluções exatas tanto para a métrica interna quanto para a externa. Comecemos com a métrica externa. Trataremos o caso onde $\omega$ é complexo, de modo que a solução da Eq.(\ref{pcc4}) é dada em termos de funções de Bessel. Usando a condição de contorno dada pela Eq.(\ref{mq4}) percebemos que a combinação de funções de Bessel que apresenta o comportamento assintótico desejado em $r\rightarrow\infty$ é da forma
\begin{eqnarray}
\label{mqncc2}
R_{ext}(r)=C_{1}\ H^{1}_{\nu}(pr)+C_{2}\ H^{2}_{\nu}(pr)\quad,
\end{eqnarray}
onde temos $p^2=\omega^2-k^2-\beta^2$ e o índice $\nu^2=m^2/(1-4\mu)^2$. Essas são as conhecidas, primeira e segunda funções de Hankel, que têm um comportamento de onda viajante quando $r\rightarrow \infty$. Detalhes são apresentado no Apêndice \ref{apen1}. Como queremos apenas ondas escapando para o infinito e nenhuma vindo dele, devemos proceder da seguinte forma:
\begin{itemize}
\label{mqncc3a}
\item  Se $\omega_{R}>0$ então $C_{2}=0\quad \Longrightarrow R_{ext}(r)=C_{1}\ H^{1}_{\nu}(pr)$.

\item  Se $\omega_{R}<0$ então $C_{1}=0\quad \Longrightarrow R_{ext}(r)=C_{2}\ H^{2}_{\nu}(pr)$.
\end{itemize}
Assim, ao calcularmos os MQN's devemos estar atentos pois, para cada ramo de $\omega_{R}$, devemos usar uma função $R(r)$ diferente para garantir que tenhamos apenas ondas escapando para o infinito.


Na região interior também teremos $\omega$ complexo. Neste caso a Eq.(\ref{pcc7}) terá como solução as funções associadas de Legendre. Sua forma geral é dada por
\begin{eqnarray}
\label{mqncc4}
R_{int}(r)=\tilde{C}_{1}\ P_{n}^{m}(\cos(r/r_{0}))+\tilde{C}_{2}\ Q^{n}_{m}(\cos(r/r_{0}))\quad,
\end{eqnarray}
onde o índice $n$ é dado por
\begin{eqnarray}
\label{mqncc4a}
n=-\frac{1}{2}+\frac{\sqrt{4(\omega^2-k^2-\beta^2)r_{0}^{2}+1}}{2}\quad.
\end{eqnarray}
Impondo que a condição de contorno dada pela Eq.(\ref{mq2a}) seja satisfeita pela Eq.(\ref{mqncc4}) nos obriga eliminar a função de Legendre $Q^{m}_{n}$, fazendo $\tilde{C}_{2}=0$, pois esta função apresenta um comportamento divergente em $r=0$.

A solução interior final se reduz então a
\begin{eqnarray}
\label{mqncc5}
R_{int}(r)=\tilde{C}_{1}\ P_{n}^{m}(\cos(r/r_{0}))\quad.
\end{eqnarray}
Por fim usaremos a condição de contorno dada pela Eq.(\ref{mq3}) nas Eqs.(\ref{mqncc2}, \ref{mqncc4}) resultando nas equações
\begin{eqnarray}
\label{mqncc6}
\frac{R_{int}(r_{s})}{R_{ext}(r_{b})}=\frac{C_{1,2}}{\tilde{C_{1}}}\quad, &\qquad& \frac{R'_{int}(r_{s})}{R'_{ext}(r_{b})}=\frac{C_{1,2}}{\tilde{C_{1}}}\quad,\nonumber\\
\nonumber\\
\frac{R'_{int}(r_{s})}{R_{int}(r_{s})}&-&\frac{R'_{ext}(r_{b})}{R_{ext}(r_{b})}=0\quad,
\end{eqnarray}
onde as constantes $C_{1,2}$ estão relacionadas a qual ramo de $\omega_{R}$ se está calculando. Agora que já impusemos as condições de contorno apropriadas ao nosso problema podemos calcular os MQN's do cilindro estático. Escolhemos uma densidade de energia $\rho$ e um raio da fronteira $r_{s}^{int}$. Isso fixa as outras constantes do problema e dessa forma podemos obter as freqüências $\omega$ calculando os zeros da Eq.(\ref{mqncc6}) para valores específicos de $k$, $\beta$ e $m$. Os zeros da Eq.(\ref{mqncc6}) podem ser obtidos numericamente. A seguir, podemos observar algumas frequências do campo escalar obtidas usando o programa {\it Maple}. Detalhes do cálculo das frequências são apresentados no Apêndice \ref{Maple}.

\begin{table}[!h]
\caption{Modos quasinormais para o cilindro estático com parâmetros $r_{s}=5.0$, $\rho=1.0\times 10^{-3}$, $\bar{r}_{0}=6.3$, $\mu=0.1$, $r_{b}=6.4$, $\beta=0$, $k=0$. Na tabela da esquerda apresentamos o modo fundamental ($n=1$) e os sobretons ($n>1$) para $m=0$. Na tabela da direita apresentamos o modo fundamental ($n=1$) para vários valores de $m$.}
\vspace{0,5cm}
\label{tab1}
\begin{tabular}{|c|c|}
\hline
$n$ & $\omega=\omega_{R}+i\ \omega_{I}$  \\
\hline
 1 &   -0.66 - $i$ 0.54 \ \ , \ \  +0.66 - $i$ 0.54    \\

 2 &   -1.33 - $i$ 0.65 \ \ , \ \  +1.33 - $i$ 0.65    \\

 3 &   -1.98 - $i$ 0.72 \ \ , \ \  +1.98 - $i$ 0.72    \\

 4 &   -2.62 - $i$ 0.78 \ \ , \ \  +2.62 - $i$ 0.78    \\

 5 &   -3.25 - $i$ 0.82 \ \ , \ \  +3.25 - $i$ 0.82    \\

 6 &   -3.89 - $i$ 0.85 \ \ , \ \  +3.89 - $i$ 0.85    \\

 7 &   -4.52 - $i$ 0.88 \ \ , \ \  +4.52 - $i$ 0.88    \\

 8 &   -5.15 - $i$ 0.91 \ \ , \ \  +5.15 - $i$ 0.91    \\
\hline
\end{tabular}
\begin{tabular}{|c|c|}
\hline
$m$ & $\omega=\omega_{R}+i\ \omega_{I}$ \\
\hline
 0 &   -0.66 - $i$ 0.54 \ \ , \ \  +0.66 - $i$ 0.54  \\

 1 &   -0.21 - $i$ 0.42 \ \ , \ \  +0.21 - $i$ 0.42  \\

 2 &   -0.49 - $i$ 0.51 \ \ , \ \  +0.49 - $i$ 0.51  \\

 3 &   -0.76 - $i$ 0.58 \ \ , \ \  +0.76 - $i$ 0.58  \\

 4 &   -1.01 - $i$ 0.63 \ \ , \ \  +1.01 - $i$ 0.63  \\

 5 &   -1.26 - $i$ 0.68 \ \ , \ \  +1.26 - $i$ 0.68  \\

 6 &   -1.51 - $i$ 0.72 \ \ , \ \  +1.51 - $i$ 0.72  \\

 7 &   -1.76 - $i$ 0.76 \ \ , \ \  +1.76 - $i$ 0.76  \\
\hline
\end{tabular}
\end{table}
Como podemos observar na parte superior da tabela (\ref{tab1}), tanto a parte real quanto a parte imaginária de $\omega$ crescem quando passamos a considerar os sobretons $n>1$. Isso significa que os sobretons mais altos decaem mais rápido que o modo fundamental. Vemos também que existe uma simetria entre o ramo positivo e negativo de $\omega_{R}$ em relação ao eixo imáginário que representa $\omega_{I}$. Essa simetria permanece para qualquer valor de $m$ e pode ser entendida levando em conta a invariância da equação de movimento em relação a aplicação do complexo conjugado e que as funções de Hankel se transformam uma na outra com a troca de sinal de $\omega_{R}$ sob a mesma aplicação. Isso explica a simetria $\omega\to -\omega^{*}$. Observando a parte de baixo da mesma tabela vemos que o modo fundamental para $m=0$ é maior que $m=1$. Para os outros modos há um crescimento de $\omega$ com o aumento de $m$.

Quando consideramos $k\neq0$ e $\beta\neq0$ observamos apenas um deslocamento positivo no modo fundamental, ou seja, o modo fundamental acontece em valores maiores de $\omega$. Isso significa que esses modos fundamentais e todos os sobretons decaem mais rapidamente que o caso $k=0$ e $\beta=0$.

Finalmente o aspecto mais significativo observado é que $\omega_{I}$ é sempre negativo indicando que o cilindro estático é estável quando submetido a uma perturbação escalar.

\section{Modos quasinormais da corda cósmica estática e em rotação}

O estudo dos MQN's para o campo escalar nos espaços-tempos da corda estática e da corda em rotação foi agrupado em uma única seção pois ambos apresentaram o mesmo comportamento: ausência de modos quasinormais de oscilação. 

Nesses dois casos, devemos impor apenas as condições de contorno dadas pelas Eqs.(\ref{mq2a}, \ref{mq4}) já que não há uma região interior. Quando aplicamos a condição de contorno dada pela Eq.(\ref{mq4}) nas soluções radiais das equações de movimento do campo escalar em cada uma das métricas o resultado é que elas devem ser funções de Hankel. Entretanto, ao testarmos a condição de regularidade das soluções em $r=0$, exigida pela Eq.(\ref{mq2a}), descobrimos que ambas as funções de Hankel são divergentes neste ponto. Assim a única solução possível seria impormos $C_{1}=C_{2}=0$. Já que uma das condições de contorno não é satisfeita, exceto no caso da solução trivial $\Phi=0$ concluímos que não há modos quasinormais de oscilação do campo escalar nestes espaços-tempos.

\section{Modos quasinormais do cilindro cósmico em rotação}
\label{mqncordarotacao}
Trabalharemos agora com o cilindro em rotação. Como comentamos na Seção (\ref{cordarotacao}) temos duas soluções internas. Iniciaremos o cálculo dos MQN's com a solução interna tipo ``vaso de flor" pois ela possui solução exata para o campo escalar. Em seguida calcularemos os MQN's da solução interna tipo ``caneta esferográfica" através de soluções aproximadas obtidas em dois limites distintos.

\subsection{Solução Vaso de flor}

Primeiramente definiremos os parâmetros do nosso problema. Neste caso os parâmetros livres são o raio da superfície $r_{s}$, ou seja, a posição radial da casca cilíndrica infinitamente fina, o momento angular por unidade de comprimento $J$ da casca e a densidade linear $\mu$ que está relacionada com a solução exterior. Essas constantes conectarão as regiões interior e exterior do espaço-tempo fixando a constante $r_{0}$  através da relação
\begin{eqnarray}
\label{mqnccrotvaso1}
r_{0}=r_{0}^{\pm}=\pm\sqrt{\frac{r^2_{s}+16J^2}{(1-4\mu)^2}}-r_{s}\quad.
\end{eqnarray}
Contudo, ainda temos a liberdade de escolher qual ramo de $r_{0}$ queremos explorar. Lembrando da discussão apresentada no capítulo \ref{cap7analisecausal} a respeito da presença de curvas fechadas tipo-tempo e a escolha $r_{0}^{\pm}$ teremos o cuidado de explorar cada um dos casos. Novamente trabalharemos apenas com valores de $\mu<1/4$.

Para esse espaço-tempo temos soluções exatas para o campo tanto para a métrica interna quanto para a externa. Comecemos com a métrica externa. Trataremos o caso onde $\omega$ é complexo de modo que a solução da Eq.(\ref{pccr4}) é dada em termos de funções de Bessel. Aplicando a condição de contorno dada pela Eq.(\ref{mq4}) nas funções de Bessel percebemos que a combinação dessas funções que apresenta o comportamento assintótico desejado em $r\rightarrow\infty$ é da forma
\begin{eqnarray}
\label{mqnccrotvaso2}
R_{ext}(r+r_{0})=C_{1}\ H^{1}_{\nu}(p(r+r_{0}))+C_{2}\ H^{2}_{\nu}(p(r+r_{0}))\quad,
\end{eqnarray}
onde $p^2=\omega^2-k^2-\beta^2$ e o índice $\nu^2=(m+4J\omega)^2/(1-4\mu)^2$. Como queremos apenas ondas escapando para o infinito e nenhuma vindo dele, devemos impor que
\begin{itemize}
\label{mqnccrotvaso3a}
\item  Se $\omega_{R}>0$ então $C_{2}=0\quad \Longrightarrow R_{ext}(r)=C_{1}\ H^{1}_{\nu}(p(r+r_{0}))$.

\item  Se $\omega_{R}<0$ então $C_{1}=0\quad \Longrightarrow R_{ext}(r)=C_{2}\ H^{2}_{\nu}(p(r+r_{0}))$.
\end{itemize}
Para a região interior também queremos que $\omega$ seja complexo. Neste caso, a Eq.(\ref{pccri12}) terá como solução as funções de Bessel dadas por
\begin{eqnarray}
\label{mqnccrotvaso4}
R_{int}(r)=\tilde{C}_{1}\ J_{m}(\tilde{p}r)+\tilde{C}_{2}\ N_{m}(\tilde{p}r)\quad,
\end{eqnarray}
onde o índice $m$ é um número inteiro e a constante $p$ satisfaz a relação
\begin{eqnarray}
\label{mqnccrotvaso4a}
\tilde{p}=\sqrt{\left(\frac{1}{\sigma_{1}^2}\left(\omega-\frac{4Jm}{r^2_{s}}\right)^2-k^2-\beta^2\right)}\quad.
\end{eqnarray}
A imposição de que o campo seja regular sobre a origem $r=0$, condição exigida pela Eq.(\ref{mq2}), nos obriga eliminar a função $N_{m}(\tilde{p}r)$, fazendo $\tilde{C}_{2}=0$.

A solução interior final se reduz então a
\begin{eqnarray}
\label{mqnccrotvaso5}
R_{int}(r)=\tilde{C}_{1}\ J_{m}(\tilde{p}r)\quad.
\end{eqnarray}
Por fim, usaremos a condição de contorno dada pela Eq.(\ref{mq3}) nas Eqs.(\ref{mqnccrotvaso2}, \ref{mqnccrotvaso5}) resultando nas equações
\begin{eqnarray}
\label{mqnccrotvaso6}
\frac{R_{int}(r_{s})}{R_{ext}(r_{s})}=\frac{C_{1,2}}{\tilde{C_{1}}}\quad, &\qquad& \frac{R'_{int}(r_{s})}{R'_{ext}(r_{s})}=\frac{C_{1,2}}{\tilde{C_{1}}}\quad,\nonumber\\
\nonumber\\
\frac{R'_{int}(r_{s})}{R_{int}(r_{s})}&-&\frac{R'_{ext}(r_{s})}{R_{ext}(r_{s})}=0\quad,
\end{eqnarray}
onde as constantes $C_{1,2}$ são novamente relacionadas à escolha do ramo positivo ou negativo de $\omega_{R}$. Agora que já impusemos as condições de contorno apropriadas ao nosso problema podemos calcular os MQN's desta solução.

Escolheremos valores para $J$, $r_{s}$ e $\mu$ e então escolheremos o ramo de $r_{0}$ que queremos analisar. Isso fixará as outras constantes do problema e dessa forma podemos obter as freqüências $\omega$ calculando os zeros da Eq.(\ref{mqnccrotvaso6})  para valores específicos de $k$, $\beta$ e $m$. Aplicando o mesmo procedimento desenvolvido para o cilindro estático, calculamos alguns modos quasinormais do campo escalar para o cilindro em rotação.

\begin{table}[!h]
\caption{Modos quasinormais para o cilindro em rotação com solução interior do tipo ``vaso de flor" com os parâmetros $J=1.0$, $\mu=1.0\times10^{-2}$, $r_{s}=5.0$, $\beta=0$, $k=0$, $r_{0}=-12$ (topo) e $r_{0}=1.7$ (abaixo). Nas tabelas localizadas à esquerda apresentamos o modo fundamental ($n=1$) e os sobretons ($n>1$) para $m=0$, e nas tabelas localizadas à direita apresentamos o modo fundamental ($n=1$) para vários valores de $m$. O símbolo $*$ indica os modos instáveis.}
\vspace{0,5cm}
\begin{tabular}{|c|c|c|}
\hline
$n$ & \multicolumn{2}{|c|}{ $\omega=\omega_{R}+i\ \omega_{I}$  } \\
\hline
1  &  0.06 + $i$\ 0.04 $^{*}$  &  -0.06 + $i$\ 0.04 $^{*}$   \\

2  &  0.75 - $i$\ 0.23 \ \     &  -0.21 + $i$\ 0.04 $^{*}$     \\

3  &  1.57 - $i$\ 0.32 \ \     &  -0.79 + $i$\ 0.01 $^{*}$     \\

4  &  2.39 - $i$\ 0.36 \ \     &  -0.90 + $i$\ 0.01 $^{*}$     \\

5  &  3.20 - $i$\ 0.40 \ \     &  -1.46 + $i$\ 0.03 $^{*}$     \\

6  &  4.01 - $i$\ 0.43 \ \     &  -1.57 - $i$\ 0.32 \ \      \\
\hline
\hline
$n$ & \multicolumn{2}{|c|}{ $\omega=\omega_{R}+i\ \omega_{I}$  }\\
\hline
1  &  0.46 - $i$\ 0.45 \ \   &   -0.46 - $i$\ 0.45   \\

2  &  1.23 - $i$\ 0.57 \ \   &   -1.23 - $i$\ 0.57   \\

3  &  2.02 - $i$\ 0.63 \ \   &   -2.02 - $i$\ 0.63   \\

4  &  2.83 - $i$\ 0.67 \ \   &   -2.82 - $i$\ 0.67   \\

5  &  3.63 - $i$\ 0.70 \ \   &   -3.63 - $i$\ 0.70   \\

6  &  4.43 - $i$\ 0.73 \ \   &   -4.43 - $i$\ 0.73   \\
\hline
\end{tabular}
\begin{tabular}{|c|c|c|}
\hline
$m$ & \multicolumn{2}{|c|}{ $\omega=\omega_{R}+i\ \omega_{I}$  } \\
\hline
0  &  0.06 + $i$\ 0.04 $^{*}$  &   -0.06 + $i$\ 0.04 $^{*}$   \\

1  &  0.03 - $i$\ 0.05 \ \     &   -0.03 + $i$\ 0.03 $^{*}$   \\

2  &  0.05 + $i$\ 0.07 $^{*}$  &   -0.56 + $i$\ 0.03 $^{*}$   \\

3  &  0.15 - $i$\ 0.22 \ \     &   -0.19 + $i$\ 0.31 $^{*}$    \\

4  &  0.22 - $i$\ 0.31 \ \     &   -0.37 + $i$\ 0.04 $^{*}$  \\

5  &  0.29 - $i$\ 0.39 \ \     &   -0.44 + $i$\ 0.59 $^{*}$   \\
\hline
\hline
$m$ & \multicolumn{2}{|c|}{ $\omega=\omega_{R}+i\ \omega_{I}$  }\\
\hline
0  &  0.46 - $i$\ 0.45 \ \  &   -0.46 - $i$\ 0.45 \\

1  &  0.94 - $i$\ 0.39 \ \  &   -0.83 - $i$\ 0.57  \\

2  &  1.43 - $i$\ 0.39 \ \  &   -0.12 - $i$\ 0.15  \\

3  &  1.90 - $i$\ 0.40 \ \  &   -0.07 - $i$\ 0.26  \\

4  &  0.30 - $i$\ 1.90 \ \  &   -0.17 - $i$\ 0.30  \\

5  &  0.83 - $i$\ 2.09 \ \  &   -0.26 - $i$\ 0.33  \\
\hline
\end{tabular}
\label{tab6}
\end{table}

Da tabela (\ref{tab6}) podemos observar que para o cilindro em rotação, diferentemente do cilindro estático, a simetria $\omega\to -\omega^{*}$ é quebrada, exceto quando $m=0$, não importando o valor escolhido para $r_{0}$. Essa quebra de simetria de $\omega$ pode ser entendida se observarmos o potencial efetivo da equação de movimento do campo escalar. Quando $m=0$ recuperamos o potencial efetivo do caso estático. Agora, quando $m\neq0$ temos um termo proporcional a $\omega$ aparecendo na ordem da função de Hankel que quebra o argumento apresentado anteriormente.

Na parte superior da tabela (\ref{tab6}), onde apresentamos os modos quasinormais obtidos, quando o parâmetro $r_{0}^{-}$ é escolhido na condição (\ref{flcond1}) observamos a presença de frequencias com $\omega_{I}>0$ indicando instabilidade desse espaço-tempo quando submetido a perturbações escalares. É importante lembrar neste momento que a escolha de $r_{0}^{-}$ implica na presença de curvas fechadas tipo tempo no exterior desse espaço-tempo.

Na parte inferior da tabela observamos apenas frequências com $\omega_{I}<0$. Essa tabela apresentadas as frequências obtidas quando escolhemos $r_{0}^{+}$ na condição (\ref{flcond1}). Isso nos indica que para essa escolha o espaço-tempo resultante é estável em relação à perturbações escalares. Neste caso, também observamos que os sobretons decaem mais rapidamente que o modo fundamental e a presença de $k\neq0$ e $\beta\neq0$ desloca positivamente o modo fundamental e seus sobretons fazendo-os decair mais rapidamente.

A relação entre $r_{0}^{\pm}$ e a estabilidade do espaço-tempo será  melhor discutida na conclusão desta tese.

\subsection{Solução Caneta esferográfica }

Neste caso calcularemos os MQN's do espaço-tempo em questão em dois limites distintos.
Os parâmetros livres para a solução interna tipo ``caneta esferográfica" serão o raio da superfície $r_{s}$, ou seja, a posição radial da fronteira do cilindro, a constante $\lambda$ que se relaciona com a densidade linear de energia e a constante $\alpha$ relacionada ao momento angular por unidade de comprimento do cilindro. Essas constantes conectam as regiões interior e exterior do espaço-tempo fixando a constante $r_{0}$  através da relação
\begin{eqnarray}
\label{mqnccrotpenest0}
r_{0}=r_{0}^{\pm}=\left[\pm \frac{\sqrt{1-(1-4\mu)^2}}{(1-4\mu)\arccos(1-4\mu)}-1\right]r_{s}\quad.
\end{eqnarray}
Novamente trataremos ambos os ramos de $r_{0}$ e trabalharemos apenas com valores de $m_{0}<1/4$.

Para a região exterior vale o resultado obtido anteriormente para a solução tipo ``vaso de flor" onde a solução final, após a imposição da condição de contorno, é dada pela Eq.(\ref{mqnccrotvaso2}) e a restrição imposta por $\omega_{R}$.

Para a região interior, soluções exatas aproximadas são obtidas em dois limites:
\begin{itemize}
\item limite quasi-estático quando $\sqrt{\lambda}r_{s}\ll 1$
\item limite de altos valores de $m$
\end{itemize}

\subsubsection{Limite quasi-estático}

O limite quasi-estático do cilindro em rotação é alcançado quando seu raio $r_{s}$ é muito menor que sua densidade de energia, ou seja, quando
\begin{eqnarray}
\label{mqnccrotpenest1}
r_{s}\ll\frac{1}{\sqrt{\lambda}}\quad.
\end{eqnarray}
Neste limite, podemos expandir as funções da métrica em série de Taylor para $r$ pequeno desprezando termos de $\mathcal{O}(\sqrt{\lambda}r)^3$. Assim procedendo temos que
\begin{eqnarray}
\label{mqnccrotpenest2}
\sin(\sqrt{\lambda}r)&\sim& \sqrt{\lambda}r+\mathcal{O}(\sqrt{\lambda}r)^3\quad,\\
\nonumber\\
\cos(\sqrt{\lambda}r)&\sim& 1 - \frac{(\sqrt{\lambda}r)^2}{2} + \mathcal{O}(\sqrt{\lambda}r)^3\quad,\\
\nonumber\\
f&\sim&  0 + \mathcal{O}(\sqrt{\lambda}r)^3\quad,\\
\nonumber\\
J&\sim&  0 + \mathcal{O}(\sqrt{\lambda}r)^3\quad,\\
\nonumber\\
\mu&\sim& \frac{\lambda r_{s}^2}{8}+ \mathcal{O}(\sqrt{\lambda}r)^3\quad.
\end{eqnarray}
Descartaremos a contribuição da função $f$ no potencial efetivo do campo escalar na região interior, mas ainda manteremos a relação dada pela Eq.(\ref{cr6a}) para calcular o valor de $J$ e substituí-lo no potencial efetivo na região exterior. Adotamos esse procedimento pois, mesmo $J$ sendo muito pequeno, ele deve ser não nulo para permitir a possibilidade de curvas fechadas tipo tempo no exterior do espaço-tempo. Por esses motivos a Eq.(\ref{pccri22}) torna-se
\begin{eqnarray}
\label{mqnccrotpenest3}
x^2\frac{d^2R}{dx^2}+ x\frac{dR}{dx} + R\left[\frac{\left(\omega^2-k^2-\beta^2\right)}{\lambda}x^2-m^2\right]=0.
\end{eqnarray}
Se supusermos que $\omega$ é complexo, a Eq.(\ref{mqnccrotpenest3}) terá solução exata em termos de funções de Bessel,
\begin{eqnarray}
\label{mqnccrotpenest4}
R_{int}(r)=\tilde{C}_{1}\ J_{m}(\tilde{p}\ r)+\tilde{C}_{2}\ N_{m}(\tilde{p}\ r)\quad,
\end{eqnarray}
onde o índice $m$ é um número inteiro e $\tilde{p}=\sqrt{\frac{\left(\omega^2-k^2-\beta^2\right)}{\lambda}}$. Em seguida, faremos a constante de integração $\tilde{C}_{2}=0$ já que a condição de contorno dada pela Eq.(\ref{mq2}) exige regularidade do campo em $r=0$. Isso implica que a solução final para o campo escalar na região interior no limite quasi-estático será
\begin{eqnarray}
\label{mqnccrotpenest5}
R_{int}(r)=\tilde{C}_{1}\ J_{m}(\tilde{p}\ r)\quad.
\end{eqnarray}
De modo semelhante ao caso anterior aplicaremos as duas condições de contorno restantes dadas pela Eq.(\ref{mq3}) às soluções interior e exterior para calcular as equações
\begin{eqnarray}
\label{mqnccrotvaso7}
\frac{R_{int}(r_{s})}{R_{ext}(r_{s})}=\frac{C_{1,2}}{\tilde{C_{1}}}\quad, &\qquad& \frac{R'_{int}(r_{s})}{R'_{ext}(r_{s})}=\frac{C_{1,2}}{\tilde{C_{1}}}\quad,\nonumber\\
\nonumber\\
\frac{R'_{int}(r_{s})}{R_{int}(r_{s})}&-&\frac{R'_{ext}(r_{s})}{R_{ext}(r_{s})}=0\quad.
\end{eqnarray}
O próximo passo será escolher valores para $\lambda$, $r_{s}$ e $\tilde{\alpha}$ e então escolher o ramo de $r_{0}$ que queremos analisar. Repetindo o procedimento adotado nos outros casos, calculamos alguns modos quasinormais para o campo escalar se propagando no cilindro em rotação no limite quasi-estático.

\begin{table}[!h]
\caption{Modos quasinormais para o cilindro em rotação com solução interior tipo ``caneta esferográfica" no limite quasi-estático, com parâmetros $J=2.0 \times 10^{-5}$, $\mu=1.0 \times 10^{-4}$, $r_{s}=1.0$, $\beta=0$, $k=0$, $\tilde{\alpha}= 0.2$, $\lambda=1.0 \times 10^{-3}$, $r_{0}= -2.0 $ (topo) e $r_{0}=3.0 \times 10^{-4}$ (abaixo). Nas duas tabelas localizadas à esquerda apresentamos o modo fundamental ($n=1$) e os sobretons ($n>1$) para $m=0$. Nas duas tabelas à direita apresentamos o modo fundamental ($n=1$) para vários valores de $m$. O símbolo $*$ indica os modos instáveis.}
\vspace{0,5cm}
\begin{tabular}{|c|c|c|}
\hline
 $n$ & \multicolumn{2}{|c|}{ $\omega=\omega_{R}+i\ \omega_{I}$  }\\
\hline
 1  & 0.30 + $i$\ 0.41 $^{*}$   &  -0.30 + $i$\ 0.41 $^{*}$  \\

 2  & 2.96 - $i$\ 0.94      &  -2.96 - $i$\ 0.94   \\

 3  &  6.16 - $i$\ 1.28      &  -3.83 + $i$\ 0.18 $^{*}$  \\

 4  &  9.33 - $i$\ 1.47      &  -5.52 + $i$\ 0.17 $^{*}$  \\

 5  &  12.5 - $i$\ 1.62      &  -6.16 - $i$\ 1.28   \\

 6  &  15.6 - $i$\ 1.73      &  -7.01 + $i$\ 0.17 $^{*}$ \\
\hline
\hline
 $n$ & \multicolumn{2}{|c|}{ $\omega=\omega_{R}+i\ \omega_{I}$  }\\
\hline
 1  &  2.61 - $i$\ 5.54      &  -2.61 - $i$\ 5.54   \\

 2  &  5.94 - $i$\ 5.72      &  -5.94 - $i$\ 5.72   \\

 3  &  9.17 - $i$\ 5.87      &  -9.17 - $i$\ 5.87   \\

 4  &  12.4 - $i$\ 6.00      &  -12.4 - $i$\ 6.00   \\

 5  &  15.5 - $i$\ 6.10      &  -15.5 - $i$\ 6.10   \\

 6  &  18.7 - $i$\ 6.18      &  -18.7 - $i$\ 6.18   \\
\hline
\end{tabular}
\begin{tabular}{|c|c|c|}
\hline
$m$ & \multicolumn{2}{|c|}{ $\omega=\omega_{R}+i\ \omega_{I}$  }\\
\hline
0  &  0.30 + $i$\ 0.41 $^{*}$      &    -0.30 + $i$\ 0.41 $^{*}$  \\

1  &  1.84 + $i$\ 0.25 $^{*}$      &    -1.84 + $i$\ 0.25 $^{*}$  \\

2  &  0.34 + $i$\ 0.47 $^{*}$     &    -3.06 + $i$\ 0.27 $^{*}$   \\

3  &  1.10 + $i$\ 0.70 $^{*}$      &    -1.11 + $i$\ 0.70 $^{*}$  \\

4  &  1.92 + $i$\ 0.86 $^{*}$      &    -1.92 + $i$\ 0.86 $^{*}$  \\

5  &  2.76 + $i$\ 0.99 $^{*}$      &    -2.76 + $i$\ 0.98 $^{*}$   \\
\hline
\hline
$m$ & \multicolumn{2}{|c|}{ $\omega=\omega_{R}+i\ \omega_{I}$  }\\
\hline
0  &   2.61 - $i$\ 5.54        &   -2.61 - $i$\ 5.54   \\

1  &   1.07 - $i$\ 5.47        &   -0.66 - $i$\ 5.46   \\

2  &   2.74 - $i$\ 5.54        &   -2.21 - $i$\ 5.60   \\

3  &   1.00 - $i$\ 5.86        &   -0.53 - $i$\ 5.90   \\

4  &   2.37 - $i$\ 6.15        &   -1.94 - $i$\ 6.27   \\

5  &   0.85 - $i$\ 6.77        &   -0.49 - $i$\ 6.81   \\
\hline
\end{tabular}
\label{tab3}
\end{table}

Inspecionando a tabela (\ref{tab3}) observamos as mesmas características apresentadas pelos modos quasinormais do cilindro em rotação com interior do tipo ``vaso de flor". Novamente a escolha de $r_{0}^{-}$ implicou no aparecimento de modos instáveis com $\omega_{I}>0$. De modo semelhante, quando $r_{0}^{+}$ foi escolhido tais modos instáveis desapareceram.

Novamente o parâmetro $r_{0}$ ditou a presença ou ausência de modos instáveis para o campo escalar neste espaço-tempo.

\subsubsection{Limite de altos valores de $m$ }

Se analisarmos a Eq.(\ref{pccri22}) veremos que
\begin{eqnarray}
\label{largem0}
m+2\tilde{\alpha}\omega f
\end{eqnarray}
é o termo que dificulta a obtenção de uma solução exata devido à forma complicada de $f(r)$. Podemos então nos perguntar qual é a contribuição do termo $2\alpha\omega f$ no potencial efetivo do campo escalar. Uma rápida inspeção da função $f(r)$ nos revela que ela é limitada no intervalo $0<r\leq r_{s}$. As constantes $\tilde{\alpha}$ e $\omega$ são quaisquer em princípio. Portanto, se pudermos garantir que
\begin{eqnarray}
\label{largem1}
m \gg \max_{r \in [0,r_{s}]} |2\tilde{\alpha}\omega\ f(r)| \quad,
\end{eqnarray}
o termo $2\tilde{\alpha}\omega f$ pode ser desprezado em relação a $m$. A condição expressa na Eq.(\ref{largem1}) é a relação que caracteriza o limite de altos valores de $m$. Desta forma, se supormos que tal limite seja satisfeito a Eq.(\ref{pccri22}) se reduzirá à equação
\begin{eqnarray}
\label{largem2}
\frac{d^2R}{dx^2}+ \frac{\cos(x)}{\sin(x)}\frac{dR}{dx} +R\left\{\frac{\left(\omega^2-k^2-\beta^2\right)}{\lambda}-\frac{m^2}{\sin(x)^2}\right\}=0\quad.
\end{eqnarray}
A solução geral da Eq.(\ref{largem2}) quando $\omega$ é complexo pode ser escrita em termos das funções de Legendre associadas. Neste caso a forma da solução será
\begin{eqnarray}
\label{largem3}
R_{int}(r)=\tilde{C}_{1}\ P_{n}^{m}\left[\cos\left(\sqrt{\lambda}r\right)\right]+\tilde{C}_{2}\ Q^{n}_{m}\left[\cos\left(\sqrt{\lambda}r\right)\right]\quad,
\end{eqnarray}
onde o índice $n$ é dado por
\begin{eqnarray}
\label{largem4}
n=-\frac{1}{2}+\frac{\sqrt{\frac{4(\omega^2-k^2-\beta^2)}{\lambda}+1}}{2}\quad.
\end{eqnarray}
A condição de regularidade sobre a origem imposta pela Eq.(\ref{mq2}) obriga-nos a descartar a função $Q_{n}^{m}$, portanto $\tilde{C}_{2}=0$. Assim, a forma final da solução para a região interior neste limite de altos valores de $m$ será
\begin{eqnarray}
\label{largem5}
R_{int}(r)=\tilde{C}_{1}\ P_{n}^{m}\left[\cos\left(\sqrt{\lambda}r\right)\right]\quad.
\end{eqnarray}
Aplicaremos as duas condições de contorno restantes dadas pela Eq.(\ref{mq3}), às soluções interior e exterior para chegar as equações
\begin{eqnarray}
\label{largem6}
\frac{R_{int}(r_{s})}{R_{ext}(r_{s})}=\frac{C_{1,2}}{\tilde{C_{1}}}\quad, &\qquad& \frac{R'_{int}(r_{s})}{R'_{ext}(r_{s})}=\frac{C_{1,2}}{\tilde{C_{1}}}\quad,\nonumber\\
\nonumber\\
\frac{R'_{int}(r_{s})}{R_{int}(r_{s})}&-&\frac{R'_{ext}(r_{s})}{R_{ext}(r_{s})}=0\quad.
\end{eqnarray}
Escolheremos então, valores para $\lambda$, $r_{s}$, $\tilde{\alpha}$ e o ramo de $r_{0}$ que queremos analisar e calcularemos numericamente os modos quasinormais do campo escalar neste limite usando o procedimento apresentado anteriormente.

\begin{table}[!h]
\caption{Modos quasinormais para o cilindro em rotação com solução interior tipo ``caneta esferográfica" no limite de altos valores de $m$ com parâmetros $J=0.1$, $\mu=1.0\times 10^{-2} $, $r_{s}=10$, $\beta=0$, $k=0$, $\tilde{\alpha}= 1.0$, $\lambda=1.0 \times 10^{-3}$, $r_{0}=-20 $ (topo) e  $r_{0}=0.4 $ (abaixo). Nas duas tabelas localizadas à esquerda apresentamos o modo fundamental ($n=1$) e os sobretons ($n>1$) para $m=70$. Nas duas tabelas à direita apresentamos o modo fundamental ($n=1$) para vários valores de $m$. Os símbolos $*$ e $\dagger$ indicam os modos instáveis e os modos puramente reais, respectivamente.}
\vspace{0,5cm}
\begin{tabular}{|c|c|c|}
\hline
$n$ & \multicolumn{2}{|c|}{ $\omega=\omega_{R}+i\ \omega_{I}$  }\\
\hline
1 &   2.20  $^{\dagger}$         &   -2.20  $^{\dagger}$    \\

2 &   2.47 + $i$\ 3.99 $^{*}$    &   -2.31 + $i$\ 4.30 $^{*}$   \\

3 &   2.80 + $i$\ 3.86 $^{*}$    &   -2.84 + $i$\ 4.08 $^{*}$   \\

4 &   2.96 + $i$\ 3.78 $^{*}$    &   -3.02 + $i$\ 3.99 $^{*}$    \\

5 &   3.12 + $i$\ 3.70 $^{*}$    &   -3.20 + $i$\ 3.90 $^{*}$    \\

6 &   3.29 + $i$\ 3.61 $^{*}$    &   -3.38 + $i$\ 3.80 $^{*}$    \\
\hline
\hline
$n$ & \multicolumn{2}{|c|}{ $\omega=\omega_{R}+i\ \omega_{I}$  }\\
\hline
1 &   0.03 - $i$\ 5.00   &   -0.15 - $i$\ 4.70  \\

2 &   0.21 - $i$\ 5.00   &   -0.32 - $i$\ 4.70  \\

3 &   0.39 - $i$\ 4.99   &   -0.48 - $i$\ 4.69 \\

4 &   0.57 - $i$\ 4.98   &   -0.64 - $i$\ 4.67  \\

5 &   0.75 - $i$\ 4.96   &   -0.81 - $i$\ 4.66  \\

6 &   0.93 - $i$\ 4.94   &   -0.97 - $i$\ 4.63  \\
\hline
\end{tabular}
\begin{tabular}{|c|c|c|}
\hline
$m$ & \multicolumn{2}{|c|}{ $\omega=\omega_{R}+i\ \omega_{I}$  }\\
\hline
60  &  1.88 $^{\dagger}$              &   -1.88 $^{\dagger}$  \\

62  &  1.94 $^{\dagger}$              &   -1.94 $^{\dagger}$  \\

65  &  2.04 $^{\dagger}$              &   -2.04 $^{\dagger}$  \\

67  &  2.10 $^{\dagger}$              &   -2.10 $^{\dagger}$  \\

70  &  2.20 $^{\dagger}$              &   -2.20 $^{\dagger}$  \\

72  &  2.26 $^{\dagger}$              &   -2.26 $^{\dagger}$  \\
\hline
\hline
$m$ & \multicolumn{2}{|c|}{ $\omega=\omega_{R}+i\ \omega_{I}$  }\\
\hline
60  &  0.18 - $i$\ 4.30     &   -0.10 - $i$\ 4.05  \\

62  &  0.19 - $i$\ 4.44     &   -0.11 - $i$\ 4.18   \\

65  &  0.11 - $i$\ 4.65     &   -0.04 - $i$\ 4.38   \\

67  &  0.11 - $i$\ 4.79     &   -0.06 - $i$\ 4.51   \\

70  &  0.03 - $i$\ 5.00     &   -0.16 - $i$\ 4.71   \\

72  &  0.04 - $i$\ 5.14     &   -0.17 - $i$\ 4.84   \\
\hline
\end{tabular}
\label{tab4}
\end{table}

No limite de altos valores $m$ vemos novamente as mesmas características apresentadas pelos modos quasinormais da corda em rotação com interior do tipo ``vaso de flor". Outra vez a escolha de $r_{0}^{-}$ implicou no aparecimento de modos instáveis com $\omega_{I}>0$ além do aparecimento de modos puramente reais. E como antes, quando $r_{0}^{+}$ foi escolhido os modos instáveis e os puramente reais desapareceram.




\chapter{Conclusões}

Já no seu início, a teoria da Relatividade Geral apresentou novas questões a cerca da natureza do espaço e do tempo. Uma das mais intrigantes, dessas questões, perguntava sobre a possibilidade de se realizar uma viagem ao passado por meio de uma máquina do tempo.
E as primeiras respostas não tardaram em aparecer. Uma grande quantidade de soluções das equações de Einstein que admitiam curvas fechadas tipo tempo foram encontradas forçando-nos a considerar, ainda que teoricamente, a possibilidade da cons-trução de uma máquina do tempo.

Ignorando todas as implicações filosóficas da construção de uma máquina do tempo concluímos que o passo seguinte, para a concretização deste feito, seria estabelecer as condições necessárias para garantir a estabilidade da máquina e a segurança do viajante. Em outras palavras, precisaríamos analisar se a simples presença do viajante do tempo não a tornaria instável inviabilizando sua volta no tempo e se a máquina não ofereceria algum risco ao viajante.

Portanto, nossa análise ao longo deste trabalho foi desenvolvida afim de responder duas questões cruciais:
\begin{enumerate}
\item \emph{Pode o viajante acessar à máquina do tempo sem sofrer algum dano?}

\item \emph{Ainda que seja possível a construção de uma máquina do tempo, ela seria estável?}
\end{enumerate}

Para os espaços-tempos dos cilindros cósmicos em rotação fomos capazes de responder as duas questões de maneira conclusiva.

A primeira pergunta foi respondida usando os resultados obtidos na análise do comportamento das geodésicas, dos tipos de curvas fechadas tipo tempo e ainda de possíveis forças de maré na região não-causal.
Pudemos observar que, a região não-causal, apesar da característica exótica, mostrou-se bastante acessível tanto a partículas massivas quanto a fótons. Diferentes tipos de curvas fechadas tipo tempo foram encontrados mostrando a variedade de maneiras de se voltar no tempo e nenhuma força de maré foi encontrada na vizinhaça da região não-causal. Portanto, se essa classe de máquinas do tempo for estável, ela também será acessível e segura ao viajante do tempo.

Para responder a segunda pergunta tivemos que usar os dados obtidos sobre o comportamento dos modos quasinormais de campos escalares e sobre a análise da região não-causal. Nossos resultados mostraram a exis-tência de uma relação estreita entre curvas fechadas tipo tempo e a instabilidade do campo escalar. Observamos que, quando os parâmetros do espaço-tempo eram escolhidos de modo a garantir a presença de modos quasinormais estáveis $\omega_{I}<0$, a análise da região não-causal revelava a ausência de curvas fechadas tipo tempo. Por outro lado, se os parâmetros escolhidos permitissem a presença de curvas fechadas tipo tempo nesses espaços-tempos, os modos quasinormais do campo escalar apresentavam modos instáveis $\omega_{I}>0$.

Esse resultado nos levou a dividir os espaços-tempos dos cilindros em rotação em dois grupos: o $grupo\ E$ caracterizado pela escolha de $r_{0}^{+}$ nas Eqs. (\ref{cr4}, \ref{cr8}) é composto pelos espaços-tempos cujo campo escalar é estável; e o $grupo\ I$, caracterizado pela escolha de $r_{0}^{-}$ nas Eqs. (\ref{cr4}, \ref{cr8}), é composto pelos espaços-tempos cujo o campo escalar é instável. Portanto, é exatamente o parâmetro $r_{0}^{\pm}$ que regula a presença das curvas fechadas tipo tempo e, consequentemente, a instabilidade do campo escalar propagando-se nos espaços-tempos dos cilindros cósmicos em rotação. As propriedades dos dois grupos podem ser esquematicamente representadas como

\vspace{0.5cm}

\begin{center}
\begin{tabular}{|c|c|}
\hline
                       &                          \\
Estabilidade do    &      Presença de curvas     \\
campo escalar      &      fechadas tipo tempo\\
                       &\\
\framebox[2cm][c]{$\omega_{I}<0$}       &      \framebox[2cm][c]{$g_{\phi\phi}<0$}     \\
$\Downarrow$        &         $\Downarrow$      \\
&\\
grupo E:            &        grupo I:           \\
\hspace{0.5cm}
 $r_{0}^{+}$ nas Eqs. (\ref{cr4}, \ref{cr8})
\hspace{0.5cm}
                    &
\hspace{0.5cm}
$r_{0}^{-}$ nas Eqs. (\ref{cr4}, \ref{cr8})
\hspace{0.5cm}
                                                \\
&\\
$\Downarrow$        &         $\Downarrow$      \\
Ausência de curvas       &    Instabilidade do     \\
fechadas tipo tempo & campo escalar\\                                        &\\
  \framebox[2cm][c]{$g_{\phi\phi}>0$}   &        \framebox[2cm][c]{$\omega_{I}>0$}     \\
                    &                           \\
\hline
\end{tabular}
\end{center}

\vspace{0.5cm}

Podemos notar que o espaço-tempo da corda em rotação não pertence a nenhum dos dois grupos pois o parâmetro $r_{0}$ é nulo neste caso. Na verdade, ele pode ser interpretado justamente como a fronteira que divide os dois grupos uma vez que ele é o caso limite dos cilindros em rotação. O caso do cilindro estático também corrobora nossas conclusões já que curvas fechadas tipo tempos são ausentes e os modos quasinormais do campo escalar indicam estabilidade.
O único caso que nada podemos concluir e que portanto deve ser analisado por outros métodos é a corda estática. Contudo, suspeitamos que nenhuma instabilidade seja encontrada pois tal espaço-tempo não admite curvas fechadas tipo tempo.
Resumindo nossa análise da estabilidade do campo escalar para essa classe de máquinas do tempo concluímos que é bem provável que a simples presença do viajante do tempo possa desestabilizá-la inviabilizando a viagem.

Finalmente, podemos nos perguntar se essa relação entre a presença de curvas fechadas tipo tempo e as instabilidades de campos de matéria pode ser entendida como uma manifestação de instabilidades do próprio espaço-tempo. Dito de outro modo, será que uma possível instabilidade do espaço-tempo devido a presença de curvas fechadas tipo tempo pode ser considerada a origem dessas instabilidades dos campos de matéria?

Como acreditamos em uma resposta afirmativa para essa questão propomos uma generalização da nossa  última conclusão apresentando a seguinte conjectura:
\begin{quote}

\emph{A presença de curvas fechadas tipo tempo em espaços-tempos do tipo $\mathcal{C}_{t}$ implicará na instabilidade linear clássica desta classe de geometrias.}

\end{quote}
Essa conjectura tem algum suporte em outro resultado anteriormente obtido tal como a estabilidade do buraco negro de Schwarzschild-G\"{o}del no regime de baixas rotações \cite{Abdalla}.


Os resultados principais desta tese são apresentados no artigo \cite{Alan}.

\chapter{Desenvolvimentos futuros}

Na tentativa de fornecer mais resultados que confirmem nossa conjectura, apresentamos alguns tópicos que serão alvo de investigações futuras:
\begin{enumerate}

\item Perturbações eletromagnética, gravitacional, Dirac e Weyl em espaços-tempos $\mathcal{C}_{t}$ que admitem curvas fechadas tipo tempo e o cálculo dos modos quasinormais associados a essas perturbações.

\item Perturbações escalar, eletromagnética e gravitacional e seus modos quasinormais em espaços-tempos axialmente simétrico e estacionários que admitem curvas fechadas tipo tempo.

\item Perturbações escalar, eletromagnética e gravitacional, Dirac e Weyl em espaços-tempos do tipo buraco de minhoca que admitem curvas fechadas tipo tempo.
\end{enumerate}

\appendix

\chapter{Soluções que admitem curvas fechadas tipo tempo}
\label{solucao_CTC}

Neste apêndice nos devotaremos à apresentação de algumas outras soluções das equações de Einstein que admitem curvas fechadas tipo tempo. A condição de existência de uma região não-causal é investigada em cada caso.

\section*{Solução de Van Stockum}
A primeira solução das equações de Einstein que admitia curvas fechadas tipo tempo foi obtida por Van Stockum \cite{Van} em 1937. Esta solução descreve um cilindro de poeira infinito em rotação que admite curvas fechadas tipo tempo quando sua rotação é alta. Uma análise bastante detalhada desta solução foi realizada por Tipler \cite{Tipler}. Neste trabalho ele discute a possibilidade de se usar essa solução para a construção uma máquina do tempo realística com conteúdo material bem simples. No regime de altas rotações $aR>1/2$ o elemento de linha do cilindro é dado por \cite{Tipler}
\begin{eqnarray}
\label{van1}
ds^2=H(dr^2+dz^2)+L d\phi^2+F dt^2+ 2Md\phi dt\quad,
\end{eqnarray}
onde as funções $F$, $M$, $L$ e $H$ são
\begin{eqnarray}
\label{van2}
H&=&e^{-a^2R^2}\left(\frac{r}{R}\right)^{-2a^2R^2}\quad,\\
\nonumber\\
L&=&\frac{Rr\sin(3\beta+\gamma)}{2\sin(2\beta)\cos(\beta)}\quad,\\
\nonumber\\
M&=&\frac{r\sin(\beta+\gamma)}{\sin(2\beta)}\quad,\\
\nonumber\\
F&=&\frac{r\sin(\beta-\gamma)}{R\sin(\beta)}\quad.
\end{eqnarray}
com
\begin{eqnarray}
\label{van3}
\beta=\arctan\left(\sqrt{4a^2R^2-1}\right)\quad,\\
\nonumber\\
\gamma=\left(\sqrt{4a^2R^2-1}\right)\ln\left(\frac{r}{R}\right)\quad.
\end{eqnarray}
Inspecionando a função $L$, que descreve a componente $g_{\phi\phi}$, vemos que uma região não-causal surge quando
\begin{eqnarray}
\label{van4}
\sin(3\beta+\gamma)=0\quad, \quad \Longrightarrow\quad \gamma=n\pi-3\beta\\
\nonumber\\
r_{nc}=R\ e^{\frac{n\pi-3\beta}{\sqrt{4a^2R^2-1}}}\quad.
\end{eqnarray}
Neste caso, teremos um número infinito de regiões não-causais permitindo a existência de curvas fechadas tipo tempo em diferentes regiões do espaço-tempo.

\section*{Solução de Gödel}

Em 1949, Gödel \cite{Goedel} apresenta uma nova solução da equação de Einstein que permitia a presença de curvas fechadas tipo tempo. Esta solução descreve um universo homogêneo com constante cosmológica em rotação cujo elemento de linha, em coordenadas cilíndricas, é dado por
\begin{eqnarray}
\label{goedel1}
ds^2=\frac{2}{\omega^2}\left[\frac{}{}-dt^2+dr^2-\left(\sinh(r)^4-\sinh(r)^2\right)d\phi^2+\right.\nonumber\ \ \ \ \\
+\left. 2\sqrt{2}\ \sinh(r)^2dtd\phi\right]+dz^2\ ,
\end{eqnarray}
onde a constante $\omega$ é sempre positiva e se relaciona com a constante cosmológica como
\begin{eqnarray}
\label{goedel2}
\omega^2=-\Lambda\quad.
\end{eqnarray}
Neste espaço-tempo a região não-causal é dada por
\begin{eqnarray}
\label{goedel3}
r>\ln\left(1+\sqrt{2}\right)\quad.
\end{eqnarray}
Portanto, podemos afirmar que a solução de Gödel admite ao menos uma curva fechada tipo tempo do tipo $S_{c}$ circular.

\section*{Solução de Soares}

Uma outra classe de soluções cosmológicas que admitem curvas fechadas tipo tempo foi obtida por Soares \cite{Damiao}. Esta classe de soluções descreve diferentes universos inomogêneos em rotação preenchidos com poeira e campo eletromagnético. A sub-classe destas soluções que mais nos interessa admite, além de curvas fechadas tipo tempo, geodésicas fechadas tipo tempo. Não discutiremos aqui o comportamento destas geodésicas pois isto é feito com certo detalhe em \cite{Damiao}. Contudo, queremos conhecer a condição de existência da região não-causal para essa sub-classe. O elemento de linha deste espaço-tempo pode ser escrito como
\begin{eqnarray}
\label{soares1}
ds^2=A_{0}^2\left[\frac{}{}dt-2\lambda_{1}\cos(\theta)d\phi\right]^2-dr^2-B^2(r)\left(d\theta^2+\sin(\theta)^2d\phi^2\right)\quad,
\end{eqnarray}
onde a função $B^{2}(r)$ pode ser escolhida entre três opções. Se a constante de integração $\Lambda$ for negativa
\begin{eqnarray}
\label{soares1a}
B^{2}(r)=\frac{-\lambda+\sqrt{\lambda^2+2\alpha_{0}|\Lambda|} \ \sin(2\sqrt{|\Lambda|})(r-C_{0})}{2|\Lambda|}\quad,
\end{eqnarray}
onde $C_{1}$ é uma constante de integração tal que $C_{0}\leq r$. Por outro lado se $\Lambda=0$ então
\begin{eqnarray}
\label{soares1b1}
B^{2}(r)=-\left(\frac{1}{\lambda}\right)\left[(r-C_{0})^2-\frac{\alpha_{0}}{2}\right]\quad.
\end{eqnarray}
Agora, se $\Lambda>0$ e $\lambda=1=2\alpha_{0}\Lambda$ então
\begin{eqnarray}
\label{soares1b}
B^{2}(r)=\frac{\left[1-e^{-2\sqrt{\Lambda}(r-C_{0})}\right]}{2\Lambda}\quad.
\end{eqnarray}
Neste caso, a condição de existência da região não-causal pode ser obtida inspecionando-se a componente $g_{\phi\phi}$ da métrica (\ref{soares1}) resultando na expressão
\begin{eqnarray}
\label{soares2}
4A_{0}^2\lambda_{1}^2\cos(\theta)^2-B^2(r)\sin(\theta)^2>0\quad.
\end{eqnarray}
As constantes $A_{0}$ e $\lambda_{1}$ estão relacionadas com a densidade de carga elétrica, a constante $\alpha_{0}$ se relaciona com a densidade de energia $\rho$ do espaço-tempo e $\lambda$ é proporcional à curvatura da 2-esfera $d\Sigma^2=K^2(\theta)(d\theta^2+\sin(\theta)^2d\phi^2)$. Assim, se a condição (\ref{soares2}) for satisfeita podemos construir uma curva fechada tipo tempo $S_{c}$ circular neste espaço-tempo.

\section*{Solução de Bonnor}

Observando as soluções acima apresentadas vemos que nenhuma delas é assintoticamente plana. Portanto, seria interessante analisarmos uma solução que tenha essa característica. Para tal escolheremos a solução obtida por Bonnor \cite{Bonnor} de um espaço-tempo assintoticamente plano que admite curvas fechadas tipo tempo. Esta solução descreve essencialmente uma régua finita sem massa em rotação. Seu elemento de linha pode ser escrito como
\begin{eqnarray}
\label{bonnor1}
ds^2=-f^{-1}\left[e^{\nu}(dz^2+dr^2)+r^2\left(1-\frac{f^2w^2}{r^2}\right)d\phi^2\right]-\nonumber\\
\nonumber\\
-2fwd\phi dt+fdt^2\quad, \ \ \ \ \
\end{eqnarray}
onde as funções $f$, $w$ e $\nu$ são
\begin{eqnarray}
\label{bonnor2}
f^{-1}&=&\cosh\left[2a\left(\frac{1}{R_{1}}-\frac{1}{R_{2}}\right)\right]\quad,\\
\nonumber\\
w&=&-2a\left(\frac{z-b}{R_{1}}-\frac{z+b}{R_{2}}\right)\quad,\\
\nonumber\\
\nu&=&-a^2\left[r^2\left(\frac{1}{R_{1}^4}-\frac{1}{R_{2}^4}\right)+\frac{R^2-b^2}{b^2R_{1}R_{2}}\right]+\frac{a^2}{b^2}\quad.
\end{eqnarray}
As constantes $a$ e $b$ estão relacionadas com o momento angular por unidade de comprimento e o tamanho da régua respectivamente. As funções $R$, $R_{1}$ e $R_{2}$ satisfazem as relações
\begin{eqnarray}
\label{bonnor2a}
R=|\sqrt{z^2+r^2}|\quad, \quad R_{1}=|\sqrt{(z-b)^2+r^2}|\quad, \quad R_{2}=|\sqrt{(z+b)^2+r^2}|\quad.
\end{eqnarray}
A inspeção da componente $g_{\phi\phi}$ da métrica (\ref{bonnor1}) mostra-nos que a região não-causal existirá se
\begin{eqnarray}
\label{bonnor3}
f^2w^2>r^2\quad.
\end{eqnarray}
Deste modo, quando a condição (\ref{bonnor3}) é satisfeita a existência de curvas fechadas tipo tempo $S_{c}$ circulares é garantida.

\chapter{Sobre as forças de maré próximas da região não-causal}
\label{forcamare}


Explorando outros aspectos de espaços-tempos que possuem curvas fechadas tipo-tempo podemos nos perguntar se essa região não-causal produz forças de maré capazes de despedaçar um viajante. Se assim o for, essa região pode até existir, mas não será acessível a um candidato a viajante do tempo. A presença de forças de maré em um espaço-tempo curvo é calculada investigando-se o comportamento de uma congruência de geodésicas tipo-tempo através da equação do desvio geodésico. Seguindo o formalismo apresentado por D'Inverno \cite{Dinverno} a equação do desvio geodésico é dada por
\begin{eqnarray}
\label{fm0}
\frac{D^{2}\xi^{i}}{d\tau^2}=R^{a}_{\phantom{a}bcd}\  v^{b}\ v^{c}\ \xi^{d}
\end{eqnarray}
onde o quadri-vetor $\xi^{i}$  descreve como duas geodésicas vizinhas em uma congruência se relacionam, ou seja, como o espaço-tempo interfere no movimento relativo entre duas partículas seguindo trajetórias geodésicas tipo-tempo que formam uma congruencia. O vetor $v^{b}$ é o vetor tangente a uma geodésica em um ponto $P$ que descreve a quadri-velocidade de uma das partículas do par. Como estamos interessados em descrever o movimento relativo entre as duas partículas vamos conectar um referencial em uma delas e reescrever a Eq.(\ref{fm0}) em termos desse novo referencial.

Sobre o mesmo ponto $P$ da trajetória da partícula introduziremos uma base tetrada ortogonal descrita pelo vetores
\begin{eqnarray}
\label{fm01}
e^{i}_{\alpha}=(e^{i}_{0},e^{i}_{1},e^{i}_{2},e^{i}_{3}),
\end{eqnarray}
onde $e^{i}_{0}$ é um vetor tipo-tempo ortogonal aos outros três vetores que são tipo-espaço. Identificaremos esse vetor tipo-tempo com a quadri-velocidade da partícula de modo que
\begin{eqnarray}
\label{fm02}
e^{i}_{0}=v^{i}\quad.
\end{eqnarray}
A informação contida na parte espacial de $\xi^{i}$ pode ser descrita, através do vetor ortogonal de conexão $\eta^{i}$ que é definido como
\begin{eqnarray}
\label{fm03}
\eta^{i}&=&\left(\delta^{i}_{b}-v^{i}v_{b}\right)\xi^{b}\nonumber\\
\nonumber\\
\eta^{i}&=&h^{i}_{b}\xi^{b}\quad.
\end{eqnarray}
onde $h^{i}_{b}$ tem a função de projetar tensores no espaço 3-dimensional ortogonal a $v^{i}$ sobre qualquer ponto $P$. Projetando $\eta^{i}$ na base tetrada ortogonal teremos
\begin{eqnarray}
\label{fm04}
\eta^{\alpha}=e^{\alpha}_{i}\eta^{i}\quad.
\end{eqnarray}
Portanto, a equação do desvio geodésico representada na base tetrada ortogonal será obtida substituindo as Eqs.(\ref{fm01}, \ref{fm03}, \ref{fm04} ) na Eq.(\ref{fm0}) resultando em
\begin{eqnarray}
\label{fm0a}
\frac{D^{2}\eta^{\alpha}}{d\tau^2}=R^{a}_{\phantom{a}bcd}\ e_{\phantom{\alpha}a}^{\alpha}\ v^{b}\ v^{c}\ e^{d}_{\phantom{d}\alpha}\ \eta^{\alpha}\quad.
\end{eqnarray}
Deste modo, se desejamos estudar o comportamento de forças de maré nas vizinhanças da região não-causal de um espaço-tempo $\mathcal{C}_{t}$ devemos ser capazes de encontrar uma base tetrada ortogonal causalmente bem comportada nessa região. A seguir, apresentaremos uma base bem comportada e discutiremos as dificuldades de encontrá-la nestes casos.
\\
\emph{Uma base tetrada ortogonal}\\
\\
Neste apêndice da tese modificaremos a definição da assinatura da métrica adotada durante toda a tese por conveniência na operação dos cálculos. Seguiremos a definição adotada por Chandrasekhar \cite{Chandra}. Portanto, seguindo o autor anteriormente mencionado, uma métrica cilindricamente simétrica e estacionária será descrita pelo elemento de linha
\begin{eqnarray}
\label{fm1}
ds^2=\left(e^{2\nu}-\omega^2e^{2\psi}\right)\ dt^2 -e^{2\psi} d\phi^2 +2\omega e^{2\psi} dtd\phi-e^{2\mu_{r}} dr^2- e^{2\mu_{z}} dz^2,
\end{eqnarray}
onde $\mu_{r}, \mu_{z}, \nu, \psi, \omega$ dependem apenas da coordenada $r$. Associada à métrica acima podemos construir uma base tetrada formada por quatro vetores reais linearmente independentes sendo um deles tipo-tempo e os outros tipo-espaço. São eles
\begin{eqnarray}
\label{fm2}
e^{i}_{0}&=&\left( \ e^{-\nu} ,\ \omega e^{-\nu} , \ 0\ , \ 0\  \right),\\
\nonumber\\
e^{i}_{1}&=&\left( \ \ 0\ \ ,\ e^{-\psi} , \ 0\ , \ 0 \ \right),\\
\nonumber\\
e^{i}_{2}&=&\left( \ \ 0\ \  , \ 0 \ ,\ e^{-\mu_{r}} , \ 0\ \right),\\
\nonumber\\
e^{i}_{3}&=&\left( \ \ 0\ \ , \ 0\  , \ 0\  ,\ e^{-\mu_{z}} \right),
\end{eqnarray}
com o índice $i=t, \phi, r, z$ respectivamente.

Esses vetores satisfazem as relações de ortogonalidade e normalização
\begin{eqnarray}
\label{fm2a}
e^{i}_{0}e_{1 i}= e^{i}_{0}e_{2 i}=e^{i}_{0}e_{3 i}=0,\qquad e^{i}_{1}e_{0 i}= e^{i}_{1}e_{2 i}=e^{i}_{1}e_{3 i}=0, \nonumber\\
\\
e^{i}_{2}e_{0 i}= e^{i}_{2}e_{1 i}=e^{i}_{2}e_{3 i}=0,\qquad e^{i}_{3}e_{0 i}= e^{i}_{3}e_{1 i}=e^{i}_{3}e_{2 i}=0,\nonumber\\
\nonumber\\
e^{i}_{0}e_{0 i}=1,\qquad e^{i}_{1}e_{1 i}=-1,\qquad e^{i}_{2}e_{2 i}=-1, \qquad e^{i}_{3}e_{3 i}=-1.
\end{eqnarray}
Essa base (\ref{fm2}) pode ser interpretada como uma descrição do espaço-tempo em um referencial localmente inercial. Fazendo uma combinação linear desses vetores podemos construir uma nova base tetrada nula. Esses novos vetores nulos serão
\begin{eqnarray}
\label{fm3}
m^{i}&=&\frac{1}{\sqrt{2}}\left( e^{i}_{0} + e^{i}_{1} \right),\\
\nonumber\\
n^{i}&=&\frac{1}{\sqrt{2}}\left( e^{i}_{0} - e^{i}_{1} \right),\\
\nonumber\\
l^{i}&=&\frac{1}{\sqrt{2}}\left( e^{i}_{2} + i e^{i}_{3} \right),\\
\nonumber\\
\bar{l}^{i}&=&\frac{1}{\sqrt{2}}\left( e^{i}_{2} - i e^{i}_{3}\right).
\end{eqnarray}
Pode-se observar que os vetores $m^{i}$ e $n^{i}$ são vetores reais enquanto $l^{i}$ e $\bar{l}^{i}$ são vetores complexos. Esses vetores são construído tal que as seguintes relações de ortogonalidade e normalização são satisfeitas:
\begin{eqnarray}
\label{fmnula}
m^{i}m_{i}=0,\qquad n^{i}n_{i}=0,\qquad \bar{l}^{i}\bar{l}_{i}=0, \qquad l^{i}l_{i}=0,
\end{eqnarray}
\begin{eqnarray}
\label{fmnorm}
m^{i}l_{i}=m^{i}\bar{l}_{i}=n^{i}l_{i}=n^{i}\bar{l}_{i}=0,\\
\nonumber\\
\label{fmnorm1}
m^{i}n_{i}=1, \qquad\qquad\quad \bar{l}^{i}l_{i}=-1.
\end{eqnarray}
Se prestarmos atenção no vetor $e^{i}_{1}$ e nos vetores nulos composto por ele vemos que, se esse espaço-tempo possui curvas fechadas tipo-tempo, ou seja, se $g_{\phi\phi}(r_{ctc})=0$, esse vetor diverge exatamente quando alcançamos a região não-causal já que
\begin{eqnarray}
\label{fm4}
g_{\phi\phi}=e^{2\psi}\qquad \Longrightarrow \qquad e^{-\psi}=\frac{1}{\sqrt{g_{\phi\phi}}}.
\end{eqnarray}
Assim sendo esse conjunto de vetores que geralmente é utilizado no caso de buracos negros não serve para o caso de espaços-tempos com curvas fechadas tipo-tempo. Portanto precisamos encontrar uma nova base tetrada que também seja regular na região não-causal.

Para sanar essa dificuldade em obter uma base tetrada nula que não seja divergente nas redondezas da região não causal propomos o seguinte procedimento generalizando o resultado obtido por Özdemir \cite{Ozdemir2}.

Reescreveremos o elemento de linha (\ref{fm1}) como sendo
\begin{eqnarray}
\label{fm12}
ds^2=\left(A\ dt- G\ d\phi\right)^2-B^2\ d\phi^2- \bar{H} dr^2- \bar{S} dz^2,
\end{eqnarray}
onde as funções $A, G, B, H, S $ dependem apenas da coordenada $r$. Adaptando a mudança de assinatura à nossa definição de elemento de linha apresentada na seção \ref{cap2geo1a} teremos
\begin{eqnarray}
\label{fm12a}
ds^2=F\ dt^2 - H\ dr^2- S\ dz^2 - L\ d\phi^2 -2M\ dtd\phi .
\end{eqnarray}
Relacionando esses dois elementos de linha podemos estabelecer as seguintes relações entre as componentes das métricas
\begin{eqnarray}
\label{fm22}
\nonumber\\
A=\sqrt{F},\qquad \bar{H}=H, \qquad \bar{S}=S,\\
\nonumber\\
\label{fm23}
G=\frac{M}{\sqrt{F}}, \qquad B^2=\frac{FL+M^2}{F}=\frac{\mathcal{X}}{F}.
\end{eqnarray}
Para construirmos uma base tetrada nula precisamos de quatro vetores tipo-luz, sendo dois deles reais $m^{\mu}$ e $n^{\mu}$ e dois imaginários $l^\mu$ e $\bar{l}^\mu$. Portanto podemos seguir o mesmo procedimento anterirormente descrito tal que eles satisfaçam as condições (\ref{fmnula}), (\ref{fmnorm}) e (\ref{fmnorm1}).

Devido aos argumentos apresentados acima, estamos interessados em vetores que sejam regulares na região não-causal. Portanto eles não devem possuir termos do tipo \begin{eqnarray}
\frac{1}{L(r)}
\end{eqnarray}
que divergiriam quando $r=r_{ctc}$. Neste ponto devemos reconhecer que encontrar um conjunto de vetores que sejam bem comportados na região de nosso interesse não é uma tarefa tão simples. Por isso partirmos de um conjunto de vetores encontrados por Ozdemir \cite{Ozdemir} para os espaços-tempos das cordas cósmicas e generalizamos esses vetores para um espaço-tempo cilindricamente simétrico e estacionário que possuem curvas fechadas tipo-tempo descrito pela métrica \ref{fm12a}. Os vetores que propomos para a construção de uma base tetrada bem comportada na região não-causal são
\begin{eqnarray}
\label{vn1}
m^{\mu}&=&\frac{1}{\sqrt{2}}\left(\ \ \ \ \frac{M}{F}\ \ ,-\frac{M}{\sqrt{FH}},\ \ 0\ \ ,\ \ 0\ \ \right),\\
\nonumber\\
\label{vn2}
n^{\mu}&=&\frac{1}{\sqrt{2}}\left(\ \ \ \ \frac{1}{M}\ \ ,-\frac{\sqrt{F}}{\sqrt{H}M},\ \ 0\ \ ,\ \ 0\ \ \right),\\
\nonumber\\
\label{vn3}
l^{\mu}&=&\frac{1}{\sqrt{2}}\left(\ \ -\frac{iM}{BF},\ \ \ \ \  0 \ \ \ \ \ ,-\frac{1}{\sqrt{S}},-\frac{i}{B}\right),\\
\nonumber\\
\label{vn4}
\bar{l}^{\mu}&=&\frac{1}{\sqrt{2}}\left(\ \ \ \ \frac{iM}{BF},\ \ \ \ \  0 \ \ \ \ \ ,-\frac{1}{\sqrt{S}},\ \ \ \frac{i}{B}\right).
\end{eqnarray}
Os termos dos vetores segue a seguinte sequência de índices $\mu=(t,r,z,\phi)$ respectivamente. Esses vetores satisfazem todas as relações (\ref{fmnula}), (\ref{fmnorm}) e (\ref{fmnorm1}). Inspecionando os vetores vemos que todas as outras componentes da métrica, exceto $g_{\phi\phi}$, devem ser não nulas na região não causal. Essa imposição poderia inviabilizar o uso dessa tetrada em problemas de interesse como buracos negros uma vez que neste caso a função $H$ se anula sobre o horizonte de eventos. E a situação seria ainda pior no caso de buracos negros em rotação que possuem ergosferas pois a função $F$ se anularia antes mesmo de alcançar-se o horizonte de eventos.

Contudo, quando estudamos buracos negros a maioria das bases tetradas também divergem sobre o horizonte de eventos. Esse problema é ``evitado" nesses casos restringindo o intervalo de interesse de análise, passando a caracterizar o espaço-tempo como sendo válido do horizonte de eventos para fora. Esse procedimento é justificado pelo argumento de que nenhuma informação pode sair do horizonte de eventos portanto não afeta o exterior diretamente.
No caso das ergosferas a situação é mais complicada pois ela se localiza fora do horizonte de eventos e portanto é afetada pela sua vizinhança. Contudo não são incomuns os casos de buracos negros que possuem curvas fechadas tipo-tempo e não possuem ergosfera. Uma explicação simplificada desse fato seria que esse arrasto do espaço-tempo altera o comportamento da coordenada angular ao invés da coordenada temporal gerando a região onde curvas fechadas tipo-tempo são possíveis. Levando essa analogia ao extremo poderíamos dizer então que, a região não-causal é uma ``ergosfera" onde toda partícula é forçada a voltar no tempo.
Um exemplo de buraco negro que possui curvas fechadas tipo-tempo e não possui ergosfera é aquele obtido por Lemos \cite{Lemos}. Esse buraco negro já esta sendo analisado em um trabalho em preparação.

Convencidos de que essa base é bem comportada queremos decompor essa base em uma base tetrada localmente inercial para calcularmos as componentes das forças de maré dada pela Eq. (\ref{fm0a}). Esse processo é semelhante àquele apresentado acima. Fazendo o processo inverso somaremos os vetores (\ref{vn1}) e (\ref{vn2}) e obtemos um vetor tipo-tempo dado por
\begin{eqnarray}
e^{\mu}_{0}=\frac{1}{\sqrt{2}}\left(m^{\mu}+n^{\mu}\right).
\end{eqnarray}
Fazendo agora a subtração do mesmos vetores obtemos um vetor tipo-espaço dado por
\begin{eqnarray}
e^{\mu}_{1}=\frac{1}{\sqrt{2}}\left(m^{\mu}-n^{\mu}\right)
\end{eqnarray}
Repetindo o procedimento para os vetores (\ref{vn3}) e (\ref{vn4}) obteremos os outros dois vetores tipo-espaço dados por
\begin{eqnarray}
e^{\mu}_{2}=\frac{1}{\sqrt{2}}\left(l^{\mu}+\bar{l}^{\mu}\right)\\
\nonumber\\
e^{\mu}_{3}=\frac{1}{\sqrt{2}}\left(l^{\mu}-\bar{l}^{\mu}\right)
\end{eqnarray}
para completar a base localmente inercial. Agora já podemos usar essa base para calcular as forças de maré que cordas cósmicas exercem sobre um par de partículas em queda livre. Como apresentado acima identificaremos o vetor tipo-tempo $e^{\mu}_{0}$ com a velocidade da partícula $v^{\mu}$ tal que
\begin{eqnarray}
\label{velo}
v^{\mu}=e^{\mu}_{0},
\end{eqnarray}
e os outros vetores com a hipersuperfície ortogonal a ele. Usando as Eqs. (\ref{Riemann1}) calculamos as componentes não nulas do tensor de Riemann e vemos que todas elas são nulas para o exterior das cordas e dos cilindros. Como $R_{abcd}=0$ neste caso podemos afirmar que nenhuma força de maré é exercida sobre as partículas nestes espaços-tempos. Esse até seria o resultado esperado uma vez que a presença da corda ou do cilindro não curva o espaço-tempo ao seu redor. Contudo, nosso trabalho não é perdido por dois motivos. Primeiro, essa base ortogonal pode ser usada para analisar o comportamento de forças de maré em outros espaços-tempos $\mathcal{C}_{t}$ que admitem curvas fechadas tipo tempo cujo o tensor de Riemann é não-nulo. E segundo, a base tetrada nula encontrada pode ser usada para calcularmos perturbações em soluções que possuem curvas fechadas tipo-tempo usando o formalismo de Newman-Penrose \cite{Newman}.

\chapter{Um exemplo do cálculo dos MQN's usando o {\it Maple}}
\label{Maple}
\noindent
Neste apêndice apresentaremos, como exemplo, a maneira pela qual obtivemos os modos quasinormais do campos escalar se progando no espaço-tempo do cilindro cósmico estático usando o programa {\it Maple}.

O ponto central do cálculo é usar o comando $fsolve$ para resolver a Eq.(\ref{mqncc6}) varrendo o plano complexo em busca dos modos quasinormais.
\\
\begin{center}
******************************************************************\\
MQM's  para a corda cósmica estática                                     \\
******************************************************************\\
\end{center}
$> restart:$\\
$> with(DEtools):$\\
$> with(plots):$\\
\begin{center}
\newpage
******************************************************************\\
SOLUÇÃO INTERNA \\
******************************************************************\\
\end{center}
$> eq_{in}:= diff(R(r),r,r)+1/(r0*tan(r/r0))*diff(R(r),r)+R(r)*((\omega^2-k^2-\beta^2)-m^2/(r0^2*sin(r/r0)^2)):$\\
$> sol_{in}:=dsolve(eq_in,R(r)):$\\
\begin{center}
******************************************************************\\
COMO QUEREMOS REGULARIDADE NA ORIGEM  C1=0      \\
******************************************************************\\
\end{center}
$> R_{in}:=C2*LegendreP(1/2*(1+(-4*k^2-4*\beta^2+4*\omega^2)*r0^2)^{(1/2)}-1/2,m,cos(r/r0)):$\\
$> derivR_{in}:=diff(R_{in},r):$\\
$> cond_{in}:=subs(r=rs,derivR_{in}/R_{in}):$\\
\begin{center}
******************************************************************\\
SOLUÇÃO EXTERNA \\
******************************************************************\\
\end{center}
$> eq_{ext}:=r^2*diff(R(r),r,r)+r*diff(R(r),r)+R(r)*(r^2*(\omega^2-k^2-\beta^2)-m^2/(1-4*\mu)^2):$\\
$> sol_{ext}:=dsolve(eq_{ext},R(r)):$\\
\begin{center}
\newpage
******************************************************************\\
COMO QUEREMOS ONDA VIAJANTE NO INFINITO USAMOS FUNÇÕES DE HANKEL APENAS OUTGOING   \\
******************************************************************\\
\end{center}
\begin{center}
******************************************************************\\
$\omega_{R} : RAMO\ POSITIVO$ \\
******************************************************************\\
\end{center}
$> R_{ext}:=C2*HankelH1(((-1+4*\mu)^2)^{(1/2)}/(-1+4*\mu)^{2}*m,(\omega^2-k^2-\beta^2)^{(1/2)}*r):$\\
$> derivR_{ext}:=diff(R_{ext},r):$\\
$> cond_{ext}:=subs(r=rb,derivR_{ext}/R_{ext}):$\\
\begin{center}
******************************************************************\\
$\omega_{R} : RAMO\ NEGATIVO$ \\
******************************************************************\\
\end{center}
$> R_{ext1}:=C2*HankelH2(((-1+4*\mu)^2)^{(1/2)}/(-1+4*\mu)^2 *m,(\omega^2-k^2-\beta^2)^{(1/2)}*r):$\\
$> derivR_{ext1}:=diff(R_ext1,r):$\\
$> cond_{ext1}:=subs(r=rb,derivR_{ext1}/R_{ext1}):$\\
\begin{center}
******************************************************************\\
CALCULANDO OS MQN'S  \\
******************************************************************\\
\end{center}
$> MQN:=cond_{ext}-cond_{in}:$\\
$> MQN1:=cond_{ext1}-cond_{in}:$\\
\begin{center}
\newpage
******************************************************************\\
DADOS DO PROBLEMA ESTUDADO \\
******************************************************************\\
\end{center}
$> \rho:=1/1000;$\\
$> \rho:=evalf(rho);$\\
$> rs:=5;$\\
\begin{center}
\begin{eqnarray}
\rho :&=& 1/1000 \nonumber\\
\rho :&=& 0.001000\nonumber \\
rs :&=& 5\nonumber
\end{eqnarray}
******************************************************************\\
\end{center}
$> r0:=evalf(sqrt(1/(8*\pi*\rho)));$\\
$> \theta1:=rs/r0;$\\
$> \theta_{c}:=evalf(\pi/2);$\\
$> \mu:=1/4*(1-cos(theta1));$\\
$> \mu_{c}:=0.25;$\\
$> rb:=r0*sin(\theta1)/(1-4*\mu);$\\
\begin{center}
\begin{eqnarray}
r0 :&=& 6.307831306\nonumber\\
\theta1 :&=& 0.7926654594\nonumber\\
\theta_{c} :&=& 1.570796327\nonumber\\
\mu :&=& 0.0745126500\nonumber\\
\mu_{c} :&=& 0.25\nonumber\\
rb :&=& 6.400185852\nonumber
\end{eqnarray}
\end{center}
\begin{center}
*****************************************************************\\
PARA CAMPOS NÃO MASSIVO     $\beta=0$  e  k=0  \\
*****************************************************************\\
\end{center}
$> m:=7:$\\
$> k:=0:$\\
$> \beta:=0:$\\
$> w0:=MQN:$\\
$> w01:=MQN1:$\\
\begin{center}
*********************************************************************\\
RAMO NEGATIVO   \\
*********************************************************************\\
\end{center}
$> \omega01:=fsolve(w01,\omega,-1.9-5*I..-0.0001+5*I, complex, fulldigits, \\
maxsols=1);$\\
$> \nu:=m/(1-4*\mu);$\\
\begin{center}
\emph{$\omega01 := -1.759014218 - 0.7623314167*I$}\\
\emph{$         \nu := 9.972228767$}\\
\end{center}
$> z1:=sqrt(\omega01^2-k^2)*r:$\\
$> limH2:=simplify(exp(-I*(z1-(\nu+1/2)*\pi/2))*exp(-I*\omega01*t));$\\
\begin{center}
\emph{ $ limH2 := exp(-1.759014218*I* r + 0.7623314168* r + 16.44973848* I - 0.7623314167* t + 1.759014218* I* t)$}\\
\end{center}
\newpage
\begin{center}
*********************************************************************\\
RAMO POSITIVO  \\
*********************************************************************\\
\end{center}
$> \omega02:=fsolve(w0, \omega, 0.0001-5*I..1.9+5*I, complex, fulldigits, \\
maxsols=1);$\\
$> \nu:=m/(1-4*\mu);$\\
\begin{center}
\emph{$\omega02 := 1.759014218 - 0.7623314167*I$}\\
\emph{ $ \nu := 9.972228767$}\\
\end{center}
$> z2:=sqrt(\omega02^2-k^2)*r:$\\
$> limH1:=simplify(exp(I*(z2-(\nu+1/2)*\pi/2))*exp(-I*\omega02*t));$\\
\begin{center}
\emph{$  limH1 := exp(0.6606589885* I* r + 0.5421910867* r - 0.7853981635* I - 0.5421910867* t - 0.6606589885* I* t)$}\\
\end{center}

\chapter{Expansões Assintóticas das Funções de Bessel e Hankel}
\label{apen1}

As funções de Hankel, Bessel e Neumann são soluções da equação diferencial de segunda ordem
\begin{eqnarray}
r^2\frac{d^2R_{\nu}(kr)}{dr^2}+r \frac{dR_{\nu}(kr)}{dr}+R_{\nu}(kr)\left[k^2r^2-\nu^2\right]=0.
\end{eqnarray}
Muitas vezes o comportamento assintótico dessas soluções são fundamentais na solução de problemas de contorno. Para detalhes sobre as funções citadas ver \cite{Arfken}.
Para descrever o comportamento dessas funções para grandes valores de $z=kr$ definimos as seguintes funções
\begin{eqnarray}
P_{\nu}(z)&\sim& 1-\frac{(\mu-1)(\mu-9)}{2!(8z)^2}+\frac{(\mu-1)(\mu-9)(\mu-25)(\mu-49)}{4!(8z)^4} ...\ \\
\nonumber\\
Q_{\nu}(z)&\sim& \frac{(\mu-1)}{1!(8z)}-\frac{(\mu-1)(\mu-9)(\mu-25)}{3!(8z)^3} ...
\end{eqnarray}
com o parâmetro $\mu=4\nu^2$. Essas duas funções tendem a $P_{\nu}\sim 1$ e $Q_{\nu}\sim 0$ quando $z\rightarrow\infty$.
O comportamento para $z$ grande da função de Bessel é
\begin{eqnarray}
J_{\nu}(z)=\sqrt{\frac{2}{\pi z}}\left[P_{\nu}(z)\cos\left(z-\left(\nu+\frac{1}{2}\right)\frac{\pi}{2}\right)-\right.\nonumber\\
\nonumber\\
\left.-\ Q_{\nu}(z)\sin\left(z-\left(\nu+\frac{1}{2}\right)\frac{\pi}{2}\right)\right],\\
\nonumber\\
-\pi<\textrm{arg}(z)<\pi.\nonumber
\end{eqnarray}
O comportamento para $z$ grande da função de Neumann é
\begin{eqnarray}
N_{\nu}(z)=\sqrt{\frac{2}{\pi z}}\left[P_{\nu}(z)\sin\left(z-\left(\nu+\frac{1}{2}\right)\frac{\pi}{2}\right)+\right.\nonumber\\
\nonumber\\
\left.+\ Q_{\nu}(z)\cos\left(z-\left(\nu+\frac{1}{2}\right)\frac{\pi}{2}\right)\right],\\
\nonumber\\
-\pi<\textrm{arg}(z)<\pi.\nonumber
\end{eqnarray}
O comportamento para $z$ grande da primeira função de Hankel é
\begin{eqnarray}
H_{\nu}^{1}(z)=\sqrt{\frac{2}{\pi z}}\ e^{i\left[z-\left(\nu+\frac{1}{2}\right)\frac{\pi}{2}\right]}[P_{\nu}(z)+i\ Q_{\nu}(z)],\\
\nonumber\\
-\pi<\textrm{arg}(z)<2\pi.\nonumber
\end{eqnarray}
O comportamento para $z$ grande da segunda função de Hankel é
\begin{eqnarray}
H_{\nu}^{2}(z)=\sqrt{\frac{2}{\pi z}}\ e^{-i\left[z-\left(\nu+\frac{1}{2}\right)\frac{\pi}{2}\right]}[P_{\nu}(z)-i\ Q_{\nu}(z)],\\
\nonumber\\
-2\pi<\textrm{arg}(z)<\pi.\nonumber
\end{eqnarray}
Portanto se uma onda $\Psi$ é dada por
\begin{eqnarray}
\psi(t,z)=e^{-i\omega t}H_{\nu}^{1,2}(z)\quad,
\end{eqnarray}
então seu comportamento, para $z\rightarrow\infty$, será de uma onda cilíndrica viajante representada por
\begin{eqnarray}
\psi(t,z)=\sqrt{\frac{2}{\pi z}}\ e^{-i\left[\omega t \pm \left[z-\left(\nu+\frac{1}{2}\right)\frac{\pi}{2}\right]\right]}\quad.
\end{eqnarray}


\end{document}